\documentclass{jfm}
\usepackage{graphicx}
\usepackage{epstopdf, epsfig}
\usepackage{caption,subfigure}
\usepackage{latexsym}
\usepackage{float}

\shorttitle{Deep learning prediction of vortex shedding over a cylinder}
\shortauthor{S. Lee and D. You}

\title{Data-driven prediction of unsteady flow over a circular cylinder using deep learning}

\author{Sangseung Lee
  \and Donghyun You\corresp{\email{dhyou@postech.ac.kr}}}

\affiliation{Department of Mechanical Engineering, 
Pohang University of Science and Technology,\\ 77 Cheongam-ro, Nam-gu, Pohang, Gyeongbuk 37673, Republic of Korea}

\begin{document}

\maketitle

\begin{abstract}
Unsteady flow fields over a circular cylinder are trained and predicted using four different deep learning networks: generative adversarial networks with and without consideration of conservation laws and convolutional neural networks with and without consideration of conservation laws.
Flow fields at future occasions are predicted based on information of flow fields at previous occasions.
Predictions of deep learning networks are conducted on flow fields at Reynolds numbers that were not informed during training.
Physical loss functions are proposed to explicitly impose information of conservation of mass and momentum to deep learning networks.
An adversarial training is applied to extract features of flow dynamics in an unsupervised manner.
Effects of the proposed physical loss functions and adversarial training on predicted results are analyzed.
Captured and missed flow physics from predictions are also analyzed.
Predicted flow fields using deep learning networks are in favorable agreement with flow fields computed by numerical simulations.
\end{abstract}


\section{Introduction}\label{sec:introduction}
Observation of fluid flow in nature, laboratory experiments, and numerical simulations has provided evidence of existence of flow features and certain, but often complex, ordinance. 
For example, in nature, Kelvin-Helmholtz waves in the cloud~\citep{dalin2010case}, von Karman vortices in ocean flow around an island~\citep{berger1972periodic}, and swirling great red spot on the Jupiter~\citep{marcus1988numerical} are flow structures that can be classified as a certain type of vortical motions produced by distinct combination of boundary conditions and initial conditions for the governing first principles.
Similar observation also has been reported in laboratory experiments and numerical simulations~\citep{freymuth1966transition,ruderich1986experimental,wu2009direct,babucke2008dns}. 
The existence of distinct and dominant flow features have also been widely investigated by reduced order models (ROMs) using mathematical decomposition techniques such as the proper orthogonal decomposition (POD) method~\citep{sirovich1987turbulence}, the dynamic mode decomposition (DMD) method~\citep{schmid2010dynamic}, and the Koopman operator method~\citep{mezic2013analysis}.

For the sake of the existence of distinct or dominant flow features, animals such as insects, birds, and fish are reported to control their bodies adequately to better adapt the fluid dynamic environment and to improve the aero- or hydro-dynamic performance and efficiency~\citep{wu2011fish,yonehara2016flight}. This suggests a possibility that they empirically learn dominant fluid motions as well as the nonlinear correlation of fluid motions and are able to estimate  future flow based on experienced flow in their living environments. 
Such observation in nature motivates us to investigate the feasibility of predicting unsteady fluid motions by learning flow features using neural networks.

Attempts to apply neural networks to problems of fluid flow have been recently conducted by \citet{tracey2015machine}, \citet{zhang2015machine}, and \citet{singh2017machine}, who utilized shallow neural networks for turbulence modeling for Reynolds-averaged Navier-Stokes (RANS) simulations.
\citet{ling2016reynolds} employed deep neural networks to better model the Reynolds stress anisotropy tensor for RANS simulations.
\citet{guo2016convolutional} employed a convolutional neural network (CNN) to predict steady flow fields around bluff objects and reported reasonable prediction of steady flow fields with significantly reduced computational cost than that required for numerical simulations. Similarly, \citet{miyanawala2017efficient} employed a CNN to predict aerodynamic force coefficients of bluff bodies, also with notably reduced computational costs. 
Those previous studies showed high potential of deep learning techniques for enhancing simulation accuracy and reducing computational cost. 

Predicting unsteady flow fields using deep learning involves extracting both spatial and temporal features of input flow field data, which could be considered as learning videos.
Video modeling enables prediction of a future frame of a video based on information of previous video frames by learning spatial and temporal features of the video.
Although deep learning techniques have been reported to generate high quality real-world like images in image modeling areas~\citep{radford2015unsupervised,denton2015deep,van2016conditional,oord2016pixel}, it is known that, for video modeling, deep learning techniques have shown difficulties in generating high quality prediction due to blurriness caused by complexity in the spatial and temporal features in a video~\citep{srivastava2015unsupervised,ranzato2014video,mathieu2015deep,xingjian2015convolutional}.

\citet{mathieu2015deep} proposed a video modeling architecture that utilizes a generative adversarial network (GAN)~\citep{goodfellow2014generative}, which combines a fully convolutional generator model and a discriminator model. The GAN was capable of generating future video frames from input frames at previous times.
The generator model generates images and the discriminator model is employed to discriminate the generated images from real (ground truth) images.
A GAN is adversarially trained so the generator network is trained to delude the discriminator network, and the discriminator network is trained not to be deluded by the generator network.
The Nash equilibrium in the two pronged adversarial training leads the network to extract underlying low-dimensional features in an unsupervised manner, in consequence, good quality images can be generated.
The most notable advantage of using the GAN is that, once it is trained, the network is possible to generate predictions in a larger domain.
This leads to a memory efficient training of videos because the network can predict a frame with a larger size than that in training.
A recurrent neural network (RNN) based architecture lends itself to learn temporal correlation among encoded information in the past and, thereby predicting future frames.
It is also worth noting that, in the present study, application of RNNs proposed by~\cite{srivastava2015unsupervised} and by~\cite{xingjian2015convolutional} has been attempted. However, it has been found that the methods can be practical only for low resolution frames since the number of weight parameters for the RNNs increases as a function of square of  resolution of a frame.
\citet{ranzato2014video} proposed a recurrent convolutional neural network (rCNN), which is also possible to predict a frame with a larger size than that in training.
However, \citet{mathieu2015deep} reported that the GAN improves the capability for predicting future frames on a video dataset of human actions~\citep{soomro2012ucf101} compared to the rCNN, of which predictions are more static for unsteady motions.
  
Prediction of unsteady flow fields using deep learning could offer new opportunities for real-time control and guidance of aero- or hydro-vehicles, fast weather forecast, {\it etc.} As the first step towards prediction of unsteady flow fields using deep learning, in the present study, it is attempted to predict rather simple but canonical unsteady vortex shedding over a circular cylinder using four different deep learning networks: GANs with and without consideration of conservation laws and CNNs with and without consideration of conservation laws.
Consideration of conservation laws is realized as a form of loss functions.
The aim of the present study is to predict unsteady flow fields at Reynolds numbers that were not utilized in the learning process.
This differs from the aim of ROMs, which is to discover and understand low-dimensional representation of flow fields at certain Reynolds numbers by learning them~\citep{liberge2010reduced,bagheri2013koopman}.

The paper is organized as follows: the method for constructing flow field datasets and deep learning methods are explained in sections 2 and 3, respectively. The results obtained using the present deep learning networks are discussed in section 4, followed by concluding remarks in section 5.

\section{Construction of flow field datasets}
\subsection{Numerical simulations}
\begin{figure}
        \centerline{\includegraphics[width = 0.3 \linewidth,trim={0 0cm 0 0cm},clip]{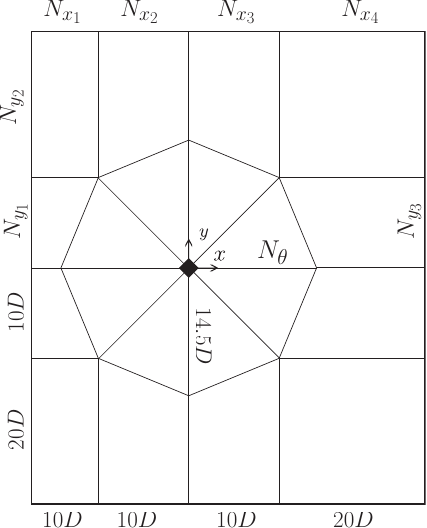}}
        \caption{The computational domain for numerical simulations. $N$ denotes the number of mesh points, where $N_{x_1}=20$,
                 $N_{x_2}=30$, $N_{x_3}=50$, $N_{x_4}=50$ $N_{y_1}=30$, $N_{y_2}=30$, $N_{y_3}=80$, and $N_\theta=150$.
                 The domain size and the number of mesh points in the spanwise direction are $6D$ ($\pi D$ for flow at $Re_D \geq 1000$) and $96$, respectively.}
        \label{fig:com_domain}
\end{figure}

Numerical simulations of flow over a circular cylinder at $Re_{D} = U_\infty D / \nu =$ $150$, $300$, $400$, $500$, $1000$, $3000$, and $3900$, where $U_\infty$, $D$, and $\nu$ are the freestream velocity, cylinder diameter, and kinematic viscosity, respectively, are conducted by solving the incompressible Navier-Stokes equations as follows:
\begin{eqnarray}
\frac{\partial u_{i}}{\partial t} + \frac{\partial u_{i}u_{j}}{\partial u_{j}} = -\frac{1}{\rho}\frac{\partial p}{\partial x_{i}} + \nu \frac{\partial^2 u_{i}}{\partial x_j \partial x_{j}}
\label{eq_momentum_diff}
\end{eqnarray}
and
\begin{eqnarray}
\frac{\partial u_{i}}{\partial x_{i}} = 0,
\label{eq_continuity_diff}
\end{eqnarray}
where $u_{i}$, $p$, and $\rho$ are the velocity, pressure, and density, respectively.
Velocity components and the pressure are non-dimensionalized by $U_{\infty}$ and $\rho U^{2}_{\infty}$, respectively.
A fully implicit fractional-step method is employed for time integration, where all terms in the Navier-Stokes equations are integrated using the Crank-Nicolson method.
Second-order central-difference schemes are employed for spatial discretization and the kinetic energy is conserved by treating face variables as arithmetic means of neighboring cells~\citep{you2008discrete}.
The computational domain consists of a block structured H-grid with an O-grid around the cylinder (figure~\ref{fig:com_domain}).
The computational domain sizes are $50D$ and $60D$ in the streamwise and the cross-flow directions, respectively, where $D$ is the cylinder diameter. 
In the spanwise direction, $6D$ is used for flow at Reynolds numbers less than 1000, while $\pi D$ is used otherwise.
The computational time-step size $\Delta t U_{\infty} / D$ of  $0.005$ is used for all simulations.
The domain size, number of grid points, and time-step sizes are determined from an extensive sensitivity study.

\subsection{Datasets}\label{sec:datasets}
Flow fields in different vortex shedding regimes are calculated for training and testing deep learning networks.
The following flow regimes and Reynolds numbers are considered: two-dimensional vortex shedding regime ($Re_{D}=150$),
three-dimensional wake transition regime ($Re_{D}=300, 400$, and $500$), and 
shear-layer transition regime ($Re_{D}= 1000, 3000$, and $3900$).
Simulation results of flow over a cylinder at each Reynolds number are collected with a time-step interval of $\delta t = 20 \Delta t U_{\infty}/D = 0.1$.
Flow variables $u_1/U_{\infty} (=u/U_{\infty})$, $u_2/U_{\infty}(=v/U_{\infty})$, $u_3/U_{\infty} (=w/U_{\infty})$, and $p/\rho U_{\infty}^{2}$ at each time-step in a square domain of $-1.5D< x < 5.5D$, $-3.5D< y < 3.5D$, $z=0D$ ($7D \times 7D$ sized domain) are interpolated into a uniform grid with $250 \times 250$ cells for all Reynolds number cases.
Thus, a dataset at each Reynolds number consists of flow fields with the size of $250 \times 250\  \hbox{(grid cells)}  \times 4\ \hbox{(flow variables)}$.

\begin{figure}
    \centering
    \subfigure[]{\includegraphics[width = 0.3\linewidth,trim={0 0cm 0 0cm},clip]{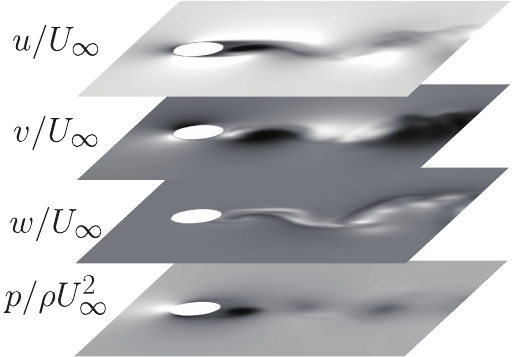} \label{fig:sample-single}}

    \subfigure[]{\includegraphics[width = 0.6\linewidth,trim={0 0cm 0 0cm},clip]{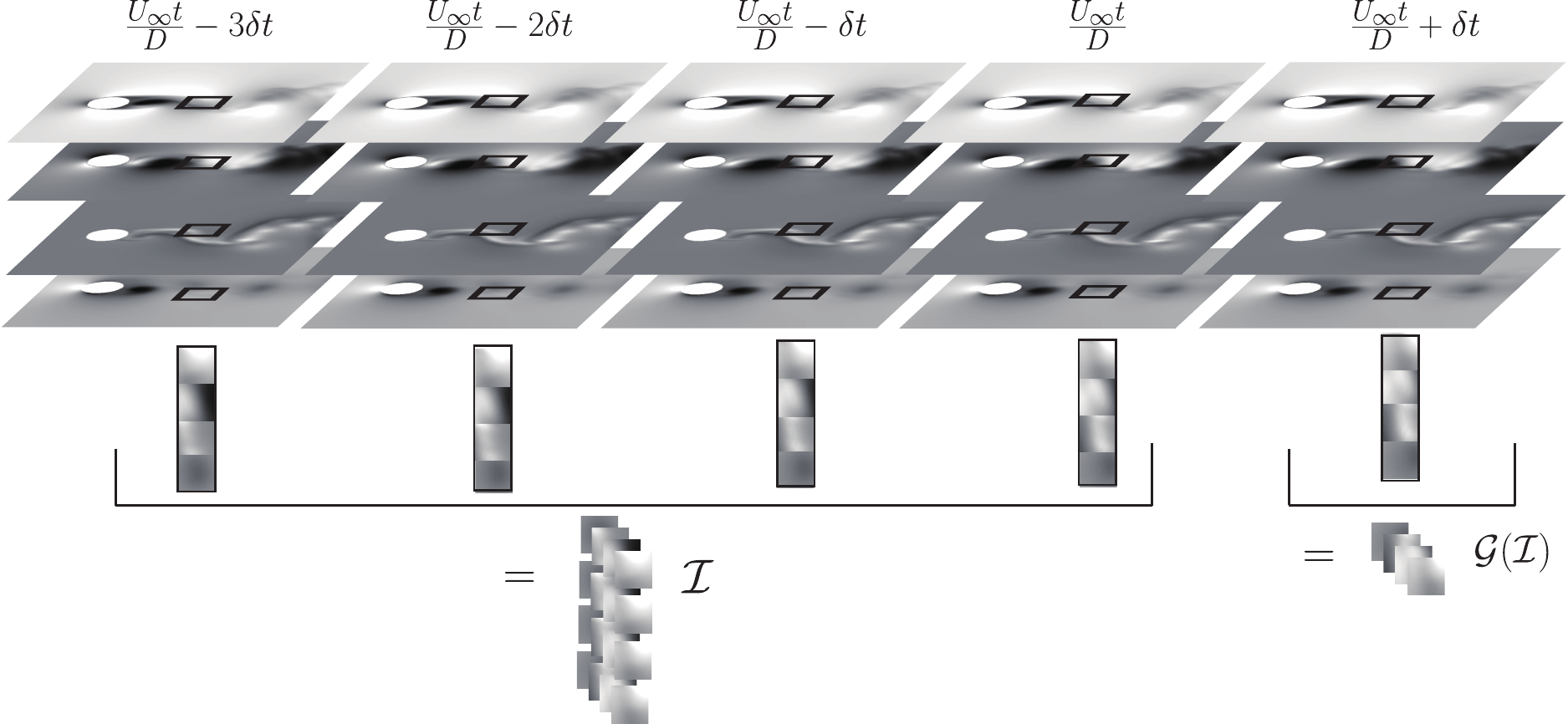} \label{fig:sample-sub}}
    \caption{\subref{fig:sample-single} Instantaneous fields of flow variables $u/U_{\infty}, v/U_{\infty}, w/U_{\infty}$, and $p/\rho U_{\infty}^{2}$ on a $7D \times 7D$ domain with $250 \times 250$ grid cells. \subref{fig:sample-sub} The procedure of subsampling five consecutive flow fields to the input ($\mathcal{I}$) and the ground truth ($\mathcal{G}(\mathcal{I})$) on a $0.896D \times 0.896D$ domain with $32 \times 32$ grid cells.}
    \label{fig:sample}
\end{figure}

The calculated datasets of flow fields are divided in training and test datasets, so that flow fields at Reynolds numbers inside the training dataset is not included in the test dataset.
Flow fields in the training dataset are randomly subsampled in time and space to five consecutive flow fields on a $0.896D \times 0.896D$ domain with $32 \times 32$ grid cells (see figure~\ref{fig:sample}).
The subsampled flow fields contain diverse types of flow such as, freestream flow, wake flow, boundary layer flow, or separating flow.
Therefore, deep learning networks are allowed to learn diverse types of flow.
The first four consecutive sets of flow fields are used as an input ($\mathcal{I}$), while the following set of flow fields is a ground truth flow field ($\mathcal{G}(\mathcal{I})$).
The pair of input and ground truth flow fields form a training sample.
In the present study, total 500,000 training samples are employed for training deep learning networks.
The predictive performance of networks are evaluated on a test dataset, which is composed of interpolated flow fields from numerical simulations on a $7D \times 7D$ domain with $250 \times 250$ grid cells.

\section{Deep learning methodology}\label{sec:deeplearning}
\subsection{Overall procedure of deep learning}
A deep learning network learns a nonlinear mapping of an input tensor and an output tensor.
The nonlinear mapping is comprised of a sequence of tensor operations and nonlinear activations of weight parameters.
The objective of deep learning is to learn appropriate weight parameters that form the most accurate nonlinear mapping of the input tensor and the output tensor that minimizes a loss function.
A loss function evaluates the difference between the estimated output tensor and the ground truth output tensor (the desired output tensor).
Therefore, deep learning is an optimization procedure for determining weight parameters that minimize a loss function.
A deep learning network is trained with the following steps:\\

1. A network estimates an output tensor from a given input through the current state of weight parameters, which is known as feed forward.

2. A loss (scalar value) is evaluated by a loss function of the difference between the estimated output tensor and the ground truth output tensor.

3. Gradients of the loss respect to each weight parameter are calculated through the chain rule of partial derivatives starting from the output tensor, which is known as back propagation.

4. The weight parameters are gradually updated in the negative direction of the gradients of the loss respect to each weight parameter.

5. Step 1 to 4 are repeated until weight parameters (deep learning network) are sufficiently updated.\\

\begin{figure}
        \centerline{\includegraphics[width = 0.4 \linewidth,trim={0 0cm 0 0cm},clip]{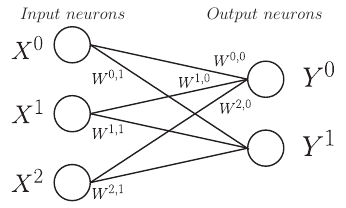}}
        \caption{Illustration of a fully connected layer.}
        \label{fig:FC}
\end{figure}

The present study utilizes two different layers that contain weight parameters: fully connected layers and convolution layers.
An illustration of a fully connected layer is shown in figure~\ref{fig:FC}.
Weight parameters of a fully connected layer are stored in connections ($W$) between layers of input ($X$) and output ($Y$) neurons, where neurons are elementary units in a fully connected layer.
Information inside input neurons is passed to output neurons through a matrix multiplication of the weight parameter matrix and the vector of input neurons as follows:
\begin{eqnarray}
Y^{i} = \sum_{j} W^{j,i} X^{j} + bias,
\end{eqnarray}
where a {\it bias} is a constant, which is also a parameter to be learned.
An output neuron of a fully connected layer collects information from all input neurons with respective weight parameters.
This provides a strength to learn a complex mapping of input and output neurons.
However, as the number of weight parameters is determined as the multiplication of the number of input and output neurons, where the number of neurons is generally in the order of hundreds or thousands, the number of weight parameters easily becomes more than sufficient.
As a result, abundant use of fully connected layers leads to inefficient learning.
Because of the reason, fully connected layers are typically used as a classifier, which collects information and classifies labels, after extracting features using convolution layers.

\begin{figure}
        \centerline{\includegraphics[width = 0.6 \linewidth,trim={0 0cm 0 0cm},clip]{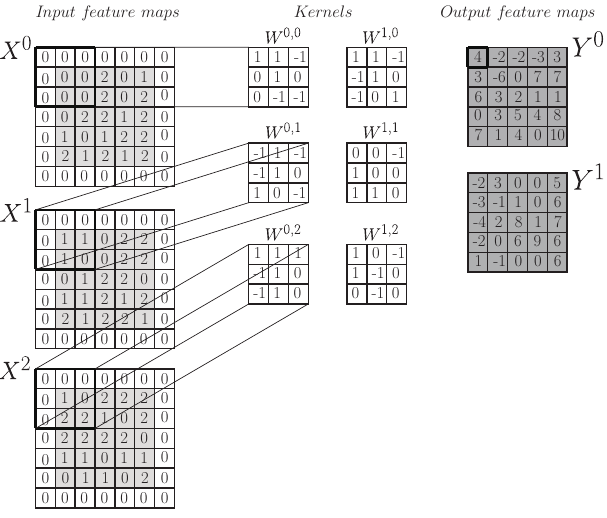}}
        \caption{Illustration of a convolution layer.}
        \label{fig:conv}
\end{figure}

An illustration of a convolution layer is shown in figure~\ref{fig:conv}.
Weight parameters ($W$) of a convolution layer are stored in kernels between input ($X$) and output ($Y$) feature maps, where feature maps are elementary units in a convolution layer.
To maintain the shape of the input after convolution operations, zeros are padded around input feature maps.
The convolution operation with padding is applied to input feature maps using kernels as follows:
\begin{eqnarray}
Y_{i,j}^{n} = \left[ \sum_{k} \sum_{c=0}^{F_{y}-1} \sum_{r=0}^{F_{x}-1} W_{r,c}^{n,k} \underbrace{X_{i+r,j+c}^{k} }_{Pad \ included}\right] + bias,
\end{eqnarray}
where $F_{x} \times F_{y}$ is the size of kernels. 
Weight parameters inside kernels are updated to extract important spatial features inside input feature maps, so an output feature map contains an encoded feature from input feature maps.
Updates of weight parameters could be affected by padding as output values near boundaries of an output feature map are calculated using parts of weight parameters of kernels, whereas values far from boundaries are calculated using all weight parameters of kernels.
However, without padding, the output shape of a feature map of a convolution layer reduces, which indicates loss of information.
Therefore, padding enables a CNN to minimize the loss of information and to be deep by maintaining the shape of feature maps, but as a trade-off it could affect updates of weight parameters.

Convolution layers contain significantly less amount of parameters to update, compared to fully connected layers, which enables efficient learning.
Therefore, convolution layers are typically used for feature extraction.

After each fully connected layer or convolution layer, a nonlinear activation function is usually applied to the output neurons or feature maps to provide nonlinearity to a deep learning network.
The hyperbolic tangent function ($f(x)=\tanh(x)$), the sigmoid function ($f(x)=1/(1+\exp^{-x})$), and the rectified linear unit (ReLU) activation function ($f(x)=\max(0,x)$)~\citep{krizhevsky2012imagenet} are examples of typically applied activation functions. In the present study, these three functions are employed as activation functions (see  section~\ref{conf} for details).

\begin{figure}
        \centerline{\includegraphics[width = 0.3 \linewidth,trim={0 0cm 0 0cm},clip]{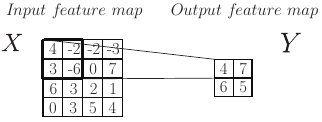}}
        \caption{Illustration of a $2\times 2$ max pooling layer.}
        \label{fig:pool}
\end{figure}

A max pooling layer is also utilized in the present study, which does not contain weight parameters but applies a max filter to non-overlapping subregions of a feature map (see figure~\ref{fig:pool}).
A max pooling layer can be connected to an output feature map of a convolution layer to extract important features.

\subsection{Configurations of deep learning networks}\label{conf}
\begin{figure}
        \centering
        \subfigure[]{\includegraphics[width = 0.8\linewidth,trim={0cm 0cm 0cm 0cm},clip]{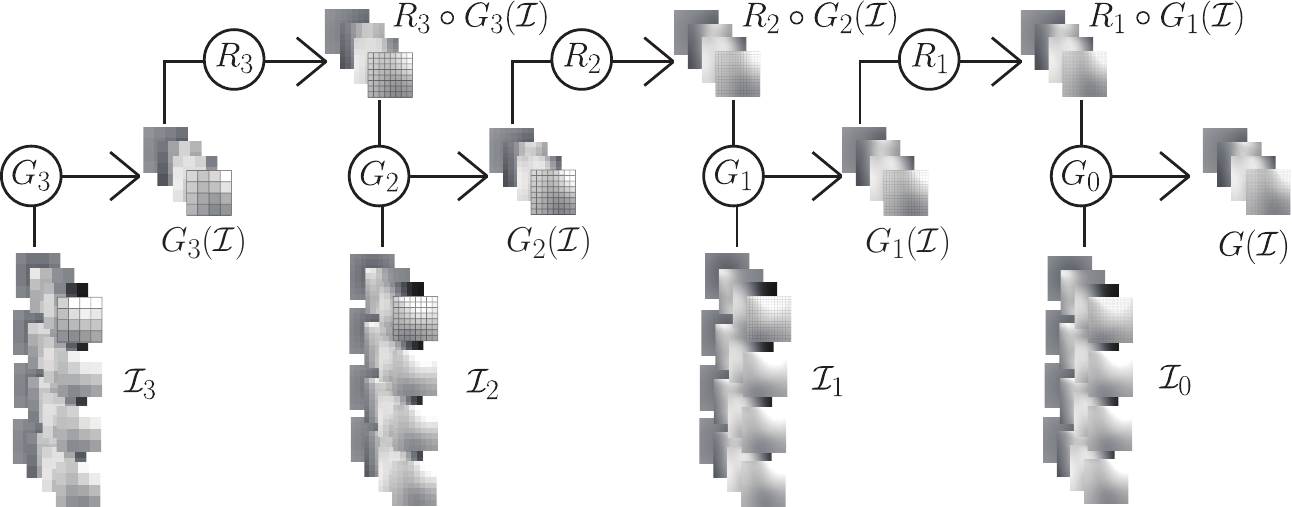}}\hspace{0mm}
        \subfigure[]{\includegraphics[width = 0.8\linewidth,trim={0cm 0cm 0cm 0cm},clip]{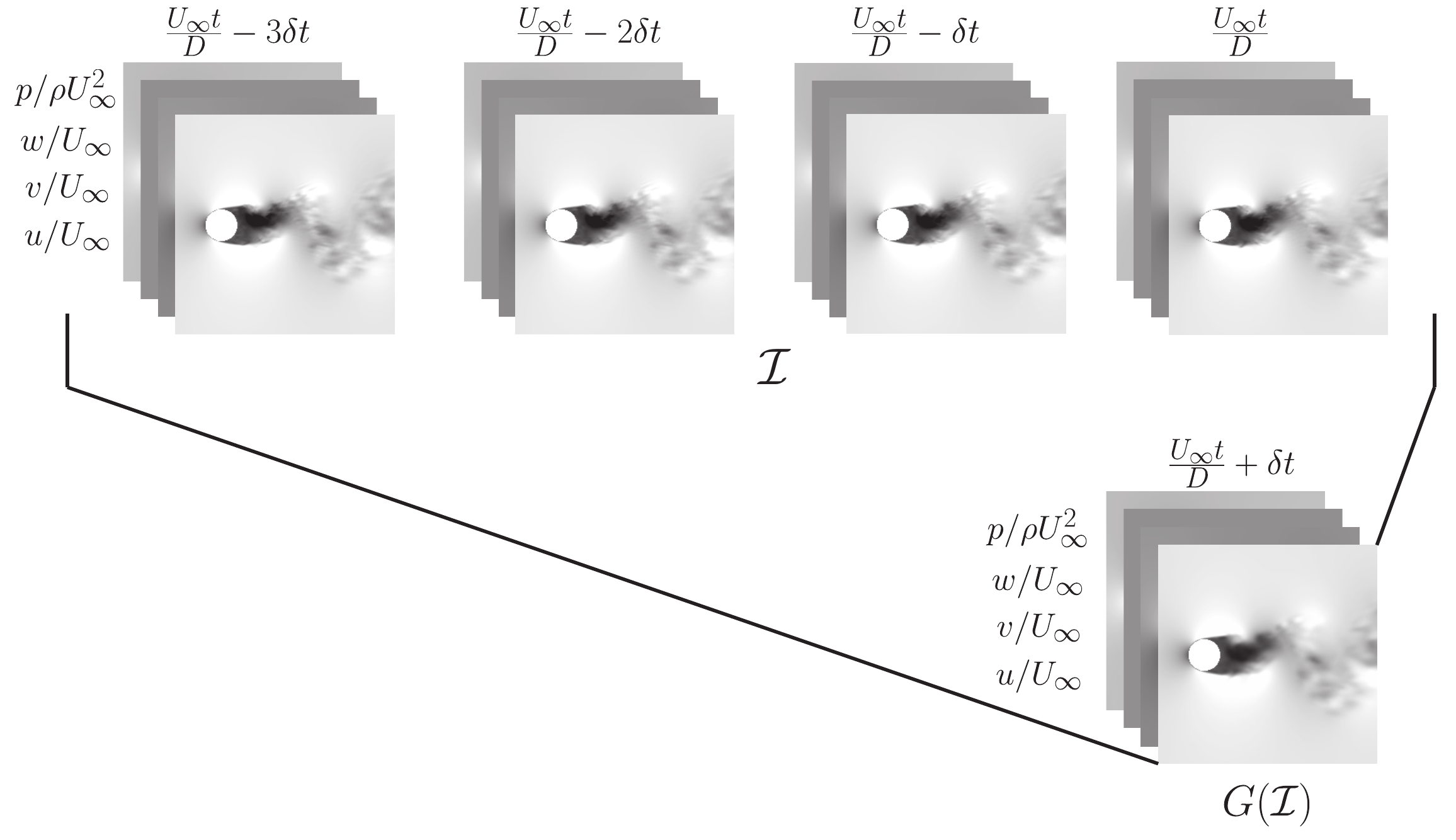}}\hspace{-1mm}
        \caption{(a) Schematic diagram of generator models. $\mathcal{I}$ is the set of input flow fields (see figure~\ref{fig:sample}) and $\mathcal{I}_{k}$ denotes interpolated input flow fields on an identical domain with $1/({2^{k} \times 2^{k}})$ coarser grid resolution. $G_{k}$ indicates a generative CNN which is fed with input $I_{k}$, while $G_{k}(\mathcal{I})$ indicates the set of predicted flow fields from the generative CNN $G_{k}$. $R_{k} \circ ()$ indicates the rescale operator, which upscales the grid size twice in both directions. (b) Example of input flow fields and the corresponding prediction of the flow field on a test data.}
        \label{fig:generator}
\end{figure}

Deep learning networks employed in the present study consists of a generator model that accepts four consecutive sets of flow fields as an input.
Each input set of flow fields is composed of flow variables of $\{u/U_{\infty},v/U_{\infty},w/U_{\infty},p/\rho U_{\infty}^{2}\}$, to take an advantage of learning correlated physical phenomena among flow variables.
The number of consecutive input flow fields is determined by a parameter study. A high number of input flow fields increases memory usage and therefore the learning time. A low number might cause a shortage of input information for the networks. Three cases with $m = 2$, 4, and 6 are trained and tested for unsteady flow fields. No significant benefit in the prediction is found with $m$ beyond 4.
The flow variables are scaled using a linear function to guarantee that all values are in -1 to 1.
This scaling supports the usage of the ReLU activation function by providing nonlinearity to networks and the hyperbolic tangent activation function by bounding predicted values.
Original values of the flow variables are retrieved by an inverse of the linear scaling.
The generator model utilized in this study is composed of a set of multi-scale generative CNNs $\{G_{0}, G_{1}, G_{2},G_{3}\}$ to learn multi-range spatial dependencies of flow structures (see table~\ref{tab:generator} and figure~\ref{fig:generator}). Details of the study for determining network parameters such as numbers of layers and feature maps are summarized in appendix~\ref{appendix:size}.

\begin{table}
        \centering
        \resizebox{0.9\textwidth}{!}{%
        \begin{tabular}{l|l|l}
            \hline \hline
            \multicolumn{3}{c}{Generator model ($G_{3}$ $\rightarrow$ $G_{2}$ $\rightarrow$ $G_{1}$ $\rightarrow$ $G_{0}$)}\\ \hline \hline
                Generative CNN & Numbers of feature maps & Kernel sizes\\ \hline
                $G_{3}$ & 16, 128, 256, 128, 4 & $3\times3$, $3\times3$, $3\times3$, $3\times3$\\
                $G_{2}$ & 20, 128, 256, 128, 4 & $5\times5$, $3\times3$, $3\times3$, $5\times5$ \\
                $G_{1}$ & 20, 128, 256, 512, 256, 128, 4& $5\times5$, $3\times3$, $3\times3$, $3\times3$, $3\times3$, $5\times5$  \\
                $G_{0}$ & 20, 128, 256, 512, 256, 128, 4 & $7\times7$, $5\times5$, $5\times5$, $5\times5$, $5\times5$, $7\times7$ \\  \hline\hline
        \end{tabular}}
        \caption{Configuration of the generator model in GANs and multi-scale CNNs (see figure~\ref{fig:generator} for connections).}
        \label{tab:generator}
\end{table}

During training, a generative CNN $G_{k}$ generates flow field predictions ($G_{k}(\mathcal{I})$) on the $0.896D \times 0.896D$ domain with resolution of $\frac{32}{2^{k}} \times \frac{32}{2^{k}}$ through padded convolution layers.
$G_{k}$ is fed with four consecutive sets of flow fields on the domain with $\frac{32}{2^{k}} \times \frac{32}{2^{k}}$ resolution ($\mathcal{I}_{k}$), which are bilinearly interpolated from the original input sets of flow fields with $32 \times 32$ resolution ($\mathcal{I}$), and a set of upscaled flow fields, which is obtained by $R_{k+1} \circ G_{k+1}(\mathcal{I})$ (see figure~\ref{fig:generator}).
$R_{k+1} \circ ()$ is an upscale operator that bilinearly interpolates a flow field on a domain with resolution of $\frac{32}{2^{k+1}} \times \frac{32}{2^{k+1}}$ to a domain with resolution of $\frac{32}{2^{k}} \times \frac{32}{2^{k}}$.
Note that domain sizes for $\frac{32}{2^{k}} \times \frac{32}{2^{k}}$ and $32 \times 32$ resolution are identical to $0.896D \times 0.896D$, where the size of the corresponding convolution kernel ranges from 3 to 7 (see table~\ref{tab:generator}).
Consequently, $G_{k}$ is possible to learn larger spatial dependencies of flow fields than $G_{k-1}$ by sacrificing resolution. 
As a result, a multi-scale CNN-based generator model enables to learn and predict flow fields with multi-scale flow phenomena.
The last layer of feature maps in each multi-scale CNN is activated with the hyperbolic tangent function to bound the output values, while other feature maps are activated with the ReLU function to provide nonlinearity to networks.

\begin{figure}
        \centerline{\includegraphics[width = 0.8 \linewidth,trim={0 0cm 0 0cm},clip]{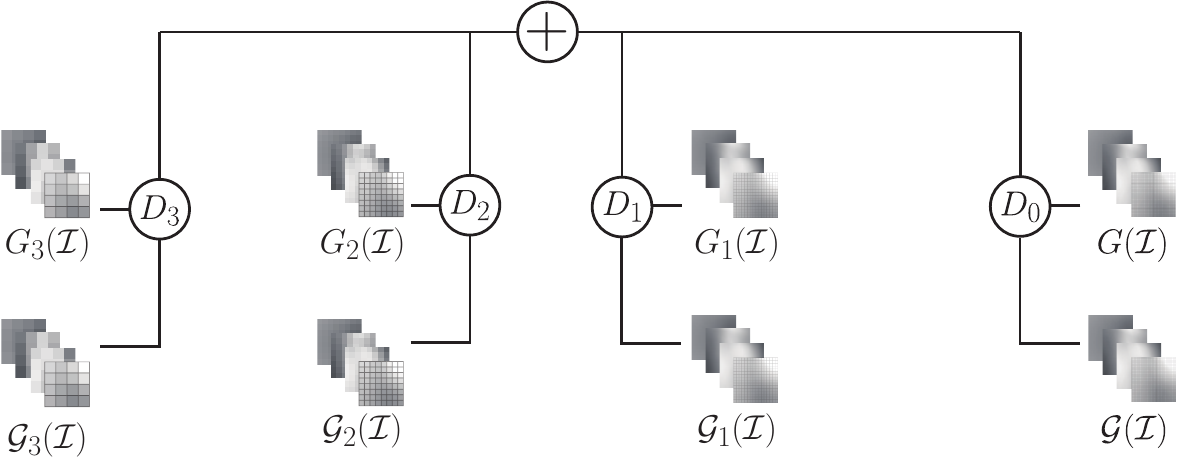}}
        \caption{Schematic diagram of the discriminator model. $D_{k}$ indicates the discriminative network which is fed with $G_{k}(\mathcal{I})$ and $\mathcal{G}_{k}(\mathcal{I})$. $G_{k}(\mathcal{I})$ indicates the set of predicted flow fields from the generative CNN $G_{k}$, while $\mathcal{G}_{k}(\mathcal{I})$ indicates the set of ground truth flow fields.}
        \label{fig:discriminator}
\end{figure}

Let $\mathcal{G}_{k}(\mathcal{I})$ be ground truth flow fields with resized resolution of $\frac{32}{2^{k}} \times \frac{32}{2^{k}}$.
The discriminator model consists of a set of discriminative networks $\{D_{0},D_{1},D_{2},D_{3}\}$ with convolution layers and fully connected layers (see table~\ref{tab:discriminator} and figure~\ref{fig:discriminator}).
A discriminative network $D_{k}$ is fed with inputs of predicted flow fields from the generative CNN ($G_{k}(\mathcal{I})$) and ground truth flow fields ($\mathcal{G}_{k}(\mathcal{I})$).
Convolution layers of a discriminative network extract low-dimensional features or representations of  predicted flow fields and ground truth flow fields through convolution operations.
$2 \times 2$ max pooling, which extracts the maximum values from each equally divided $2\times2$ sized grid on a feature map, is added after convolution layers to pool the most important features. The max pooling layer outputs feature maps with resolution of $\frac{32}{2^{k+1}} \times \frac{32}{2^{k+1}}$.
The pooled features are connected to fully connected layers.
Fully connected layers compare pooled features to classify ground truth flow fields into class 1 and predicted flow fields into class 0.
The output of each discriminative network is a single continuous scalar between 0 and 1, where an output value larger than a threshold (0.5) is classified into class 1 and an output value smaller than the threshold is classified into class 0.
Output neurons of the last fully connected layer of each discriminative network $D_{k}$ are activated using the sigmoid function to bound the output values from 0 to 1, while other output neurons, including feature maps of convolution layers, are activated with the ReLU activation function.

Note that the number of neurons in the first layer of fully connected layers (see table~\ref{tab:discriminator}) is a function of square of the subsampled input resolution ($32 \times 32$); as a result, parameters to learn are increased in the order of square of the subsampled input resolution.
Training could be inefficient or nearly impossible in a larger input domain size with the equivalent resolution (for example, $250 \times 250$ resolution on the domain size of $7D\times 7D$) due to the fully connected layer in the discriminator model depending on computing hardware.
On the other hand, parameters in the generator model (fully convolutional architecture with padded convolutions) do not depend on the size and resolution of the subsampled inputs.
This enables the generator model to predict flow fields in a larger domain size ($7D\times 7D$ domain with $250 \times 250$ resolution) compared to the subsampled input domain size ($0.896D\times 0.896D$ domain with $32 \times 32$ resolution).
\begin{table}
	\centering
	\resizebox{0.9\textwidth}{!}{%
	    \begin{tabular}{c|c|c|c}
                        \hline \hline
			\multicolumn{4}{c}{Discriminator model (convolution layers$\rightarrow$max pooling layer$\rightarrow$fully connected layers)}\\ \hline \hline 
			$D_{0}$ & $D_{1}$ & $D_{2}$ & $D_{3}$\\ \hline\hline
			\multicolumn{4}{c}{Convolution layers (top row: numbers of feature maps, bottom row: kernel sizes)}\\ \hline
			4,128,256,512,128 & 4,128,256,256 & 4,64,128,128 & 4,64 \\
			$7\times7$, $7\times7$, $5\times5$, $5\times5$ & $5\times5$, $5\times5$, $5\times5$ & $3\times3$, $3\times3$, $3\times3$ & $3\times3$ \\ \hline\hline
			\multicolumn{4}{c}{$2\times2$ max pooling layer}\\ \hline\hline
			\multicolumn{4}{c}{Fully connected layers (neuron numbers)}\\ \hline
			$16\times16\times128$,$1024$,$512$,$1$ & $8\times8\times256$,1024,512,1 & $4\times4\times128$,1024,512,1 & $2\times2\times64$,512,256,1 \\ \hline \hline
	    \end{tabular}}
	\caption{Configuration of the discriminator model inside the GAN.}
	\label{tab:discriminator}
\end{table}

The generator model is trained with the Adam optimizer, which is known to efficiently train a network particularly in regression problems~\citep{adam14}.
This optimizer computes individual learning rates, which are updated during training, for different weight parameters in a network.
The maximum learning rate of the parameters in the generator model is limited to $4 \times 10^{-5}$.
However, the Adam optimizer is reported to perform worse than a gradient descent method with a constant learning rate for a classification problem using CNNs~\citep{wilson2017marginal}.
As the discriminator model performs classification using convolutional neural networks, the discriminator model is trained with the gradient descent method along with a constant learning rate of $0.02$.
The same optimization method and learning rate are also utilized in the discriminator model by~\cite{mathieu2015deep}.
Networks are trained up to $6 \times 10^{5}$ iterations with a batch size of 8. Training of networks is observed to be sufficiently converged without overfitting as shown in figure~\ref{fig:iter} in section~\ref{appendix:size}.

\subsection{Conservation principles}\label{sub:cons}
Let $\Omega$ be an arbitrary open, bounded, and connected domain in $\mathbb{R}^{3}$, $\partial \Omega$ be a surface of which an outward unit normal vector can be defined as  $\hat{n}=(n^{1},n^{2},n^{3})$.
Also let $\rho(t,\vec{x})$ be the density, $\vec{u}(t,\vec{x})=(u_{1},u_{2},u_{3})$ be the velocity vector, $p(t,\vec{x})$ be the pressure, and $\tau(t,\vec{x})$ be the shear stress tensor ($\tau_{ij} = \rho \nu \frac{\partial u_{j}}{\partial x_{i}}$) of ground truth flow fields as a function of time $t$ and space $\vec{x} \in \mathbb{R}^{3}$.
Then conservation laws for mass and momentum can be written as follows:
\begin{eqnarray}
\frac{d}{dt} \int_{\Omega} \rho dV = - \int_{\partial \Omega} \rho u_{j} {n}^{j} dS
\end{eqnarray}
and
\begin{eqnarray}
\frac{d}{dt} \int_{\Omega} \rho u_{i} dV = - \int_{\partial \Omega} (\rho u_{i}) u_{j} {n}^{j} dS - \int_{\partial \Omega} (p\delta_{ij}) {n}^{j} dS + \int_{\partial \Omega} \tau_{ji} n^{j} dS,
\end{eqnarray}
where $\delta_{ij}$ is the Kronecker delta.
The present study utilizes subsets of three-dimensional data (two-dimensional slices). Therefore, the domain $\Omega$ becomes a surface in $\mathbb{R}^{2}$ and the surface $\partial \Omega$ becomes a line in $\mathbb{R}^{1}$.
Exact mass and momentum conservation can not be calculated because derivatives in the spanwise direction are not available in two-dimensional slice data.
Instead, conservation principles of mass and momentum in a flow field predicted by deep learning are considered in a form that compares the difference between predicted and ground truth flow fields in a two-dimensional space ($\mathbb{R}^{2}$). 

Extension of the present deep learning methods to three-dimensional volume flow fields is algorithmically straightforward. However,  the increase of the required memory space and the operation counts is significant, making the methods impractical. For example, the memory space and the operation counts for $32 \times 32 \times 32$ sized volume flow fields are estimated to increase two-orders of magnitudes than those required for the $32\times 32$ two-dimensional flow fields.

\subsection{Loss functions}
For a given set of input and ground truth flow fields, the generator model predicts flow fields that minimize a total loss function which is a combination of specific loss functions as follows:
\begin{eqnarray}
\mathcal{L}_{generator} = \frac{1}{N \lambda_{\sum}} \sum_{k=0}^{N-1} \{ \lambda_{l2} \mathcal{L}_{2}^{k} + \lambda_{gdl}\mathcal{L}_{gdl}^{k} + 
\lambda_{phy}(\mathcal{L}_{c}^{k} + \mathcal{L}_{mom}^{k})+ \lambda_{adv} \mathcal{L}_{adv}^{G,k} \},
\label{eqn: L_gen}
\end{eqnarray}
where $N(=4)$ is the number of scales of the multi-scale CNN and $\lambda_{\sum} =  \lambda_{l2} + \lambda_{gdl} + \lambda_{phy}+ \lambda_{adv}$.
Contributions of each loss function can be controlled by tuning coefficients $\lambda_{l2}$, $\lambda_{gdl}$, $\lambda_{phy}$, and $\lambda_{adv}$.

$\mathcal{L}_{2}^{k}$ minimizes the difference between predicted flow fields and ground truth flow fields (see equation~(\ref{eqn:loss_l2})).
$\mathcal{L}_{gdl}^{k}$ is applied to sharpen flow fields by directly penalizing gradient differences between predicted flow fields and ground truth flow fields (see equation~(\ref{eqn:loss_lgdl})).
Loss functions $\mathcal{L}_{2}^{k}$ and $\mathcal{L}_{gdl}^{k}$ provide prior information to networks that predicted flow fields should resemble ground truth flow fields.
These loss functions support networks to learn fluid dynamics that corresponds to the flow field resemblance, by extracting features in a supervised manner.

$\mathcal{L}_{c}$ enables networks to learn mass conservation by minimizing the total absolute sum of differences of mass fluxes in each cell in an $x-y$ plane as defined in equation~(\ref{eqn:loss_lc}).
$\mathcal{L}_{mom}$ enables networks to learn momentum conservation by minimizing the total absolute sum of differences of momentum fluxes due to convection, pressure gradient, and shear stress in each cell in an $x-y$ plane as defined in equation~(\ref{eqn:loss_lmom}).
Loss functions $\mathcal{L}_{c}$ and $\mathcal{L}_{mom}$, which are denoted as physical loss functions, provide explicit prior information of physical conservation laws to networks, and support networks to extract features including physical conservation laws in a supervised manner. Consideration of conservation of kinetic energy can also be realized using a loss function, while it is not included in the present study since the stability of flow fields predicted by the present networks are not affected by the conservation of kinetic energy.

$\mathcal{L}_{adv}^{G}$ is a loss function with purpose to delude the discriminator model to classify generated flow fields as ground truth flow fields (see equation~(\ref{eqn:loss_ladvg})).
The loss function $\mathcal{L}_{adv}^{G}$ provides knowledge in a concealed manner that features of the predicted and the ground truth flow fields should be indistinguishable.
This loss function supports networks to extract features of underlying fluid dynamics in an unsupervised manner.

The loss function of the discriminator model is defined as follows:
\begin{equation}
\mathcal{L}_{discriminator} = \frac{1}{N}\sum_{k=0}^{N-1} \left[ L_{bce}(D_{k}(\mathcal{G}_{k}(\mathcal{I})),1) + L_{bce}(D_{k}(G_{k}(\mathcal{I})),0)\right],
\label{eqn: ladvd}
\end{equation}
where $L_{bce}$ is the binary cross entropy loss function defined as
\begin{equation}
L_{bce}(a,b) = -b \log(a) - (1-b) \log(1-a),
\label{eqn: lbce}
\end{equation}
for scalar values $a$ and $b$ between 0 and 1.
$\mathcal{L}_{discriminator}$ is minimized so that the discriminator model appropriately classifies  ground truth flow fields into class 1 and predicted flow fields into class 0.
The discriminator model learns flow fields in a low-dimensional feature space.

\section{Results}\label{sec:Results}
\subsection{Comparison of deep learning networks}\label{sec:comparison}
Four deep learning networks with different combinations of coefficients for loss functions are discussed in the present section.
Case A employs a GAN with physical loss functions ($\lambda_{l2}=\lambda_{gdl}=1.0$, $\lambda_{phy}=1.0$, and $\lambda_{adv}=0.1$);
Case B employs a GAN without physical loss functions ($\lambda_{l2}=\lambda_{gdl}=1.0$, $\lambda_{phy}=0.0$, and $\lambda_{adv}=0.1$);
Case C employs a multi-scale CNN with physical loss functions ($\lambda_{l2}=\lambda_{gdl}=1.0$, $\lambda_{phy}=1.0$, and $\lambda_{adv}=0.0$);
Case D employs a multi-scale CNN without physical loss functions ($\lambda_{l2}=\lambda_{gdl}=1.0$, $\lambda_{phy}=0.0$, and $\lambda_{adv}=0.0$).
See appendix~\ref{appendix:adv} and~\ref{appendix:phy} for the determination of the weight parameters $\lambda_{adv}$ and $\lambda_{phy}$, respectively.
All deep learning cases (Cases A-D) are trained with flow fields at $Re_{D}=300$ and $500$, which are in the three-dimensional wake transition regime, and tested on flow fields at $Re_{D}=150$ (the two-dimensional vortex shedding regime), $400$ (the same flow regime with training) and $3900$ (the shear-layer transition regime).

\begin{figure}
    \centering
    \subfigure{\includegraphics[width=0.19\linewidth,trim={0.5cm 0.5cm 0.5cm 0.5cm},clip]{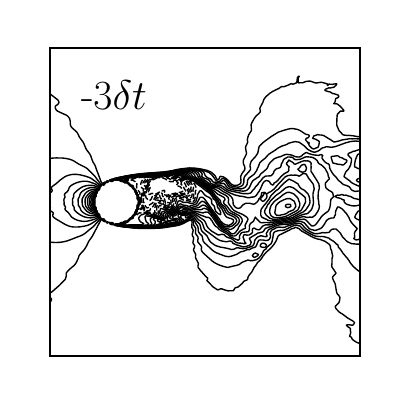}}
    \subfigure{\includegraphics[width=0.19\linewidth,trim={0.5cm 0.5cm 0.5cm 0.5cm},clip]{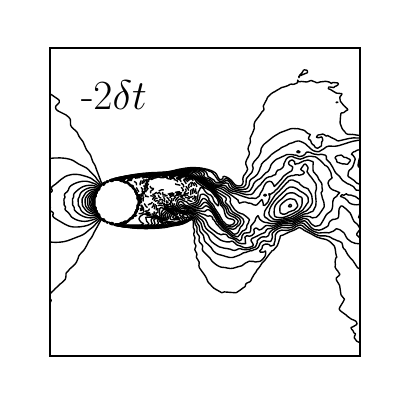}}
    \setcounter{subfigure}{0}%
    \subfigure[]{}\hspace{-1mm}
    \subfigure{\includegraphics[width=0.19\linewidth,trim={0.5cm 0.5cm 0.5cm 0.5cm},clip]{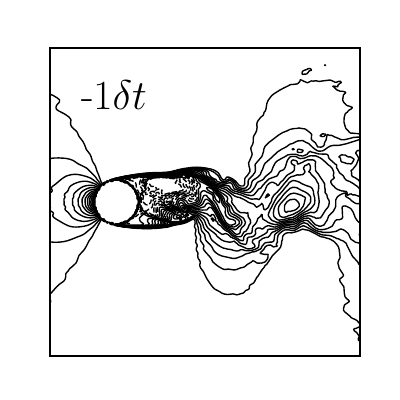}}
    \subfigure{\includegraphics[width=0.19\linewidth,trim={0.5cm 0.5cm 0.5cm 0.5cm},clip]{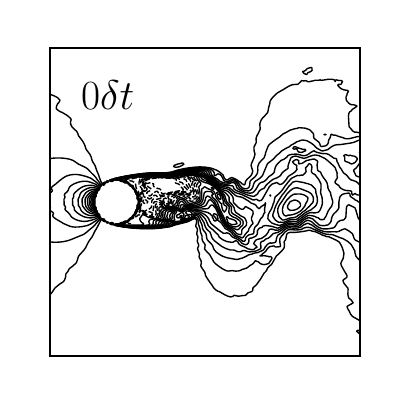}}

    \subfigure{\includegraphics[width=0.19\linewidth,trim={0.5cm 0.5cm 0.5cm 0.5cm},clip]{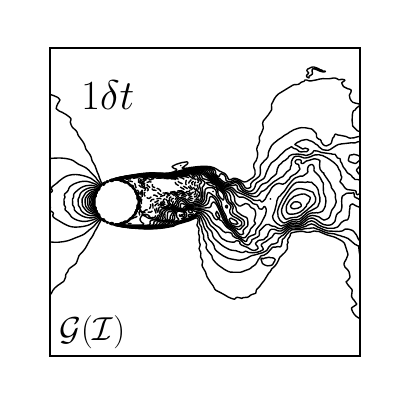}}
    \subfigure{\includegraphics[width=0.19\linewidth,trim={0.5cm 0.5cm 0.5cm 0.5cm},clip]{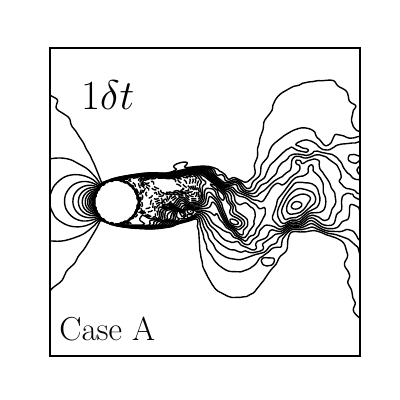}}
    \setcounter{subfigure}{1}%
    \subfigure[]{\includegraphics[width=0.19\linewidth,trim={0.5cm 0.5cm 0.5cm 0.5cm},clip]{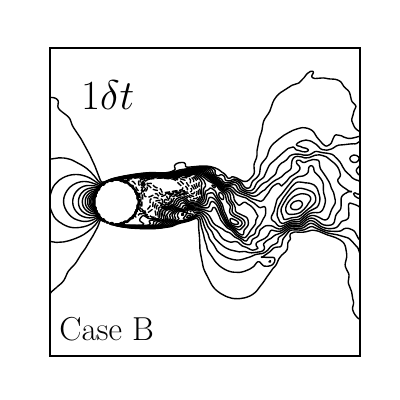}}
    \subfigure{\includegraphics[width=0.19\linewidth,trim={0.5cm 0.5cm 0.5cm 0.5cm},clip]{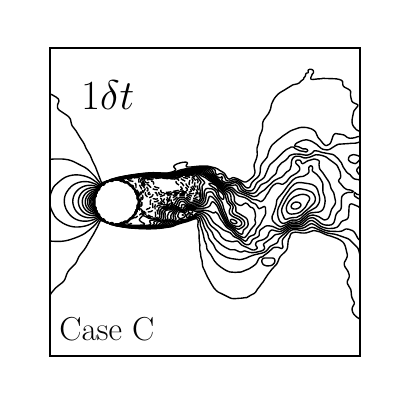}}
    \subfigure{\includegraphics[width=0.19\linewidth,trim={0.5cm 0.5cm 0.5cm 0.5cm},clip]{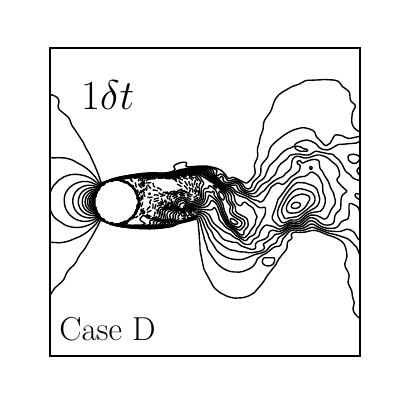}}

    \subfigure{\includegraphics[width=0.19\linewidth,trim={0.5cm 0.5cm 0.5cm 0.5cm},clip]{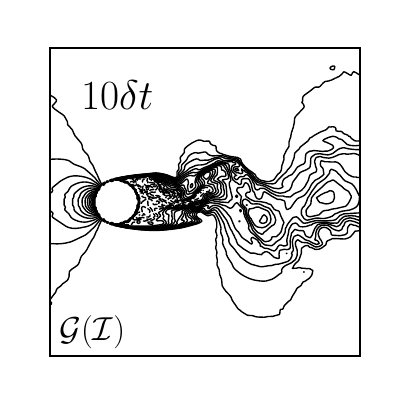}}
    \subfigure{\includegraphics[width=0.19\linewidth,trim={0.5cm 0.5cm 0.5cm 0.5cm},clip]{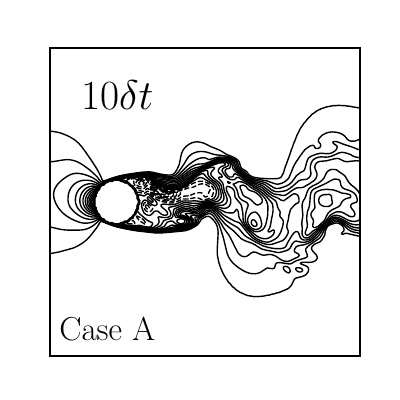}}
    \setcounter{subfigure}{2}%
    \subfigure[]{\includegraphics[width=0.19\linewidth,trim={0.5cm 0.5cm 0.5cm 0.5cm},clip]{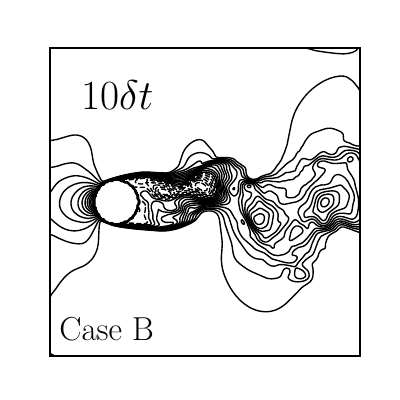}}
    \subfigure{\includegraphics[width=0.19\linewidth,trim={0.5cm 0.5cm 0.5cm 0.5cm},clip]{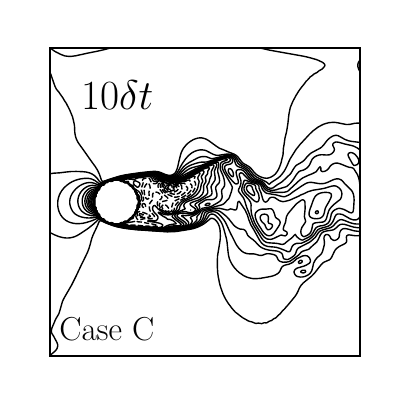}}
    \subfigure{\includegraphics[width=0.19\linewidth,trim={0.5cm 0.5cm 0.5cm 0.5cm},clip]{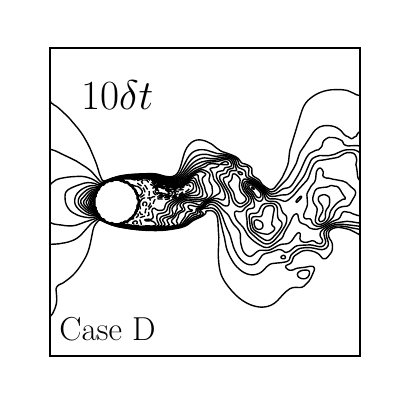}}
    \caption{
    Comparison of the streamwise velocity ($u/ U_{\infty}$) at $Re_{D}=3900$ predicted in Cases A-D.
    (a) Input set; after (b) a single prediction step ($1\delta t$), and (c) 9 more recursive prediction steps ($10\delta t$).
    20 contour levels from -0.5 to 1.0 are shown.
     Solid lines and dashed lines indicate positive and negative contour levels, respectively.
    }
 \label{fig:u-comparison}
\end{figure}

Predicted flow fields at $Re_D=3900$ from Cases A-D are shown in figure~\ref{fig:u-comparison}.
Flow fields after time steps larger than $\delta t$ are predicted recursively by utilizing flow fields predicted prior time-steps as parts of the input.
Flow fields predicted after a single time-step ($1 \delta t$) are found to agree well with ground truth flow fields for all deep learning cases, even though the trained networks have not seen such small-scale flow structures at a higher Reynolds number.
Note that the time step size for network prediction $\delta t$ corresponds to 20 times of the simulation time-step size. Differences between the predicted and the ground truth flow fields increase as the number of the recursive step increases because errors from the previous predictions are accumulated to the next time-step prediction.
Particularly, dissipation of small-scale flow-structures in the wake region is observed, while large-scale vortical motions characterizing the Karman vortex shedding are favorably predicted.

\begin{figure}
    \centering
    \subfigure{\includegraphics[width=0.23\linewidth,trim={0.5cm 0.5cm 0.5cm 0.5cm},clip]{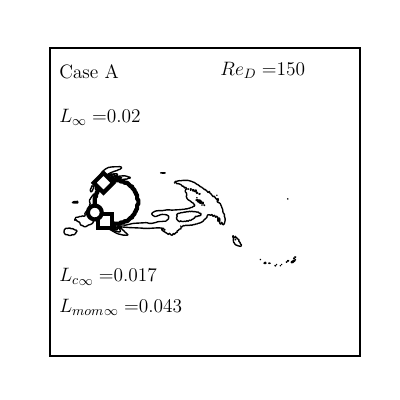}}
    \subfigure{\includegraphics[width=0.23\linewidth,trim={0.5cm 0.5cm 0.5cm 0.5cm},clip]{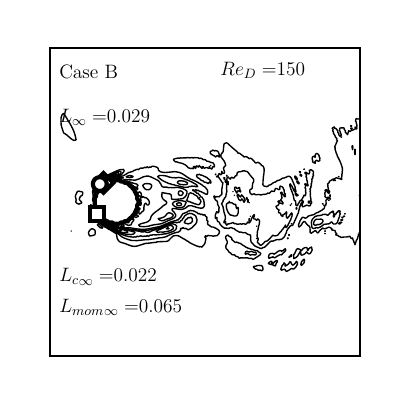}}
    \hspace{-1.0mm}%
    \setcounter{subfigure}{0}%
    \subfigure[]{}
    \hspace{-1.0mm}
    \subfigure{\includegraphics[width=0.23\linewidth,trim={0.5cm 0.5cm 0.5cm 0.5cm},clip]{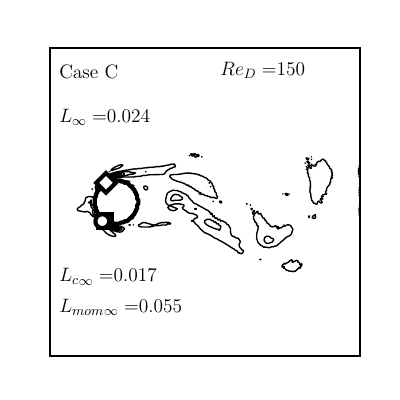}}
    \subfigure{\includegraphics[width=0.23\linewidth,trim={0.5cm 0.5cm 0.5cm 0.5cm},clip]{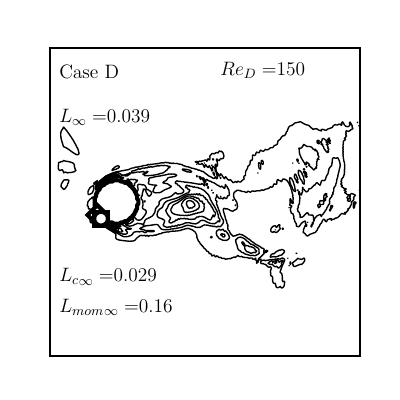}}

    \subfigure{\includegraphics[width=0.23\linewidth,trim={0.5cm 0.5cm 0.5cm 0.5cm},clip]{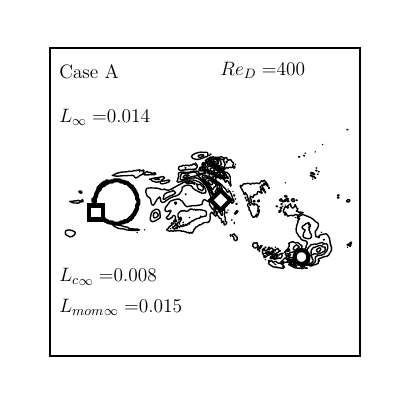}}
    \subfigure{\includegraphics[width=0.23\linewidth,trim={0.5cm 0.5cm 0.5cm 0.5cm},clip]{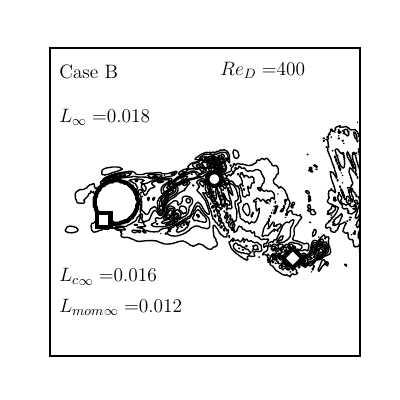}}
    \hspace{-1.0mm}%
    \setcounter{subfigure}{1}%
    \subfigure[]{}
    \hspace{-1.0mm}
    \subfigure{\includegraphics[width=0.23\linewidth,trim={0.5cm 0.5cm 0.5cm 0.5cm},clip]{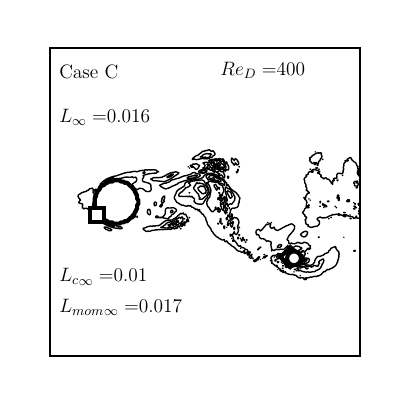}}
    \subfigure{\includegraphics[width=0.23\linewidth,trim={0.5cm 0.5cm 0.5cm 0.5cm},clip]{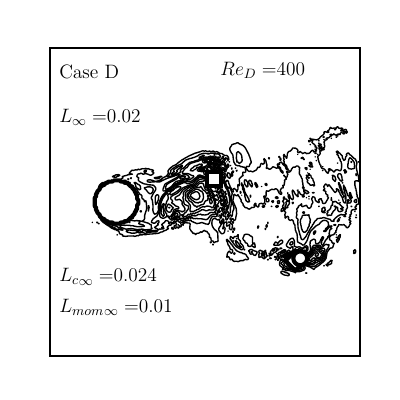}}

    \subfigure{\includegraphics[width=0.23\linewidth,trim={0.5cm 0.5cm 0.5cm 0.5cm},clip]{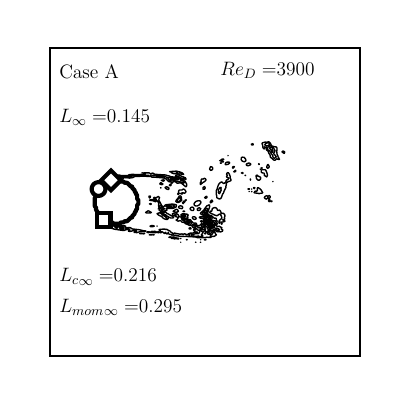}}
    \subfigure{\includegraphics[width=0.23\linewidth,trim={0.5cm 0.5cm 0.5cm 0.5cm},clip]{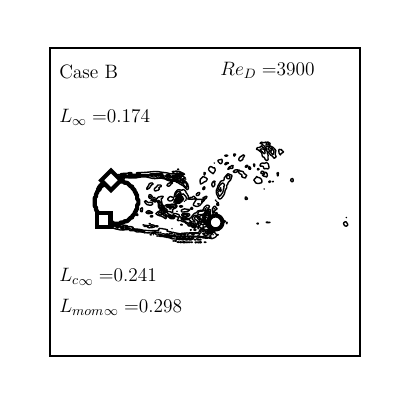}}
    \hspace{-1.0mm}%
    \setcounter{subfigure}{2}%
    \subfigure[]{}
    \hspace{-1.0mm}
    \subfigure{\includegraphics[width=0.23\linewidth,trim={0.5cm 0.5cm 0.5cm 0.5cm},clip]{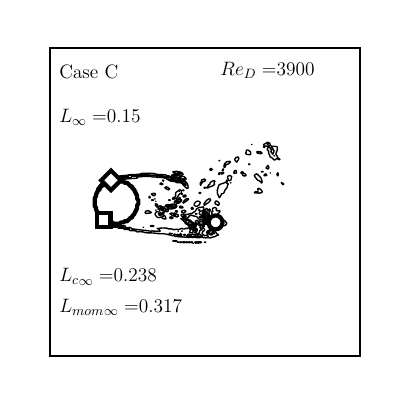}}
    \subfigure{\includegraphics[width=0.23\linewidth,trim={0.5cm 0.5cm 0.5cm 0.5cm},clip]{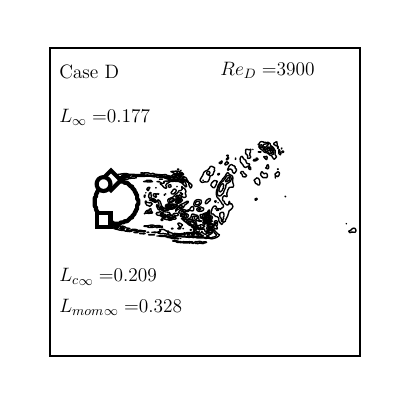}}
    \caption{Local distributions of errors for $u/U_{\infty}$ after a single prediction step at (a) $Re_{D}=150$ (10 contour levels from 0 to 0.04), (b) $Re_{D}=400$ (10 contour levels from 0 to 0.04), and (c) $Re_{D}=3900$ (10 contour levels from 0.0 to 0.33). Locations of $L_{\infty}$, ${L_c}_\infty$ (maximum error in mass conservation), and ${L_{mom}}_\infty$ (maximum error in momentum conservation) are indicated by $\circ$, $\diamond$, and $\Box$, respectively.}
    \label{fig:err-single}
\end{figure}
%
\begin{figure}
    \centering
    \subfigure{\includegraphics[width = 0.345\linewidth,trim={0.3cm 0.3cm 0.3cm 0.3cm},clip]{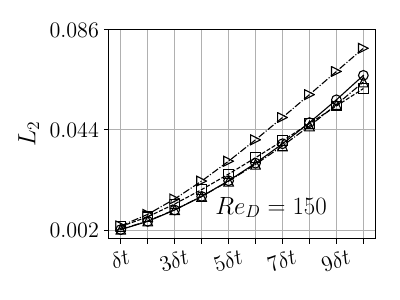}}
    \setcounter{subfigure}{0}%
    \subfigure[]{\includegraphics[width = 0.315\linewidth,trim={1.10cm 0.3cm 0.3cm 0.3cm},clip]{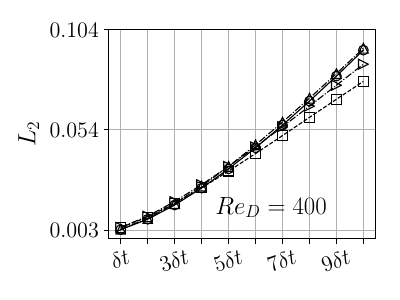}}
    \subfigure{\includegraphics[width = 0.315\linewidth,trim={1.10cm 0.3cm 0.3cm 0.3cm},clip]{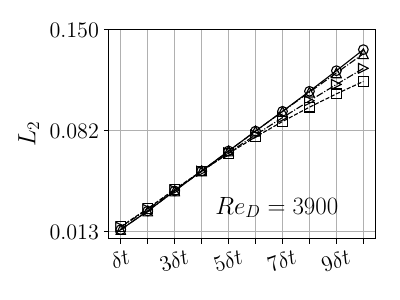}}
    \vspace{-0.2cm}

    \subfigure{\includegraphics[width = 0.345\linewidth,trim={0.3cm 0.3cm 0.3cm 0.3cm},clip]{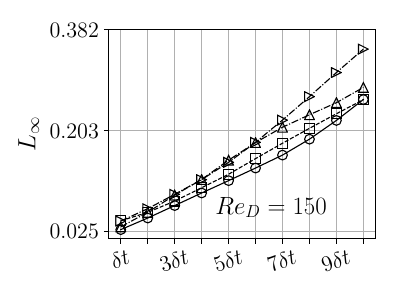}}
    \setcounter{subfigure}{1}%
    \subfigure[]{\includegraphics[width = 0.315\linewidth,trim={1.10cm 0.3cm 0.3cm 0.3cm},clip]{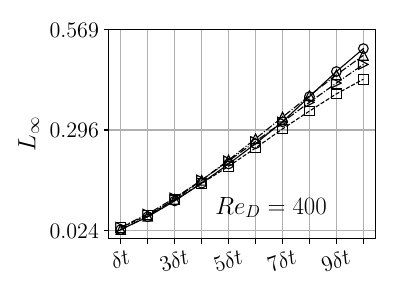}}
    \subfigure{\includegraphics[width = 0.315\linewidth,trim={1.10cm 0.3cm 0.3cm 0.3cm},clip]{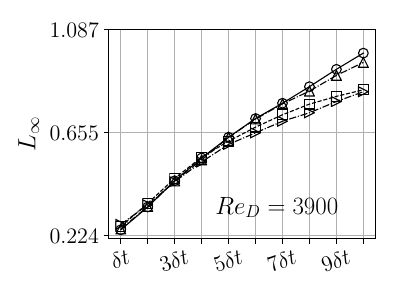}}
    \vspace{-0.2cm}

    \subfigure{\includegraphics[width = 0.345\linewidth,trim={0.3cm 0.3cm 0.3cm 0.3cm},clip]{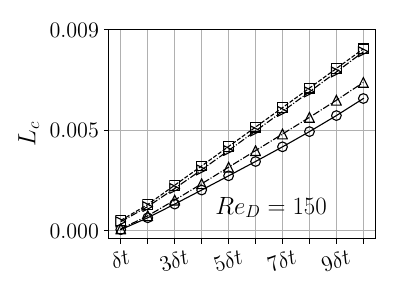}}
    \setcounter{subfigure}{2}%
    \subfigure[]{\includegraphics[width = 0.315\linewidth,trim={1.10cm 0.3cm 0.3cm 0.3cm},clip]{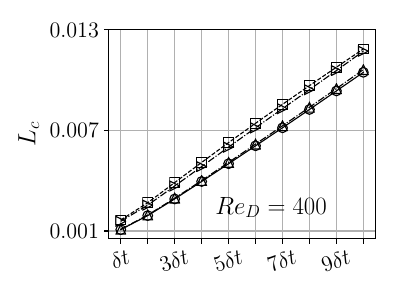}}
    \subfigure{\includegraphics[width = 0.315\linewidth,trim={1.10cm 0.3cm 0.3cm 0.3cm},clip]{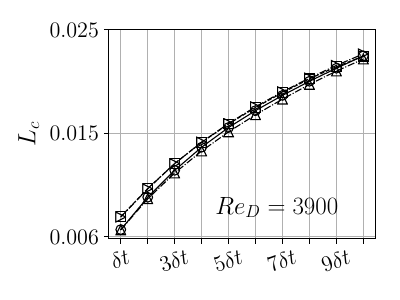}}
    \vspace{-0.2cm}

    \subfigure{\includegraphics[width = 0.345\linewidth,trim={0.3cm 0.3cm 0.3cm 0.3cm},clip]{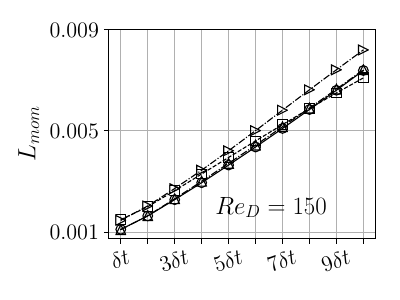}}
    \setcounter{subfigure}{3}%
    \subfigure[]{\includegraphics[width = 0.315\linewidth,trim={1.10cm 0.3cm 0.3cm 0.3cm},clip]{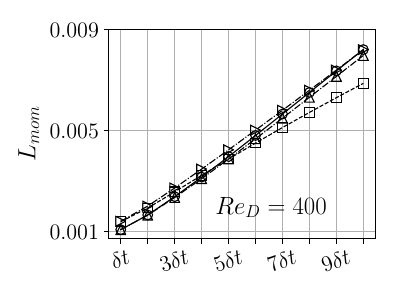}}
    \subfigure{\includegraphics[width = 0.315\linewidth,trim={1.10cm 0.3cm 0.3cm 0.3cm},clip]{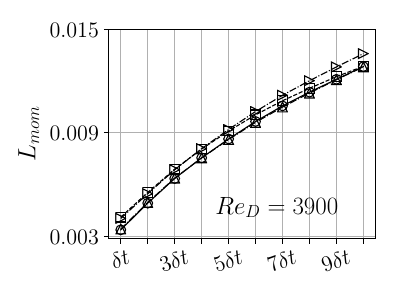}}
    \vspace{-0.2cm}    \caption{Comparisons of $L_2$, $L_\infty$, $L_c$, and $L_{mom}$ errors for Cases A-D. See appendix~\ref{appendix:err} for definitions of the errors.
    The time-step interval between flow fields is $\delta t = 20 \Delta t U_{\infty}/D = 0.1$.
    $\circ$ and solid line: Case A; $\square$ and dashed line: Case B; $\triangle$ and dash-dotted line: Case C; $\rhd$ and dotted line: Case D.
    }
    \label{fig:COM}
\end{figure}

Local distributions of errors for the streamwise velocity after a single time-step for four deep learning cases are compared in figure~\ref{fig:err-single}, while global errors such as $L_{2}$, $L_{\infty}$, $L_{c}$, and $L_{mom}$ errors as a function of the recursive time-step are compared in figure~\ref{fig:COM}. See appendix~\ref{appendix:err} for definitions of  errors.
All networks show that the maximum errors are located accelerating boundary layers on the cylinder wall or in the bridal region in the wake. 
Steep velocity gradients captured with relatively coarse resolution in the deep-learning prediction is considered as the cause for relatively high errors in accelerating boundary layers. 
Magnitudes of the maximum errors at $Re_{D}=400$ are found to be the smallest (see figure~\ref{fig:err-single}(b)), while magnitudes of the maximum errors at $Re_{D}=150$ (figure~\ref{fig:err-single}(a)) and $3900$ (figure~\ref{fig:err-single}(c)) are larger than those at $Re_D=400$. 
This result implies that a network best performs on predicting flow fields at a regime that has been utilized during training, 
while the network shows relatively large errors in predicting flow fields at the flow regime with higher complexity.

Interestingly, unlike errors at $1 \delta t$, as the recursive prediction step advances, errors at $Re_{D}=150$ are observed to increase more slowly than those at $Re_{D}=400$ (see figure~\ref{fig:COM}).
This implies that deep learning networks are capable of effectively learning large-scale or mainly two-dimensional vortex shedding physics from flow at three-dimensional wake transition regimes ($Re_{D}=300$ and $500$), thereby accurately predicting two-dimensional vortex shedding at $Re_D=150$, of which flow fields are not included in the training dataset. 

As also shown in figure~\ref{fig:COM}, the multi-scale CNN with physical loss functions (Case C) shows reduction of $L_{c}$ and $L_{mom}$ errors, during recursive prediction steps, compared to the multi-scale CNN without physical loss functions (Case D), indicating the advantage of the incorporation of physical loss functions in improving the conservation of mass and momentum.
At the same time, however,  $L_{2}$ and $L_{\infty}$ errors at $Re_{D}=400$ and $3900$ are found to increase in Cases C and D. 
Case A, which employs the GAN with physical loss functions, shows similar error trends to Case C but with smaller magnitudes of the $L_{\infty}$ error at $Re_{D}=150$.

On the other hand, the GAN without physical loss functions (Case B) shows smaller $L_{2}$ and $L_{mom}$ errors for all three Reynolds number cases than those in Case D which employs the multi-scale CNN without physical loss functions.
$L_{\infty}$ errors in Case B at $Re_{D}=150$ and $400$ are also significantly smaller than those in Case D.
These results imply that GANs (with and without physical loss functions, Case A and B) and the multi-scale CNN with physical loss functions (Case C) are better capable of extracting features related to unsteady vortex shedding physics over a circular cylinder, than the multi-scale CNN without physical loss functions (Case D).
The GAN without physical loss function (Case B) is found to consistently reduce errors associated with resemblance ($L_{2}$ and $L_{\infty}$) while error behaviors associated with conservation loss functions are rather inconsistent.   
Effects of physical loss functions on reduction of conservation errors are identifiable for networks with physical loss functions (Cases A and C).
 
\begin{figure}
    \centering
    \subfigure{\includegraphics[width=0.23\linewidth,trim={0.5cm 0.5cm 0.5cm 0.5cm},clip]{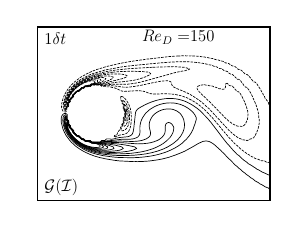}}
    \subfigure{\includegraphics[width=0.23\linewidth,trim={0.5cm 0.5cm 0.5cm 0.5cm},clip]{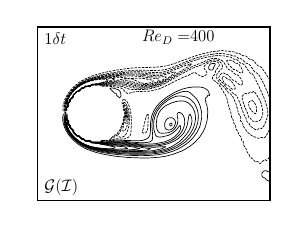}}
    \subfigure{\includegraphics[width=0.23\linewidth,trim={0.5cm 0.5cm 0.5cm 0.5cm},clip]{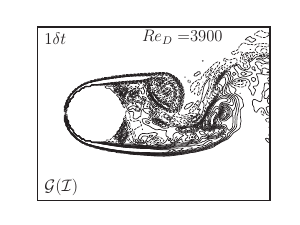}}
    \vspace{-3mm}

    \subfigure{\includegraphics[width=0.23\linewidth,trim={0.5cm 0.5cm 0.5cm 0.5cm},clip]{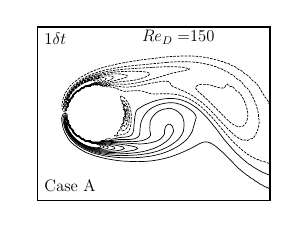}}
    \subfigure{\includegraphics[width=0.23\linewidth,trim={0.5cm 0.5cm 0.5cm 0.5cm},clip]{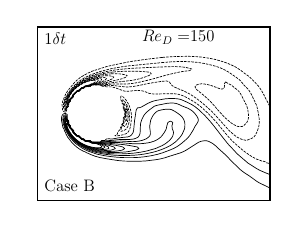}}
    \subfigure{\includegraphics[width=0.23\linewidth,trim={0.5cm 0.5cm 0.5cm 0.5cm},clip]{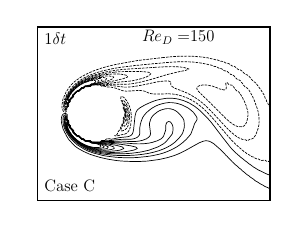}}
    \subfigure{\includegraphics[width=0.23\linewidth,trim={0.5cm 0.5cm 0.5cm 0.5cm},clip]{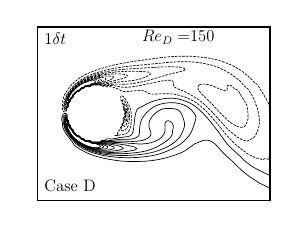}}
    \vspace{-3mm}

    \subfigure{\includegraphics[width=0.23\linewidth,trim={0.5cm 0.5cm 0.5cm 0.5cm},clip]{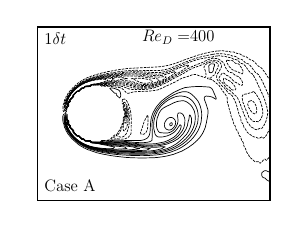}}
    \subfigure{\includegraphics[width=0.23\linewidth,trim={0.5cm 0.5cm 0.5cm 0.5cm},clip]{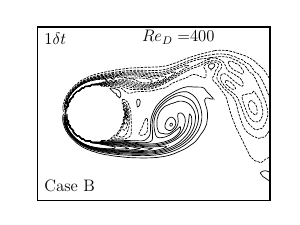}}
    \subfigure{\includegraphics[width=0.23\linewidth,trim={0.5cm 0.5cm 0.5cm 0.5cm},clip]{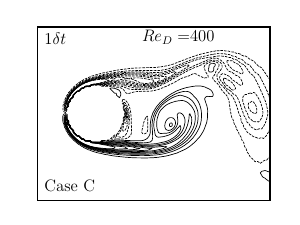}}
    \subfigure{\includegraphics[width=0.23\linewidth,trim={0.5cm 0.5cm 0.5cm 0.5cm},clip]{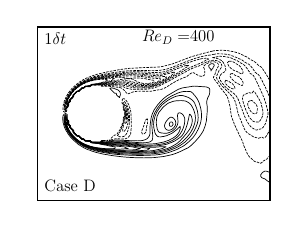}}
    \vspace{-3mm}

    \subfigure{\includegraphics[width=0.23\linewidth,trim={0.5cm 0.5cm 0.5cm 0.5cm},clip]{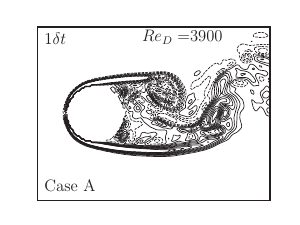}}
    \subfigure{\includegraphics[width=0.23\linewidth,trim={0.5cm 0.5cm 0.5cm 0.5cm},clip]{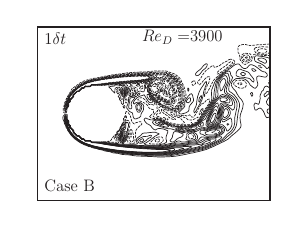}}
    \setcounter{subfigure}{0}%
    \hspace{-1.0mm}%
    \subfigure[]{} \hspace{-1.0mm}
    \subfigure{\includegraphics[width=0.23\linewidth,trim={0.5cm 0.5cm 0.5cm 0.5cm},clip]{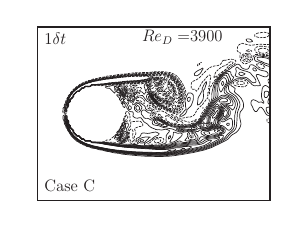}}
    \subfigure{\includegraphics[width=0.23\linewidth,trim={0.5cm 0.5cm 0.5cm 0.5cm},clip]{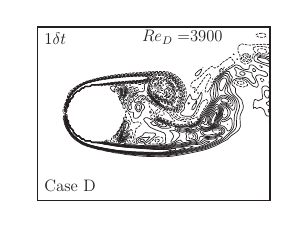}}
    \vspace{-3mm}

    \subfigure{\includegraphics[width=0.23\linewidth,trim={0.5cm 0.5cm 0.5cm 0.5cm},clip]{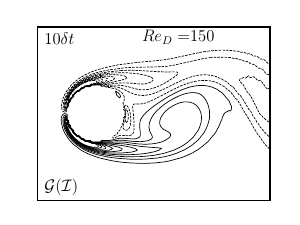}}
    \subfigure{\includegraphics[width=0.23\linewidth,trim={0.5cm 0.5cm 0.5cm 0.5cm},clip]{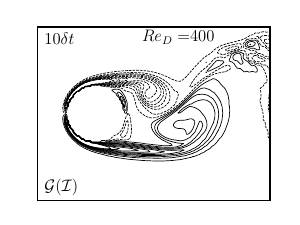}}
    \subfigure{\includegraphics[width=0.23\linewidth,trim={0.5cm 0.5cm 0.5cm 0.5cm},clip]{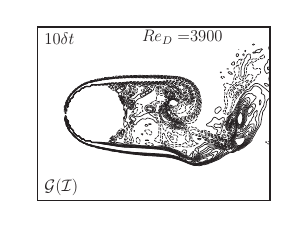}}
    \vspace{-3mm}

    \subfigure{\includegraphics[width=0.23\linewidth,trim={0.5cm 0.5cm 0.5cm 0.5cm},clip]{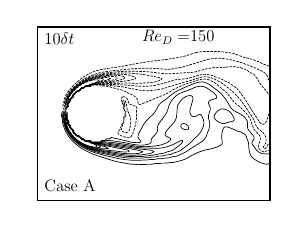}}
    \subfigure{\includegraphics[width=0.23\linewidth,trim={0.5cm 0.5cm 0.5cm 0.5cm},clip]{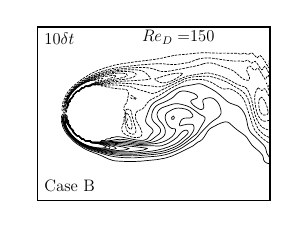}}
    \subfigure{\includegraphics[width=0.23\linewidth,trim={0.5cm 0.5cm 0.5cm 0.5cm},clip]{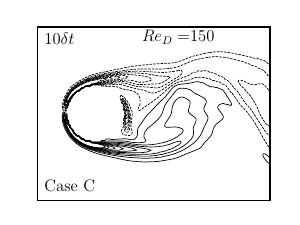}}
    \subfigure{\includegraphics[width=0.23\linewidth,trim={0.5cm 0.5cm 0.5cm 0.5cm},clip]{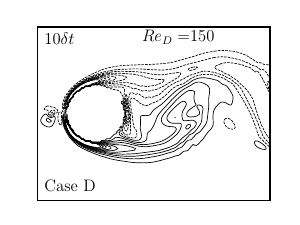}}
    \vspace{-3mm}

    \subfigure{\includegraphics[width=0.23\linewidth,trim={0.5cm 0.5cm 0.5cm 0.5cm},clip]{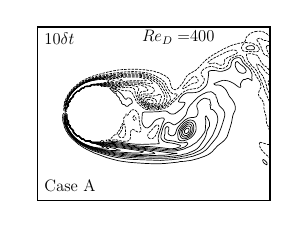}}
    \subfigure{\includegraphics[width=0.23\linewidth,trim={0.5cm 0.5cm 0.5cm 0.5cm},clip]{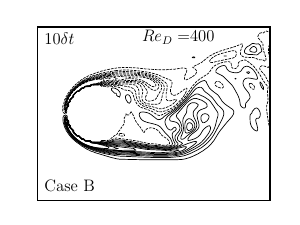}}
    \subfigure{\includegraphics[width=0.23\linewidth,trim={0.5cm 0.5cm 0.5cm 0.5cm},clip]{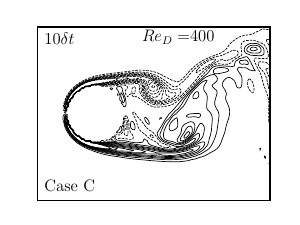}}
    \subfigure{\includegraphics[width=0.23\linewidth,trim={0.5cm 0.5cm 0.5cm 0.5cm},clip]{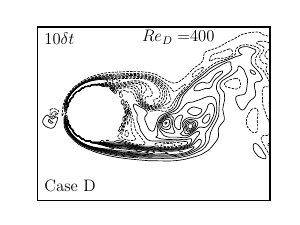}}
    \vspace{-3mm}

    \subfigure{\includegraphics[width=0.23\linewidth,trim={0.5cm 0.5cm 0.5cm 0.5cm},clip]{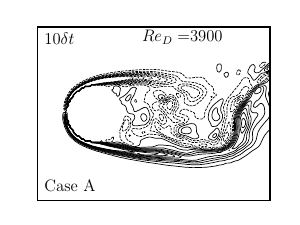}}
    \subfigure{\includegraphics[width=0.23\linewidth,trim={0.5cm 0.5cm 0.5cm 0.5cm},clip]{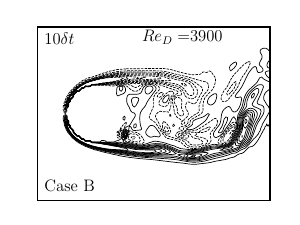}}
    \setcounter{subfigure}{1}%
    \hspace{-1.0mm}%
    \subfigure[]{} \hspace{-1.0mm}
    \subfigure{\includegraphics[width=0.23\linewidth,trim={0.5cm 0.5cm 0.5cm 0.5cm},clip]{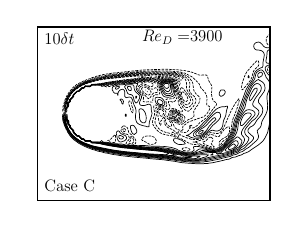}}
    \subfigure{\includegraphics[width=0.23\linewidth,trim={0.5cm 0.5cm 0.5cm 0.5cm},clip]{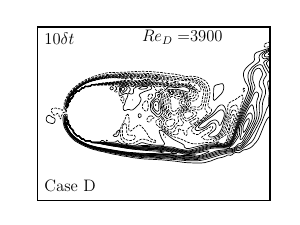}}
    \vspace{-3mm}
                \caption{Comparison of the spanwise vorticity after (a) a single prediction step  ($1\delta t$) and (b) 9 more recursive prediction steps ($10\delta t$).  
                20 contour levels from -20.0 to 20.0 are shown. Solid lines and dashed lines indicate positive and negative contour levels, respectively.
    }
    \label{fig:vor-comp}
\end{figure}

Vortical structures at each Reynolds number predicted by the present four deep learning networks appear to be similar to each other after a single prediction step as shown in figure~\ref{fig:vor-comp}(a).
However, all deep learning cases show difficulties in learning production of small scale vortical structures. At $10\delta t$, small scale vortical structures, which do not present in the ground truth flow field are found to be generated inside shedded large-scale vortices at $Re_D=150$, while many small scale vortices are missed in the wake at $Re_D=3900$ (figure~\ref{fig:vor-comp}(b)).
This observation implies that a network shows difficulty in predicting flow fields, especially in recursive predictions as errors from previous predictions are accumulated, at flow regimes which are different from the regime for training.

After a few number of recursive prediction steps, Case D, where the multi-scale CNN without physical loss functions is applied, shows unphysical vortical structures near the front stagnation point, which are not present in flow fields predicted by other cases at the three considered Reynolds numbers (figure~\ref{fig:vor-comp}~(b)).
The effect of the inaccurate prediction in Case D on errors also appears in figure~\ref{fig:COM}, of which magnitudes are larger than those in Cases A, B, and C.

All deep learning cases are found to be capable of predicting future flow fields, particularly in single-step predictions. However, networks with additional consideration of physics in either a supervised or an unsupervised manner (Cases A-C) are recommended for predicting further future flow fields with many recursive steps.
Especially, the GAN without physical loss functions (Case B) is found to be the best among the considered networks for minimizing $L_{2}$ and $L_{\infty}$ errors (see figure~\ref{fig:COM}) while also satisfying the conservation of mass and momentum favorably.

\subsection{Analysis on captured and missed flow physics}\label{sub:missed}
Discussion in the present section is focused on the GAN without physical loss functions (Case B), which is trained with flow fields at $Re_{D}=300$ and $500$ (the three-dimensional wake transition regime) and tested on flow fields at $Re_{D}=150$ (the two-dimensional vortex shedding regime), $400$ (the same flow regime with training) and $3900$ (the  shear-layer transition regime), in order to assess what flow characteristics the network captures or misses.

\begin{figure}
    \centering
    \subfigure{\includegraphics[width=0.3\linewidth,trim={0.5cm 0.5cm 0.5cm 0.5cm},clip]{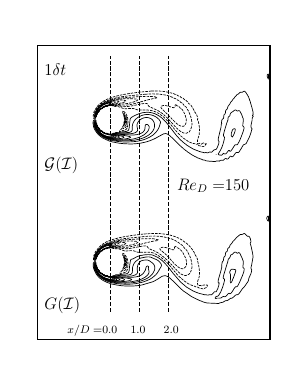}}
    \setcounter{subfigure}{0}%
    \subfigure[]{\includegraphics[width=0.3\linewidth,trim={0.5cm 0.5cm 0.5cm 0.5cm},clip]{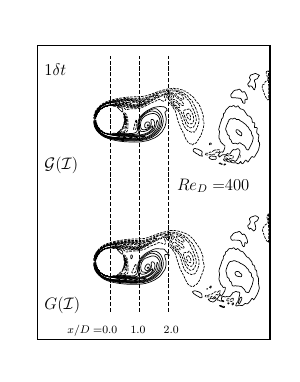}}
    \subfigure{\includegraphics[width=0.3\linewidth,trim={0.5cm 0.5cm 0.5cm 0.5cm},clip]{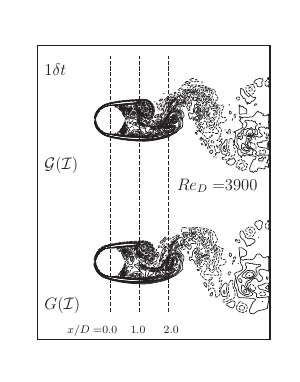}}
 
    \subfigure{\includegraphics[width=0.3\linewidth,trim={0.5cm 0.5cm 0.5cm 0.5cm},clip]{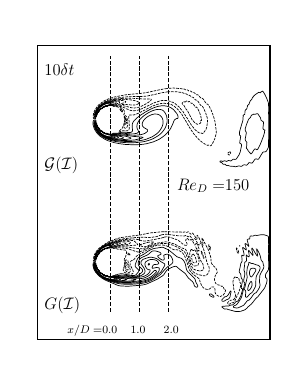}}
    \setcounter{subfigure}{1}%
    \subfigure[]{\includegraphics[width=0.3\linewidth,trim={0.5cm 0.5cm 0.5cm 0.5cm},clip]{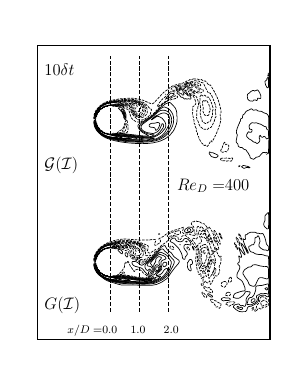}}
    \subfigure{\includegraphics[width=0.3\linewidth,trim={0.5cm 0.5cm 0.5cm 0.5cm},clip]{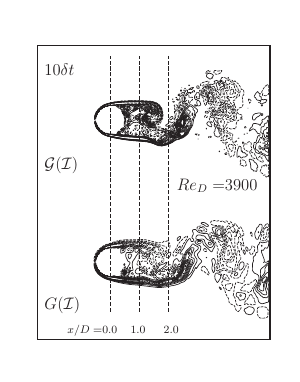}}
    \caption{Contours of the spanwise vorticity calculated using ground truth velocity fields ($\mathcal{G}(\mathcal{I})$) and velocity fields predicted by the GAN ($G(\mathcal{I})$) after (a) a single prediction step ($1\delta t$) and (b) 9 more recursive prediction steps ($10\delta t$) at $Re_{D}=150, 400,$ and $3900$.
    20 contour levels from -20.0 to 20.0 are shown.
    Solid lines and dashed lines indicate positive and negative contour levels, respectively.
    }
    \label{fig:vor-GAN}
\end{figure}
Contour plots of the spanwise vorticity calculated using ground truth velocity fields and velocity fields predicted by the GAN are compared in figure~\ref{fig:vor-GAN} for three Reynolds numbers at $1\delta t$ and $10\delta t$.
First of all, laminar flow at the frontal face of the cylinder as well as the separated laminar shear layers including lengthening of the shear layers and detachment from the wall are observed to be well captured in all three Reynolds number cases. Convection (downstream translation) and diffusion of overall large scale vortical structures in the wake are also favorably predicted at both $1\delta t$ and $10\delta t$.
However, as also mentioned in the previous section, prediction results show differences in the generation and dissipation of small-scale vortices.
After a number of recursive prediction steps, along with the non-zero spanwise velocity, unexpected smaller scale vortices than those present in the ground truth flow field are generated at $Re_{D}=150$, at which the Reynolds number regime, downstream vortical structures are expected to be laminar and two dimensional. 
Generation of smaller scale vortical structures than those in ground truth flow fields after a few recursive predictions is also noticed in the GAN prediction at $Re_D=400$.
On the other hand, it is found that the GAN fails in accurately predicting small-scale vortical structures inside large-scale vortices at $Re_{D}=3900$.
It is thought that the present results imply that the GAN is not fully trained for predicting production and dissipation of small-scale vortices. The lack of flow information along the spanwise direction is considered as a major cause for the incapability. Due to the reason mentioned in section~\ref{sub:cons}, the spanwise information in the present training dataset includes only the spanwise velocity on a two-dimensional sliced domain, therefore misses variation of flow variables along the spanwise direction.

The lack of spanwise information of flow variables seems to lead the network to miss the mechanism for generation of small-scale vortices, which can be formulated as the vortex stretching term in the spanwise vorticity ($\omega_z$) equation. The stretching term 
$\omega_{z} \frac{\partial w}{\partial z}$, which is associated with the generation of small-scale vortices, is missed in the present training. On the other hand, convection and diffusion of the spanwise vorticity are dominated by $u\frac{\partial \omega_z}{\partial x}+v\frac{\partial \omega_z}{\partial y}$ and $\frac{1}{Re_D}(\frac{\partial^2 \omega_z}{\partial x \partial x} + \frac{\partial^2 \omega_z}{\partial y \partial y})$, which can be rather easily trained using the given flow field data.

\begin{figure}
    \centering
    \subfigure{\includegraphics[width=0.335\linewidth,trim={0.3cm 0.3cm 0.3cm 0.3cm},clip]{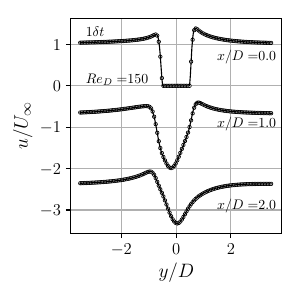}}
    \setcounter{subfigure}{0}%
    \subfigure[]{\includegraphics[width=0.31\linewidth,trim={0.85cm 0.3cm 0.3cm 0.3cm},clip]{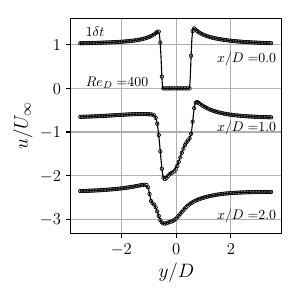}}
    \subfigure{\includegraphics[width=0.31\linewidth,trim={0.85cm 0.3cm 0.3cm 0.3cm},clip]{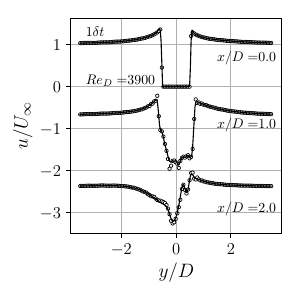}}
 
    \subfigure{\includegraphics[width=0.335\linewidth,trim={0.3cm 0.3cm 0.3cm 0.3cm},clip]{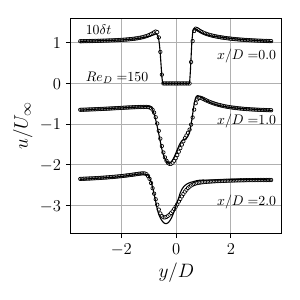}}
    \setcounter{subfigure}{1}%
    \subfigure[]{\includegraphics[width=0.31\linewidth,trim={0.85cm 0.3cm 0.3cm 0.3cm},clip]{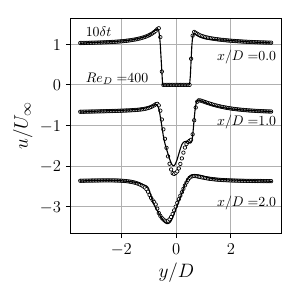}}
    \subfigure{\includegraphics[width=0.31\linewidth,trim={0.85cm 0.3cm 0.3cm 0.3cm},clip]{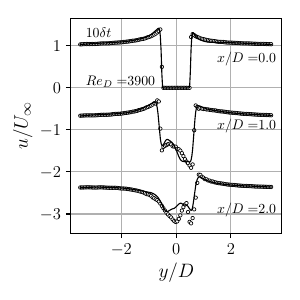}}
 
    \caption{Profiles of the streamwise velocity at three streamwise locations after (a) a single prediction step ($1\delta t$) and (b) 9 more recursive prediction steps ($10\delta t$) at $Re_{D}=150, 400,$ and $3900$.
    Circles indicate ground truth results and solid lines indicate results predicted by the GAN. Profiles at $x/D=1.0$ and 2.0 are shifted by -1.7 and -3.4 in the vertical axis, respectively.
    }
    \label{fig:u-GAN}
\end{figure}

Convection and diffusion phenomena in flow around a cylinder are investigated more quantitatively in the development of the velocity deficit. Profiles of the streamwise velocity from ground truth flow fields ($\circ$) and flow fields predicted by the GAN (solid lines) at three streamwise locations, $x/D=0.0$, $1.0$, and $2.0$, are compared in figure~\ref{fig:u-GAN}.
Velocity profiles at $x/D=0.0$ show no identifiable differences between ground truth and GAN flow fields at both $1\delta t$ and $10\delta t$ at all Reynolds numbers ($Re_{D}=150$, $400$, and $3900$). 
This is because flow at $x/D=0.0$ is laminar two-dimensional boundary layer flow of which characteristics is rather easily trained by the network. 
Noticeable differences in the velocity deficit are observed in the comparison at $10\delta t$ in the wake region, $x/D=2.0$, at $Re_D=3900$, where small scale oscillatory motions are not accurately captured by the GAN.
Recursively predicted velocity deficits at $Re_D=150$ and $400$ are in good agreement with the ground truth velocity deficit in terms of the peak, width, and shape at both streamwise locations.

\begin{figure}
    \centering
    \subfigure{\includegraphics[width=0.335\linewidth,trim={0.3cm 0.3cm 0.3cm 0.3cm},clip]{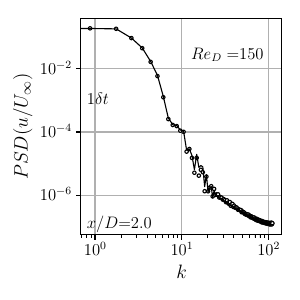}}
    \setcounter{subfigure}{0}%
    \subfigure[]{\includegraphics[width=0.31\linewidth,trim={0.85cm 0.3cm 0.3cm 0.3cm},clip]{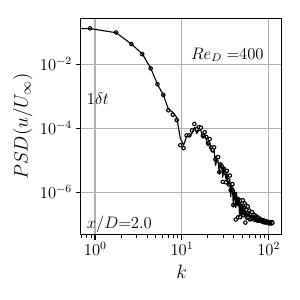}}
    \subfigure{\includegraphics[width=0.31\linewidth,trim={0.85cm 0.3cm 0.3cm 0.3cm},clip]{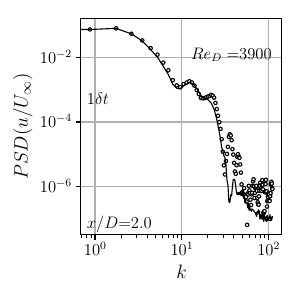}}
 
    \subfigure{\includegraphics[width=0.335\linewidth,trim={0.3cm 0.3cm 0.3cm 0.3cm},clip]{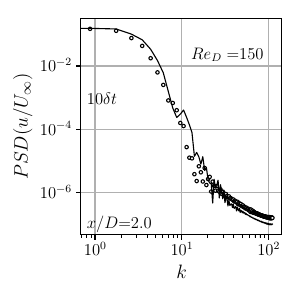}}
    \setcounter{subfigure}{1}%
    \subfigure[]{\includegraphics[width=0.31\linewidth,trim={0.85cm 0.3cm 0.3cm 0.3cm},clip]{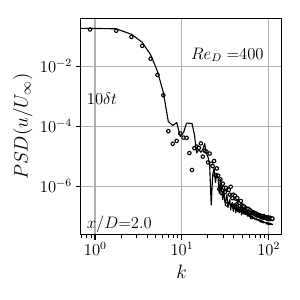}}
    \subfigure{\includegraphics[width=0.31\linewidth,trim={0.85cm 0.3cm 0.3cm 0.3cm},clip]{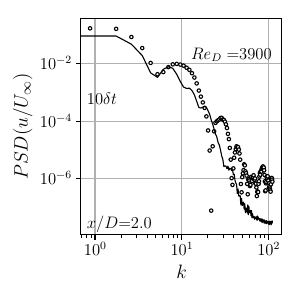}}
 
    \caption{Power spectral density of the streamwise velocity at $x/D =2.0$ after (a) a single prediction step ($1\delta t$) and (b) 9 more recursive prediction steps ($10\delta t$) at $Re_{D}=150, 400,$ and $3900$.
    Circles indicate ground truth results and solid lines indicate results predicted by the GAN.
    }
    \label{fig:PSD-GAN}
\end{figure}

Plots of the power spectral density (PSD) of the streamwise velocity along the vertical axis ($y$) in the wake region at $x/D=2.0$ are shown in figure~\ref{fig:PSD-GAN} to evaluate wavenumber contents of wake flow.
At $Re_{D}=150$ and $400$, PSDs which are produced by the GAN show good agreements with ground truth results in the single-step prediction ($1 \delta t$), while PSDs are found to be still close to ground truth PSDs with marginal deviations in the middle to high wavenumber contents ($k> 10$) after 9 recursive predictions.
On the other hand, PSDs produced by the GAN at $Re_{D}=3900$ at both $1\delta t$ and $10\delta t$ show noticeable deviations from ground truth PSDs, especially in high wavenumber contents, again indicating the difficulty in learning the mechanism for production of small scale vortices (high wavenumber contents).
\bigskip

\subsection{Training with additional data}\label{sec:comparison}
The GAN without physical loss functions is trained with additional flow field data at Reynolds numbers of $1000$ and $3000$, in order to investigate the effect of small-scale contents in training data on the prediction of small-scale vortical motions in flow in the shear-layer transition regime ($Re_D=3900$).
\begin{figure}
    \centering
    \subfigure{\includegraphics[width=0.25\linewidth,trim={0.5cm 0.5cm 0.5cm 0.5cm},clip]{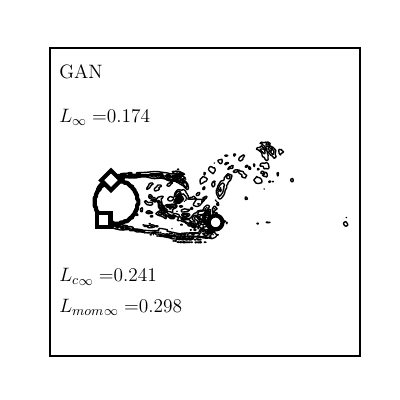}}
    \subfigure{\includegraphics[width=0.25\linewidth,trim={0.5cm 0.5cm 0.5cm 0.5cm},clip]{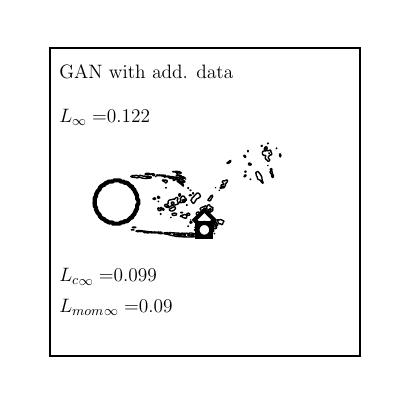}}
    \caption{Local distributions of errors for $u/U_{\infty}$ after a single prediction step at $Re_{D}=3900$ (10 contour levels from 0.0 to 0.33). Locations of $L_{\infty}$, ${L_c}_\infty$ (maximum error in mass conservation), and ${L_{mom}}_\infty$ (maximum error in momentum conservation) are indicated by $\circ$, $\diamond$, and $\Box$, respectively.}
    \label{fig:err-add-single}
\end{figure}
Local distributions of errors for the streamwise velocity after a single time-step for the GAN and the GAN with additional flow field data are compared in figure~\ref{fig:err-add-single}. Magnitudes of maximum errors, especially the mass and momentum errors, are significantly reduced by training the network with flow fields at the same flow regime to be predicted. Nevertheless, maximum errors are still larger than those at low Reynolds numbers (see figure~\ref{fig:err-single}~(a) and (b)). The lack of spanwise information in the input is considered to be the remaining cause for the errors.
\begin{figure}
    \centering
    \subfigure{\includegraphics[width=0.30\linewidth,trim={0.5cm 0.5cm 0.5cm 0.5cm},clip]{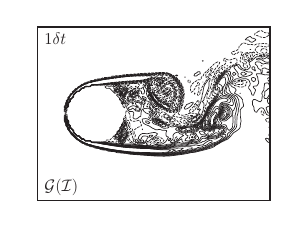}}
    \setcounter{subfigure}{0}%
    \subfigure[]{\includegraphics[width=0.30\linewidth,trim={0.5cm 0.5cm 0.5cm 0.5cm},clip]{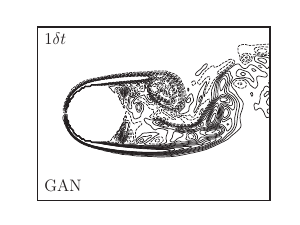}}
    \subfigure{\includegraphics[width=0.30\linewidth,trim={0.5cm 0.5cm 0.5cm 0.5cm},clip]{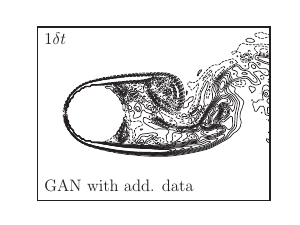}}

    \subfigure{\includegraphics[width=0.30\linewidth,trim={0.5cm 0.5cm 0.5cm 0.5cm},clip]{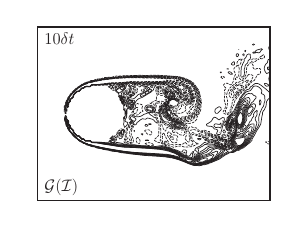}}
    \setcounter{subfigure}{1}%
    \subfigure[]{\includegraphics[width=0.30\linewidth,trim={0.5cm 0.5cm 0.5cm 0.5cm},clip]{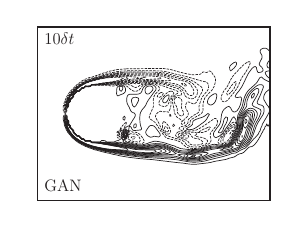}}
    \subfigure{\includegraphics[width=0.30\linewidth,trim={0.5cm 0.5cm 0.5cm 0.5cm},clip]{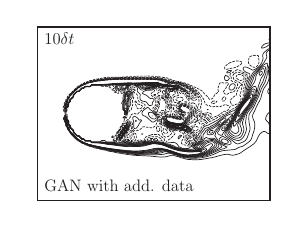}}

    \subfigure{\includegraphics[width=0.30\linewidth,trim={0.3cm 0.3cm 0.3cm 0.3cm},clip]{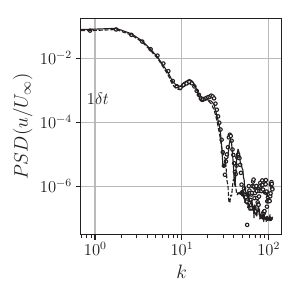}}
    \setcounter{subfigure}{2}%
    \subfigure[]{}
    \subfigure{\includegraphics[width=0.30\linewidth,trim={0.3cm 0.3cm 0.3cm 0.3cm},clip]{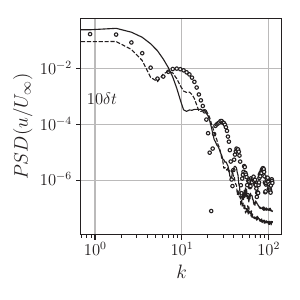}}
     \caption{Contour plots of the spanwise vorticity calculated using ground truth velocity fields and velocity fields predicted by the GANs at $Re_{D}=3900$ after (a) $1\delta t$ and (b) $10\delta t$. (c) Plots of the power spectral density at $1\delta t$ and $10\delta t$. 20 contour levels from -20.0 to 20.0 are shown. Solid lines and dashed lines denote positive and negative contour levels, respectively. Circles indicate ground truth result, while the dashed line and the solid line correspond to predicted results using the GAN and the GAN with additional data, respectively. }
    \label{fig:vor-add} 
\end{figure}
Contours of the spanwise vorticity calculated by ground truth flow fields, flow fields predicted by the GAN trained with data at $Re_D=300$ and $500$, and flow fields predicted by the GAN trained with additional data at $Re_D=1000$ and $3000$ are compared in figures~\ref{fig:vor-add}(a) and (b).
Training with additional data at the same flow regime is found to clearly improve the prediction of small-scale motions after a single-prediction step ($1\delta t$).  The spanwise vorticity predicted by the GAN which is trained with  additional data is found to much better agree with the ground truth vorticity than that predicted by the GAN which is trained with flow fields only at $Re_D=300$ and $500$ after 9 more recursive prediction steps ($10\delta t$) as shown in figure~\ref{fig:vor-add}(b). However, as discussed in the previous section (section~\ref{sub:missed}), the GAN trained with additional data also suffers from lacking of production of small-scale vortical structures. 
PSDs produced by the GAN trained for $Re_D=300$ and $500$ and the GAN trained with additional data are close to the ground truth PSD at $1\delta t$, while the GAN trained with additional data better predict small-scale high wavenumber contents. Differences among predicted and ground truth PSDs become larger  at $10\delta t$, where reduced small scale high wavenumber contents are clearly observable for both GANs (figure~\ref{fig:vor-add}(c)).
\bigskip

\subsection{Training with a large time-step interval}\label{sec:comparison}
\begin{figure}
    \centering
    \subfigure{\includegraphics[width=0.335\linewidth,trim={0.3cm 0.3cm 0.3cm 0.3cm},clip]{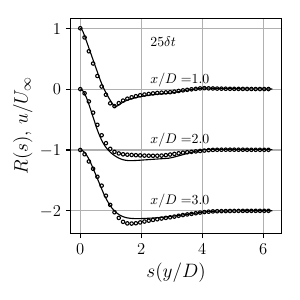}}
    \subfigure{\includegraphics[width=0.31\linewidth,trim={0.85cm 0.3cm 0.3cm 0.3cm},clip]{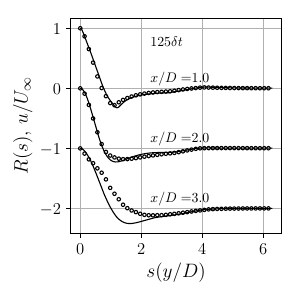}}
    \caption{Two point correlations of the streamwise velocity at three downstream locations ($x/D=1.0$, $2.0$, and $3.0$) at $Re_{D}=3900$.
    Circles indicate ground truth results and solid lines indicate predicted results by the GAN. Two point correlations at $x/D=2.0$ and $3.0$ are shifted by -1 and -2 along the vertical axis, respectively.
    }
    \label{fig:COR}
\end{figure}

To investigate the potential of using a GAN in practical applications, where predicting large-scale flow motions is important, the GAN without physical loss functions is trained with a large time-step interval of $25 \delta t = 500 \Delta t U_{\infty}/D = 2.5$. 
This time-step interval is 25 times larger than the previous deep learning time-step interval and 500 times larger than the simulation time-step interval.
Figure~\ref{fig:COR} shows plots of two point correlations of the streamwise velocity along the $y$ direction, which provide information of the large scale fluid motions at three downstream wake locations at $Re_D =  3900$. 
After a single step with $25 \delta t$, it is found that two-point correlations predicted by the GAN are in favorable agreement with correlations of the ground truth flow field. After 4 additional recursive large steps ($125\delta t$), however, small deviations of correlations from ground truth results are observed in the downstream wake region ($x/D = 3.0$).
Note that $125 \delta t$ corresponds to $2500$ time-steps of the conducted numerical simulation for the ground truth flow field.

Contour plots of the streamwise velocity predicted by the GAN at $Re_{D}=3900$ are shown in figure~\ref{fig:u-T25} (see figures~\ref{fig:v-T25}-\ref{fig:p-T25} in appendix~\ref{appendix:fields} for contour plots of the other flow variables).
Flow fields at $50 \delta t$, $75 \delta t$, $100 \delta t$, and $125 \delta t$ are recursively predicted. As shown in figure~\ref{fig:u-T25}, large-scale oscillations of the streamwise velocity behind the cylinder are favorably predicted, while small-scale flow structures are found to be rather rapidly dissipated compared to those in ground truth flow fields.
This may be partly due to the dynamics of small-scale flow structures, of which time-scales ($\tau$) are smaller than the training interval size ($t=n {D}/{U_{\infty}} \delta t $, where $n$ is an integer), are disregarded from input information.
The time scale of a small-scale flow structure can be approximated as
\begin{eqnarray}
\tau \sim (\frac{\nu}{\epsilon})^{1/2}
\end{eqnarray}
according to~\citet{tennekes1972first}, where, $\nu$ is the kinematic viscosity and $\epsilon$ is the dissipation rate per unit mass that is approximated as
\begin{eqnarray}
	\epsilon \sim \frac{u^{3}}{l} \sim \frac{U^{3}_{\infty}}{D},
\end{eqnarray}
where $u$ is the velocity scale and $l$ is the length scale of a large-scale flow motion.
The ratio of the time-scale for a small-scale flow structure to the training interval size can be derived as follows:
\begin{eqnarray}
	\frac{\tau}{t} \sim \frac{\tau U_{\infty}}{n D \delta t} \sim \frac{1}{n \delta t}(\frac{U_{\infty}{D}}{\nu})^{-1/2} = \frac{1}{n \delta t \sqrt{Re_{D}}}.
\end{eqnarray}

The ratio of the time-scale for a small-scale flow structure to the training interval size decreases as the Reynolds number and the integer $n$ increase.
Therefore, small-scale flow structures are reasonably well captured by the network trained with a small training-step interval (see figures~\ref{fig:u-R1R3R4}-\ref{fig:p-R1R3R4}), while it is found that small-scale flow structures predicted by the network trained with a large training-step interval of $25\delta t$, rapidly disappear in the wake (see figures~\ref{fig:u-T25},~\ref{fig:v-T25}-\ref{fig:p-T25}).

\begin{figure}
    \centering
    \subfigure{\includegraphics[width=0.19\linewidth,trim={0.5cm 0.5cm 0.5cm 0.5cm},clip]{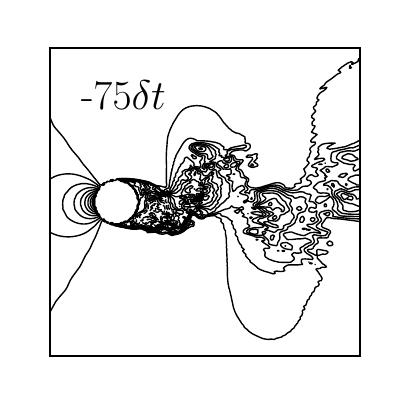}}
    \subfigure{\includegraphics[width=0.19\linewidth,trim={0.5cm 0.5cm 0.5cm 0.5cm},clip]{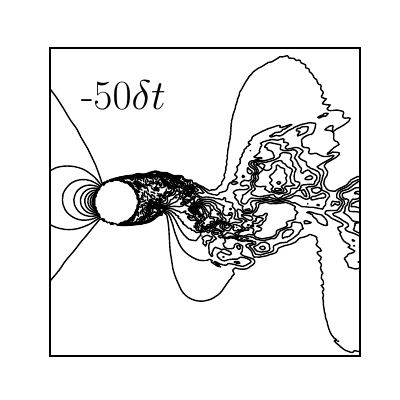}}
    \setcounter{subfigure}{0}%
    \subfigure[]{}\hspace{-1mm}
    \subfigure{\includegraphics[width=0.19\linewidth,trim={0.5cm 0.5cm 0.5cm 0.5cm},clip]{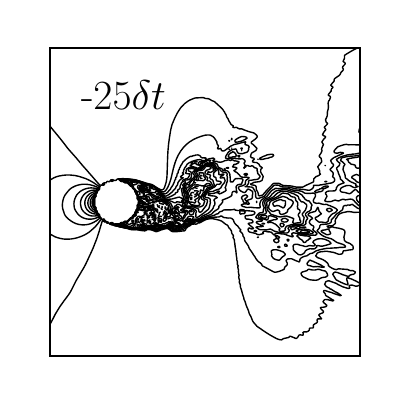}}
    \subfigure{\includegraphics[width=0.19\linewidth,trim={0.5cm 0.5cm 0.5cm 0.5cm},clip]{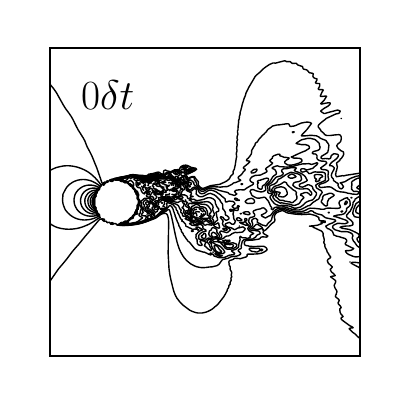}}

    \subfigure{\includegraphics[width=0.19\linewidth,trim={0.5cm 0.5cm 0.5cm 0.5cm},clip]{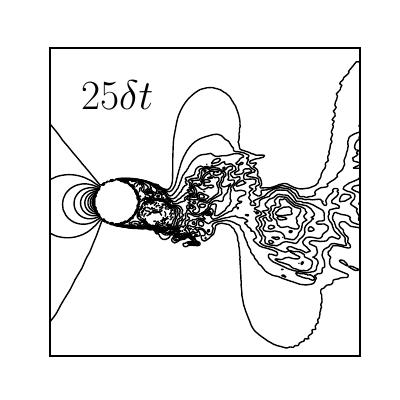}}
    \subfigure{\includegraphics[width=0.19\linewidth,trim={0.5cm 0.5cm 0.5cm 0.5cm},clip]{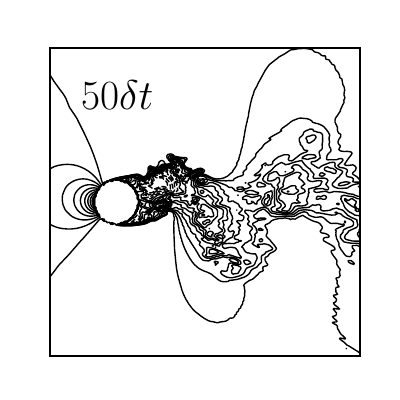}}
    \setcounter{subfigure}{1}%
    \subfigure[]{\includegraphics[width=0.19\linewidth,trim={0.5cm 0.5cm 0.5cm 0.5cm},clip]{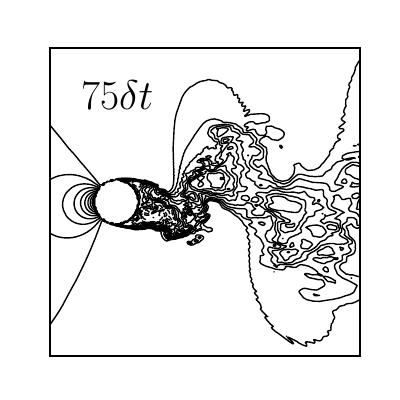}}
    \subfigure{\includegraphics[width=0.19\linewidth,trim={0.5cm 0.5cm 0.5cm 0.5cm},clip]{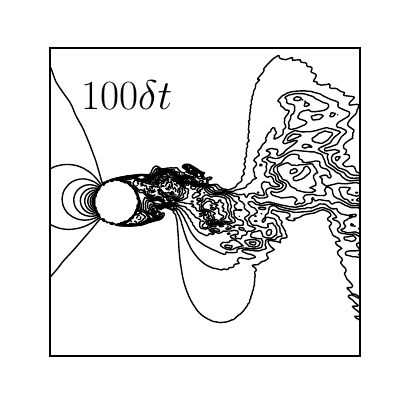}}
    \subfigure{\includegraphics[width=0.19\linewidth,trim={0.5cm 0.5cm 0.5cm 0.5cm},clip]{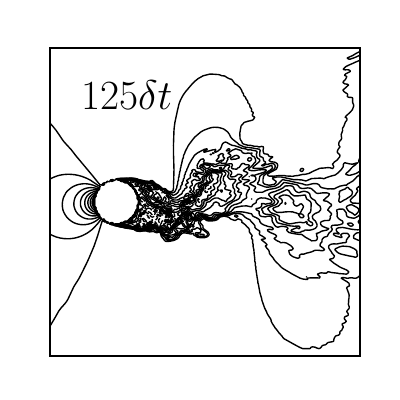}}

    \subfigure{\includegraphics[width=0.19\linewidth,trim={0.5cm 0.5cm 0.5cm 0.5cm},clip]{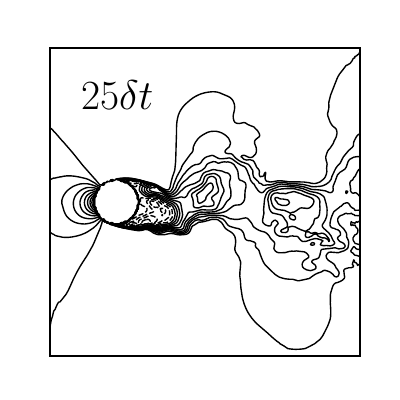}}
    \subfigure{\includegraphics[width=0.19\linewidth,trim={0.5cm 0.5cm 0.5cm 0.5cm},clip]{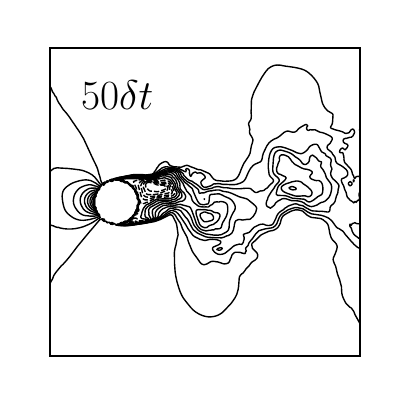}}
    \setcounter{subfigure}{2}%
    \subfigure[]{\includegraphics[width=0.19\linewidth,trim={0.5cm 0.5cm 0.5cm 0.5cm},clip]{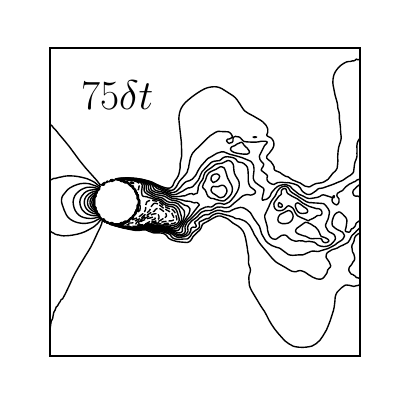}}
    \subfigure{\includegraphics[width=0.19\linewidth,trim={0.5cm 0.5cm 0.5cm 0.5cm},clip]{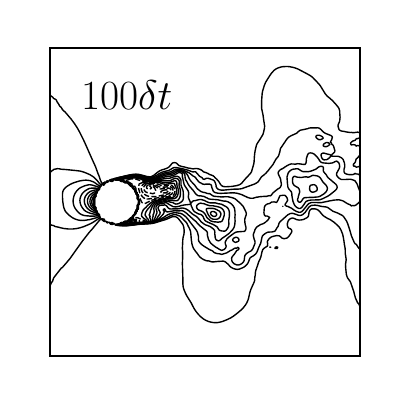}}
    \subfigure{\includegraphics[width=0.19\linewidth,trim={0.5cm 0.5cm 0.5cm 0.5cm},clip]{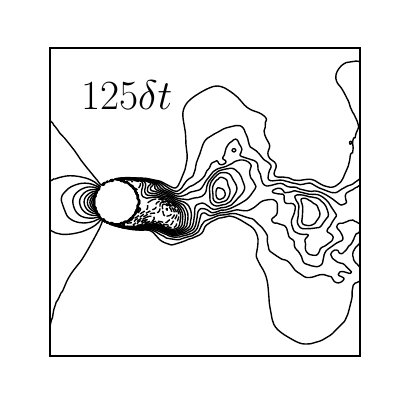}}
    \caption{
    Contour plots of the streamwise velocity ($u/ U_{\infty}$) at $Re_{D}=3900$ after $25\delta t$, $50 \delta t$, $75 \delta t$, $100 \delta t$, and $125 \delta t$, where $1 \delta t = 20 \Delta t U_{\infty}/D = 0.1$. Flow fields at $50 \delta t$, $75 \delta t$, $100 \delta t$, and $125 \delta t$ are recursively predicted (utilizing flow fields predicted prior time-steps as parts of the input).
    (a) Input set, (b) ground truth flow fields, and (c) flow fields predicted by the GAN.
    14 contour levels from -0.5 to 1.0 are shown.
     Solid lines and dashed lines indicate positive and negative contour levels, respectively.
    }
    \label{fig:u-T25}
\end{figure}
Regardless of the rapid loss of small-scale flow structures in the wake, flow fields predicted after a single large prediction-step interval of $25\delta t$ exhibits lower errors compared to flow fields recursively predicted at 25 small prediction-steps of $25 \times 1\delta t$ (see table~\ref{tab:com}). The reduction of errors implies that predicting with a network trained with a large time-step interval enables the network to focus more on energetic large-scale flow motions by disregarding small-scale flow motions.

\begin{table}
  \begin{center}
\def~{\hphantom{0}}
  \begin{tabular}{ccccc}
      \hline\hline
      Variable & Time-step interval & Number of recursive steps & $L_\infty$ & $L_2$ \\ \hline
      $u/U_{\infty}$ & $\delta t$ & 25 & $1.74 \pm 0.28$ &  $0.062 \pm 0.002$  \\
                           & $25 \delta t$ & 1 & $0.93 \pm 0.19$ &$0.025 \pm 0.001$\\
       $v/U_{\infty}$ & $\delta t$ & 25 & $1.72 \pm 0.39$ & $0.064 \pm 0.003$\\
                           & $25 \delta t$ & 1 & $0.95 \pm 0.15$ & $0.032 \pm 0.002$\\
       $w/U_{\infty}$ & $\delta t$ & 25 & $0.91 \pm 0.19$ & $0.030 \pm 0.003$\\
                            & $25 \delta t$ & 1 & $0.74 \pm 0.11$ & $0.015 \pm 0.001$\\
       $p/\rho U^{2}_{\infty}$ & $\delta t$ & 25 & $0.94 \pm 0.17$ & $0.040 \pm 0.004$\\
                                         & $25 \delta t$ & 1 & $0.56 \pm 0.11$ & $0.012 \pm 0.001$\\ \hline\hline
  \end{tabular}
  \caption{Comparison of errors for each flow variable at $Re_D=3900$ from predictions obtained after 25 small time-step intervals of $1\delta t$ and after a single large time-step interval of $25\delta t$. $1 \delta t = 20 \Delta t U_{\infty} /D = 0.1$.  Errors are composed of the mean and standard deviations determined by 32 independent prediction results.}
  \label{tab:com}
  \end{center}
\end{table}
\section{Conclusion}
Unsteady flow fields around a circular cylinder at Reynolds numbers that were not informed during training were predicted using deep learning techniques.
Datasets of flow fields have been constructed using numerical simulations in three different flow regimes: two-dimensional laminar vortex shedding regime, three-dimensional wake transition regime, and shear-layer transition regime.
The present deep learning techniques are found to well predict convection and diffusion of large-scale vortical structures, while the mechanism for production of small-scale vortical structures is difficult to account for.
Depending on the training scheme, the present deep learning techniques are found to be also capable of successfully predicting large-scale flow motions with large time-step interval sizes, which can be two to three-orders of magnitude larger than the time-step interval size for the conventional unsteady numerical simulations. Predictions using the present deep learning networks can be conducted with significantly lower computational cost than numerical simulations regardless of the Reynolds number. Wall-clock time of 0.3 seconds is required for a time-step advance using a single graphic processing unit (NVIDIA Titan Xp).

Four deep learning networks, GANs with and without physical loss functions and multi-scale CNNs with and without physical loss functions, have been trained and compared for the predictive performance.
The physical loss functions proposed in the present study inform the networks with, explicitly, the conservation of mass and momentum.
Adversarial training in the GAN allows the deep learning network to extract various flow features in an unsupervised manner.
All four deep learning techniques are shown to be capable of predicting flow fields in the immediate future.
However, in the long-term prediction using a recursive technique, which employs the predicted flow fields as parts of the input dataset, GANs and the multi-scale CNN with physical loss functions are shown to be more predictive than the multi-scale CNN without physical loss functions.
It has been found that the GAN without physical loss functions is the best for achieving resemblance to the ground truth flow field during recursive predictions. Especially, GAN-based networks take advantage of unsupervised training, so they can be applied to problems where underlying physics are unknown a priori.
The present deep learning methods are expected to be useful in many practical applications, such as real-time flow control and guidance of aero- or hydro-vehicles, fast weather forecast, \textit{etc.}, where fast prediction of energetic large-scale flow motions is important.

\section{Acknowledgements}\label{sec:Acknowledgements}
This work was supported by the Samsung Research Funding Center of Samsung Electronics under Project Number SRFC-TB1703-01 and National Research Foundation of Korea (NRF) under Grant Number NRF-2017R1E1A 1A03070514.

\pagebreak
\bibliographystyle{jfm}
\bibliography{SLEE_JFM_revised}

\begin{thebibliography}{35}
\expandafter\ifx\csname natexlab\endcsname\relax\def\natexlab#1{#1}\fi
\def\au#1{#1} \def\ed#1{#1} \def\yr#1{#1}\def\at#1{#1}\def\jt#1{\textit{#1}}
  \def\bt#1{#1}\def\bvol#1{\textbf{#1}} \def\vol#1{#1} \def\pg#1{#1}
  \def\publ#1{#1}\def\arxiv#1{#1}\def\org#1{#1}\def\st#1{\textit{#1}}

\bibitem[Babucke {\em et~al.\/}(2008)Babucke, Kloker \& Rist]{babucke2008dns}
{\sc \au{Babucke, Andreas}, \au{Kloker, Markus} \& \au{Rist, Ulrich}} \yr{2008}
   \at{{DNS} of a plane mixing layer for the investigation of sound generation
  mechanisms}.  \jt{Computers \& Fluids}  \bvol{37}~(4),  \pg{360--368}.

\bibitem[Bagheri(2013)]{bagheri2013koopman}
{\sc \au{Bagheri, Shervin}} \yr{2013}  \at{Koopman-mode decomposition of the
  cylinder wake}.  \jt{Journal of Fluid Mechanics}  \bvol{726},  \pg{596--623}.

\bibitem[Berger \& Wille(1972)]{berger1972periodic}
{\sc \au{Berger, Eberhard} \& \au{Wille, Rudolf}} \yr{1972}  \at{Periodic flow
  phenomena}.  \jt{Annual Review of Fluid Mechanics}  \bvol{4}~(1),
  \pg{313--340}.

\bibitem[Dalin {\em et~al.\/}(2010)Dalin, Pertsev, Frandsen, Hansen, Andersen,
  Dubietis \& Balciunas]{dalin2010case}
{\sc \au{Dalin, P}, \au{Pertsev, N}, \au{Frandsen, S}, \au{Hansen, O},
  \au{Andersen, H}, \au{Dubietis, A} \& \au{Balciunas, R}} \yr{2010}  \at{A
  case study of the evolution of a {K}elvin--{H}elmholtz wave and turbulence in
  noctilucent clouds}.  \jt{Journal of Atmospheric and Solar-Terrestrial
  Physics}  \bvol{72}~(14),  \pg{1129--1138}.

\bibitem[Denton {\em et~al.\/}(2015)Denton, Chintala \& Fergus]{denton2015deep}
{\sc \au{Denton, Emily~L}, \au{Chintala, Soumith} \& \au{Fergus, Rob}}
  \yr{2015} Deep generative image models using a {L}aplacian pyramid of
  adversarial networks.  \bt{In {\em Advances in {N}eural {I}nformation
  {P}rocessing {S}ystems\/}},  \pg{pp. 1486--1494}.

\bibitem[Freymuth(1966)]{freymuth1966transition}
{\sc \au{Freymuth, Peter}} \yr{1966}  \at{On transition in a separated laminar
  boundary layer}.  \jt{Journal of Fluid Mechanics}  \bvol{25}~(04),
  \pg{683--704}.

\bibitem[Goodfellow {\em et~al.\/}(2014)Goodfellow, Pouget-Abadie, Mirza, Xu,
  Warde-Farley, Ozair, Courville \& Bengio]{goodfellow2014generative}
{\sc \au{Goodfellow, Ian}, \au{Pouget-Abadie, Jean}, \au{Mirza, Mehdi}, \au{Xu,
  Bing}, \au{Warde-Farley, David}, \au{Ozair, Sherjil}, \au{Courville, Aaron}
  \& \au{Bengio, Yoshua}} \yr{2014} Generative adversarial nets.  \bt{In {\em
  Advances in {N}eural {I}nformation {P}rocessing {S}ystems\/}},  \pg{pp.
  2672--2680}.

\bibitem[Guo {\em et~al.\/}(2016)Guo, Li \& Iorio]{guo2016convolutional}
{\sc \au{Guo, X.}, \au{Li, W.} \& \au{Iorio, F.}} \yr{2016} Convolutional
  neural networks for steady flow approximation.  \bt{In {\em Proceedings of
  the 22nd ACM SIGKDD International Conference on Knowledge Discovery and Data
  Mining\/}},  \pg{pp. 481--490}. ACM.

\bibitem[Kingma \& Ba(2014)]{adam14}
{\sc \au{Kingma, Diederik~P} \& \au{Ba, Jimmy}} \yr{2014}  \at{Adam: A method
  for stochastic optimization}.  \jt{arXiv preprint arXiv:1412.6980} .

\bibitem[Krizhevsky {\em et~al.\/}(2012)Krizhevsky, Sutskever \&
  Hinton]{krizhevsky2012imagenet}
{\sc \au{Krizhevsky, Alex}, \au{Sutskever, Ilya} \& \au{Hinton, Geoffrey~E}}
  \yr{2012} Imagenet classification with deep convolutional neural networks.
  \bt{In {\em Advances in {N}eural {I}nformation {P}rocessing {S}ystems\/}},
  \pg{pp. 1097--1105}.

\bibitem[Liberge \& Hamdouni(2010)]{liberge2010reduced}
{\sc \au{Liberge, Erwan} \& \au{Hamdouni, Aziz}} \yr{2010}  \at{Reduced order
  modelling method via proper orthogonal decomposition (pod) for flow around an
  oscillating cylinder}.  \jt{Journal of fluids and structures}  \bvol{26}~(2),
   \pg{292--311}.

\bibitem[Ling {\em et~al.\/}(2016)Ling, Kurzawski \&
  Templeton]{ling2016reynolds}
{\sc \au{Ling, Julia}, \au{Kurzawski, Andrew} \& \au{Templeton, Jeremy}}
  \yr{2016}  \at{Reynolds averaged turbulence modelling using deep neural
  networks with embedded invariance}.  \jt{Journal of Fluid Mechanics}
  \bvol{807},  \pg{155--166}.

\bibitem[Marcus(1988)]{marcus1988numerical}
{\sc \au{Marcus, Philip~S}} \yr{1988}  \at{Numerical simulation of {J}upiter's
  great red spot}.  \jt{Nature}  \bvol{331}~(6158),  \pg{693--696}.

\bibitem[Mathieu {\em et~al.\/}(2015)Mathieu, Couprie \&
  LeCun]{mathieu2015deep}
{\sc \au{Mathieu, Michael}, \au{Couprie, Camille} \& \au{LeCun, Yann}}
  \yr{2015}  \at{Deep multi-scale video prediction beyond mean square error}.
  \jt{arXiv preprint arXiv:1511.05440} .

\bibitem[Mezi{\'c}(2013)]{mezic2013analysis}
{\sc \au{Mezi{\'c}, Igor}} \yr{2013}  \at{Analysis of fluid flows via spectral
  properties of the {K}oopman operator}.  \jt{Annual Review of Fluid Mechanics}
   \bvol{45},  \pg{357--378}.

\bibitem[Miyanawala \& Jaiman(2017)]{miyanawala2017efficient}
{\sc \au{Miyanawala, T.~P.} \& \au{Jaiman, R.~K.}} \yr{2017}  \at{An efficient
  deep learning technique for the {N}avier-{S}tokes equations: Application to
  unsteady wake flow dynamics}.  \jt{arXiv preprint arXiv:1710.09099} .

\bibitem[van~den Oord {\em et~al.\/}(2016{\natexlab{{\em a\/}}})van~den Oord,
  Kalchbrenner, Espeholt, Vinyals \& Graves]{van2016conditional}
{\sc \au{van~den Oord, Aaron}, \au{Kalchbrenner, Nal}, \au{Espeholt, Lasse},
  \au{Vinyals, Oriol} \& \au{Graves, Alex}} \yr{2016{\natexlab{{\em a\/}}}}
  Conditional image generation with {P}ixel{CNN} decoders.  \bt{In {\em
  Advances in Neural Information Processing Systems\/}},  \pg{pp. 4790--4798}.

\bibitem[van~den Oord {\em et~al.\/}(2016{\natexlab{{\em b\/}}})van~den Oord,
  Kalchbrenner \& Kavukcuoglu]{oord2016pixel}
{\sc \au{van~den Oord, Aaron}, \au{Kalchbrenner, Nal} \& \au{Kavukcuoglu,
  Koray}} \yr{2016{\natexlab{{\em b\/}}}}  \at{Pixel recurrent neural
  networks}.  \jt{arXiv preprint arXiv:1601.06759} .

\bibitem[Radford {\em et~al.\/}(2015)Radford, Metz \&
  Chintala]{radford2015unsupervised}
{\sc \au{Radford, Alec}, \au{Metz, Luke} \& \au{Chintala, Soumith}} \yr{2015}
  \at{Unsupervised representation learning with deep convolutional generative
  adversarial networks}.  \jt{arXiv preprint arXiv:1511.06434} .

\bibitem[Ranzato {\em et~al.\/}(2014)Ranzato, Szlam, Bruna, Mathieu, Collobert
  \& Chopra]{ranzato2014video}
{\sc \au{Ranzato, MarcAurelio}, \au{Szlam, Arthur}, \au{Bruna, Joan},
  \au{Mathieu, Michael}, \au{Collobert, Ronan} \& \au{Chopra, Sumit}} \yr{2014}
   \at{Video (language) modeling: a baseline for generative models of natural
  videos}.  \jt{arXiv preprint arXiv:1412.6604} .

\bibitem[Ruderich \& Fernholz(1986)]{ruderich1986experimental}
{\sc \au{Ruderich, R} \& \au{Fernholz, HH}} \yr{1986}  \at{An experimental
  investigation of a turbulent shear flow with separation, reverse flow, and
  reattachment}.  \jt{Journal of Fluid Mechanics}  \bvol{163},  \pg{283--322}.

\bibitem[Schmid(2010)]{schmid2010dynamic}
{\sc \au{Schmid, Peter~J}} \yr{2010}  \at{Dynamic mode decomposition of
  numerical and experimental data}.  \jt{Journal of Fluid Mechanics}
  \bvol{656},  \pg{5--28}.

\bibitem[Singh {\em et~al.\/}(2017)Singh, Medida \&
  Duraisamy]{singh2017machine}
{\sc \au{Singh, Anand~Pratap}, \au{Medida, Shivaji} \& \au{Duraisamy, Karthik}}
  \yr{2017}  \at{Machine-learning-augmented predictive modeling of turbulent
  separated flows over airfoils}.  \jt{AIAA Journal}  \pg{pp. 1--13}.

\bibitem[Sirovich(1987)]{sirovich1987turbulence}
{\sc \au{Sirovich, Lawrence}} \yr{1987}  \at{Turbulence and the dynamics of
  coherent structures part {I}: coherent structures}.  \jt{Quarterly of Applied
  Mathematics}  \bvol{45}~(3),  \pg{561--571}.

\bibitem[Soomro {\em et~al.\/}(2012)Soomro, Zamir \& Shah]{soomro2012ucf101}
{\sc \au{Soomro, Khurram}, \au{Zamir, Amir~Roshan} \& \au{Shah, Mubarak}}
  \yr{2012}  \at{Ucf101: A dataset of 101 human actions classes from videos in
  the wild}.  \jt{arXiv preprint arXiv:1212.0402} .

\bibitem[Srivastava {\em et~al.\/}(2015)Srivastava, Mansimov \&
  Salakhudinov]{srivastava2015unsupervised}
{\sc \au{Srivastava, Nitish}, \au{Mansimov, Elman} \& \au{Salakhudinov,
  Ruslan}} \yr{2015} Unsupervised learning of video representations using
  {LSTMs}.  \bt{In {\em International Conference on Machine Learning\/}},
  \pg{pp. 843--852}.

\bibitem[Tennekes \& Lumley(1972)]{tennekes1972first}
{\sc \au{Tennekes, H} \& \au{Lumley, JL}} \yr{1972} A first course in
  turbulence.

\bibitem[Tracey {\em et~al.\/}(2015)Tracey, Duraisamy \&
  Alonso]{tracey2015machine}
{\sc \au{Tracey, Brendan}, \au{Duraisamy, Karthik} \& \au{Alonso, Juan}}
  \yr{2015}  \at{A machine learning strategy to assist turbulence model
  development}.  \jt{AIAA Paper}  \bvol{1287},  \pg{2015}.

\bibitem[Wilson {\em et~al.\/}(2017)Wilson, Roelofs, Stern, Srebro \&
  Recht]{wilson2017marginal}
{\sc \au{Wilson, Ashia~C}, \au{Roelofs, Rebecca}, \au{Stern, Mitchell},
  \au{Srebro, Nati} \& \au{Recht, Benjamin}} \yr{2017} The marginal value of
  adaptive gradient methods in machine learning.  \bt{In {\em Advances in
  Neural Information Processing Systems\/}},  \pg{pp. 4148--4158}.

\bibitem[Wu(2011)]{wu2011fish}
{\sc \au{Wu, Theodore~Yaotsu}} \yr{2011}  \at{Fish swimming and bird/insect
  flight}.  \jt{Annual Review of Fluid Mechanics}  \bvol{43},  \pg{25--58}.

\bibitem[Wu \& Moin(2009)]{wu2009direct}
{\sc \au{Wu, Xiaohua} \& \au{Moin, Parviz}} \yr{2009}  \at{Direct numerical
  simulation of turbulence in a nominally zero-pressure-gradient flat-plate
  boundary layer}.  \jt{Journal of Fluid Mechanics}  \bvol{630},  \pg{5--41}.

\bibitem[Xingjian {\em et~al.\/}(2015)Xingjian, Chen, Wang, Yeung, Wong \&
  Woo]{xingjian2015convolutional}
{\sc \au{Xingjian, SHI}, \au{Chen, Zhourong}, \au{Wang, Hao}, \au{Yeung,
  Dit-Yan}, \au{Wong, Wai-Kin} \& \au{Woo, Wang-chun}} \yr{2015} Convolutional
  {LSTM} network: A machine learning approach for precipitation nowcasting.
  \bt{In {\em Advances in neural information processing systems\/}},  \pg{pp.
  802--810}.

\bibitem[Yonehara {\em et~al.\/}(2016)Yonehara, Goto, Yoda, Watanuki, Young,
  Weimerskirch, Bost \& Sato]{yonehara2016flight}
{\sc \au{Yonehara, Yoshinari}, \au{Goto, Yusuke}, \au{Yoda, Ken}, \au{Watanuki,
  Yutaka}, \au{Young, Lindsay~C}, \au{Weimerskirch, Henri}, \au{Bost,
  Charles-Andr{\'e}} \& \au{Sato, Katsufumi}} \yr{2016}  \at{Flight paths of
  seabirds soaring over the ocean surface enable measurement of fine-scale wind
  speed and direction}.  \jt{Proceedings of the National Academy of Sciences}
  \bvol{113}~(32),  \pg{9039--9044}.

\bibitem[You {\em et~al.\/}(2008)You, Ham \& Moin]{you2008discrete}
{\sc \au{You, D.}, \au{Ham, F.} \& \au{Moin, P.}} \yr{2008}  \at{Discrete
  conservation principles in large-eddy simulation with application to
  separation control over an airfoil}.  \jt{Physics of Fluids}  \bvol{20}~(10),
   \pg{101515}.

\bibitem[Zhang \& Duraisamy(2015)]{zhang2015machine}
{\sc \au{Zhang, Ze~Jia} \& \au{Duraisamy, Karthikeyan}} \yr{2015}  \at{Machine
  learning methods for data-driven turbulence modeling}.  \jt{AIAA}
  \bvol{2460},  \pg{2015}.

\end{thebibliography}
\appendix

\section{Loss functions}\label{appendix:loss}
$\mathcal{L}_{2}^{k}$ minimizes the difference between the predicted and the ground truth flow fields as follows:
\begin{equation}
\mathcal{L}_{2}^{k} = ||G_{k}(\mathcal{I}) - \mathcal{G}_{k}(\mathcal{I})||_{2}^{2}.
\label{eqn:loss_l2}
\end{equation}

$\mathcal{L}_{gdl}^{k}$ is a second-order central-difference version of the gradient difference loss function proposed by~\citet{mathieu2015deep}, which is applied to sharpen flow fields by directly penalizing gradient differences between the predicted and the ground truth flow fields as follows:
\begin{eqnarray}
\mathcal{L}_{gdl}^{k} =  \sum_{i} \sum_{j} \nonumber 
&\left| \bigg| \frac{\left(\mathcal{G}_{k}(\mathcal{I})_{(i+1,j)}-\mathcal{G}_{k}(\mathcal{I})_{(i-1,j)}\right)}{2} \bigg| - \bigg| \frac{\left(G_{k}(\mathcal{I})_{(i+1,j)} - G_{k}(\mathcal{I})_{(i-1,j)} \right)}{2}\bigg| \right| \nonumber \\
+ \sum_{i} \sum_{j} 
&\left| \bigg| \frac{\left(\mathcal{G}_{k}(\mathcal{I})_{(i,j+1)}-\mathcal{G}_{k}(\mathcal{I})_{(i,j-1)}\right)}{2} \bigg| - \bigg| \frac{\left(G_{k}(\mathcal{I})_{(i,j+1)} - G_{k}(\mathcal{I})_{(i,j-1)} \right)}{2} \bigg| \right|,
\label{eqn:loss_lgdl}
\end{eqnarray}
where the subscript $(i,j)$ indicates grid indices in the discretized flow domain, and $n_{x}$ and $n_{y}$ indicate the number of grid cells in $x$ and $y$ directions, respectively.

Let $u^{k}$, $v^{k}$, $w^{k}$, and $p^{k}$ be non-dimensionalized flow variables retrieved from ground truth flow fields ($\mathcal{G}_{k}(\mathcal{I})$) and $\widetilde{u}^{k}$, $\widetilde{v}^{k}$, $\widetilde{w}^{k}$, and $\widetilde{p}^{k}$ be non-dimensionalized flow variables retrieved from predicted flow fields ($G_{k}(\mathcal{I})$).
Flow variables on right, left, top, and bottom cell surfaces are calculated by the arithmetic mean between two neighboring cells as $\phi_{r} = \frac{1}{2}(\phi_{(i,j)} + \phi_{(i+1,j)})$, $\phi_{l} = \frac{1}{2}(\phi_{(i,j)} + \phi_{(i-1,j)})$, $\phi_{t} = \frac{1}{2}(\phi_{(i,j)} + \phi_{(i,j+1)})$, and $\phi_{b} = \frac{1}{2}(\phi_{(i,j)} + \phi_{(i,j-1)})$ for a variable $\phi$ which is a function of the grid index ($i$,$j$).
$\mathcal{L}_{c}$ enables networks to learn mass conservation by minimizing the total absolute sum of mass flux differences in each cell in an $x-y$ plane as follows:
\begin{eqnarray}
&&{\Delta {Con.}}^k_{(i,j)} =\big|\left(u^{k}_r - u^{k}_l\right) - \left(\widetilde{u}^{k}_r - \widetilde{u}^{k}_l\right) \big|
+ \big| \left(v^{k}_t - v^{k}_b\right) - \left(\widetilde{v}^{k}_t - \widetilde{v}^{k}_b\right) \big|, \cr\cr
&&\mathcal{L}_{c}^{k} = \sum_{i} \sum_{j} {\Delta {Con.}}^k_{(i,j)}.
\label{eqn:loss_lc}
\end{eqnarray}

$\mathcal{L}_{mom}$ enables networks to learn momentum conservation by minimizing the total absolute sum of  differences of momentum fluxes due to convection, pressure gradient, and shear stress in each cell in an $x-y$ plane as follows:
\begin{eqnarray}
{\Delta Mom.}^k_{(i,j)} &=& \big|  \left((u^{k}_r)^{2} - (u^{k}_l)^{2}\right) - \left((\widetilde{u}^{k}_r)^{2} - (\widetilde{u}^{k}_l)^{2}\right)   \big|  + \big| \left(u^{k}_rv^{k}_r- u^{k}_l v^{k}_l\right) - \left(\widetilde{u}^{k}_r \widetilde{v}^{k}_r- \widetilde{u}^{k}_l \widetilde{v}^{k}_l\right) \big| \cr\cr
&+& \big| \left((v^{k}_t)^{2} - (v^{k}_b)^{2}\right) - \left((\widetilde{v}^{k}_t)^{2} - (\widetilde{v}^{k}_b)^{2}\right) \big| + \big| \left(v^{k}_t u^{k}_t- v^{k}_b u^{k}_b\right) - \left(\widetilde{v}^{k}_t \widetilde{u}^{k}_t- \widetilde{v}^{k}_b \widetilde{u}^{k}_b\right) \big| \cr\cr
&+& \big| \left(p^{k}_r - p^{k}_l\right) - \left(\widetilde{p}^{k}_r - \widetilde{p}^{k}_l\right) \big| + \big| \left(p^{k}_t - p^{k}_b\right) - \left(\widetilde{p}^{k}_t - \widetilde{p}^{k}_b\right) \big| \cr\cr
&+&\frac{1}{Re_{D}} \bigg\{\bigg| \left( \frac{v^{k}_{(i+1,j)}-2 v^{k}_{(i,j)}+ v^{k}_{(i-1,j)}}{ \Delta_{x}} \right) 
-\left( \frac{\widetilde{v}^{k}_{(i+1,j)}-2\widetilde{v}^{k}_{(i,j)}+\widetilde{v}^{k}_{(i-1,j)}}{ \Delta_{x}}  \right) \bigg| \cr
&&\ \ \ \ \ \  + \bigg| \left( \frac{v^{k}_{(i,j+1)}-2v^{k}_{(i,j)}+v^{k}_{(i,j-1)}}{ \Delta_{y}}  \right) 
-\left( \frac{\widetilde{v}^{k}_{(i,j+1)}-2\widetilde{v}^{k}_{(i,j)}+\widetilde{v}^{k}_{(i,j-1)}}{ \Delta_{y}} \right) \bigg| \cr
&&\ \ \ \ \ \  +  \bigg| \left( \frac{u^{k}_{(i,j+1)}-2u^{k}_{(i,j)}+u^{k}_{(i,j-1)}}{ \Delta_{y}}  \right) 
-\left( \frac{\widetilde{u}^{k}_{(i,j+1)}-2\widetilde{u}^{k}_{(i,j)}+\widetilde{u}^{k}_{(i,j-1)}}{ \Delta_{y}}  \right) \bigg| \cr
&&\ \ \ \ \ \  +  \bigg| \left( \frac{u^{k}_{(i+1,j)}-2u^{k}_{(i,j)}+u^{k}_{(i-1,j)}}{ \Delta_{x}}  \right) 
-\left( \frac{\widetilde{u}^{k}_{(i+1,j)}-2\widetilde{u}^{k}_{(i,j)}+\widetilde{u}^{k}_{(i-1,j)}}{ \Delta_{x}}  \right) \bigg| \bigg\},\cr\cr
\mathcal{L}_{mom}^{k} &=& \sum_{i} \sum_{j} {\Delta Mom.}^k_{(i,j)}, 
\label{eqn:loss_lmom}
\end{eqnarray}
where $\Delta_{x}$ and $\Delta_{y}$ are grid spacings in $x$ and $y$ directions, respectively.

$\mathcal{L}_{adv}^{G}$ is a loss function with purpose to delude the discriminator model to classify generated flow fields to class 1 as follows:
\begin{equation}
\mathcal{L}_{adv}^{G} = L_{bce}(D_{k}(G_{k}(\mathcal{I})),1).
\label{eqn:loss_ladvg}
\end{equation}

\section{Error functions}\label{appendix:err}
Let $u$, $v$, $w$, and $p$ be non-dimensionalized flow variables retrieved from ground truth flow fields and $\widetilde{u}$, $\widetilde{v}$, $\widetilde{w}$, and $\widetilde{p}$ be non-dimensionalized flow variables retrieved from predicted flow fields. Error functions are defined as follows:

\begin{eqnarray}
&&{L}_{2}= \bigg(\frac{1}{4n_{x} n_y}\sum_{i} \sum_{j} \{(u_{(i,j)}-\widetilde{u}_{(i,j)})^{2} + (v_{(i,j)}-\widetilde{v}_{(i,j)})^{2} \cr\cr
&&\ \ \ \ \  \ \ \ \ \ \  + (w_{(i,j)}-\widetilde{w}_{(i,j)})^{2} + (p_{(i,j)}-\widetilde{p}_{(i,j)})^{2} \} \bigg)^{1/2},
\label{eqn:err_l2}\\
&&L_{\infty} = \frac{1}{4} \bigg( \max_{i,j} | u_{(i,j)} - \widetilde{u}_{(i,j)} | + \max_{i,j} | v_{(i,j)} - \widetilde{v}_{(i,j)} |   \cr\cr 
&&\ \ \ \ \  \ \ \ \ \ \ + \max_{i,j} | w_{(i,j)} - \widetilde{w}_{(i,j)} | +  \max_{i,j} | p_{(i,j)} - \widetilde{p}_{(i,j)} | \bigg),
\label{eqn:err_linf} \\
&&{L}_{c}= \frac{1}{n_x n_y}\sum_{i} \sum_{j} \Delta Con._{(i,j)},
\label{eqn:err_lc}\\
&&{L}_{mom} = \frac{1}{n_x n_y}\sum_{i} \sum_{j} \Delta Mom._{(i,j)},
\label{eqn:err_lmom}
\end{eqnarray}
where $\Delta Con._{(i,j)}$ and $\Delta Mom._{(i,j)}$ are defined in equations~(\ref{eqn:loss_lc}) and~(\ref{eqn:loss_lmom}), respectively.

The present loss functions and error functions for conservation of mass and momentum are not identical to the original forms of conservation laws, but are formulated using the triangle inequality. Therefore, the minimization of the present physical loss functions satisfies conservation of mass and momentum more strictly. In fact, smaller errors are calculated using the original forms of conservation laws, while the errors behave similarly to $L_c$ and $L_{mom}$ as a function of $\delta t$.

\section{Parameter study}
\subsection{Effects of numbers of layers and feature maps}\label{appendix:size}
\begin{table}
        \centering
        \resizebox{0.9\textwidth}{!}{%
        \begin{tabular}{l|l|l}
            \hline \hline
            \multicolumn{3}{c}{$GM_{16}$}\\ \hline 
                Generative CNN & Numbers of feature maps & Kernel sizes\\ \hline
                $G_{3}$ & 16 $\mathcal{N}_{1}$ $\mathcal{N}_{1}$ 4 & $3\times3$, $3\times3$, $3\times3$\\
                $G_{2}$ & 20 $\mathcal{N}_{1}$ $\mathcal{N}_{1}$ 4 & $5\times5$, $3\times3$, $5\times5$ \\
                $G_{1}$ & 20 $\mathcal{N}_{1}$ $\mathcal{N}_{2}$ $\mathcal{N}_{2}$ $\mathcal{N}_{1}$ 4& $5\times5$, $3\times3$, $3\times3$, $3\times3$, $5\times5$  \\
                $G_{0}$ & 20 $\mathcal{N}_{1}$ $\mathcal{N}_{2}$ $\mathcal{N}_{2}$ $\mathcal{N}_{1}$ 4 & $7\times7$, $5\times5$, $5\times5$, $5\times5$, $7\times7$ \\  \hline\hline
            \multicolumn{3}{c}{$GM_{18}$}\\ \hline 
                Generative CNN & Numbers of feature maps & Kernel sizes\\ \hline
                $G_{3}$ & 16 $\mathcal{N}_{1}$ $\mathcal{N}_{2}$ $\mathcal{N}_{1}$ 4 & $3\times3$, $3\times3$, $3\times3$, $3\times3$\\
                $G_{2}$ & 20 $\mathcal{N}_{1}$ $\mathcal{N}_{2}$ $\mathcal{N}_{1}$ 4 & $5\times5$, $3\times3$, $3\times3$, $5\times5$ \\
                $G_{1}$ & 20 $\mathcal{N}_{1}$ $\mathcal{N}_{2}$ $\mathcal{N}_{2}$ $\mathcal{N}_{1}$ 4& $5\times5$, $3\times3$, $3\times3$, $3\times3$, $5\times5$  \\
                $G_{0}$ & 20 $\mathcal{N}_{1}$ $\mathcal{N}_{2}$ $\mathcal{N}_{2}$ $\mathcal{N}_{1}$ 4 & $7\times7$, $5\times5$, $5\times5$, $5\times5$, $7\times7$ \\  \hline\hline
            \multicolumn{3}{c}{$GM_{20}$}\\ \hline 
                Generative CNN & Numbers of feature maps & Kernel sizes\\ \hline
                $G_{3}$ & 16 $\mathcal{N}_{1}$ $\mathcal{N}_{2}$ $\mathcal{N}_{1}$ 4 & $3\times3$, $3\times3$, $3\times3$, $3\times3$\\
                $G_{2}$ & 20 $\mathcal{N}_{1}$ $\mathcal{N}_{2}$ $\mathcal{N}_{1}$ 4 & $5\times5$, $3\times3$, $3\times3$, $5\times5$ \\
                $G_{1}$ & 20 $\mathcal{N}_{1}$ $\mathcal{N}_{2}$ $\mathcal{N}_{3}$ $\mathcal{N}_{2}$ $\mathcal{N}_{1}$ 4& $5\times5$, $3\times3$, $3\times3$, $3\times3$, $3\times3$, $5\times5$  \\
                $G_{0}$ & 20 $\mathcal{N}_{1}$ $\mathcal{N}_{2}$ $\mathcal{N}_{3}$ $\mathcal{N}_{2}$ $\mathcal{N}_{1}$ 4 & $7\times7$, $5\times5$, $5\times5$, $5\times5$, $5\times5$, $7\times7$ \\  \hline\hline
            \multicolumn{3}{c}{\textbf{Number sets}} \\ \hline
                $N_{32}$ & $N_{64}$ & $N_{128}$ \\ \hline 
                $\mathcal{N}_{1}=32$, $\mathcal{N}_{2}=64$, $\mathcal{N}_{3}=128$, & $\mathcal{N}_{1}=64$, $\mathcal{N}_{2}=128$, $\mathcal{N}_{3}=256$ & $\mathcal{N}_{1}=128$, $\mathcal{N}_{2}=256$, $\mathcal{N}_{3}=512$ \\ \hline
        \end{tabular}}
        \caption{Configurations ($GM_{16}$, $GM_{18}$, and $GM_{20}$) and number sets ($N_{32}$, $N_{64}$, and $N_{128}$) of generator models used in the parameter study.}
        \label{tab:generator_size}
\end{table}

\begin{figure}
    \centering
    \subfigure{\includegraphics[width = 0.345\linewidth,trim={0.3cm 0.3cm 0.3cm 0.3cm},clip]{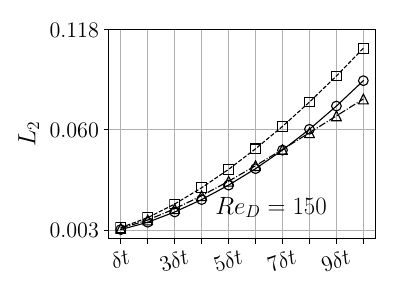}}
    \setcounter{subfigure}{0}%
    \subfigure[]{\includegraphics[width = 0.315\linewidth,trim={1.10cm 0.3cm 0.3cm 0.3cm},clip]{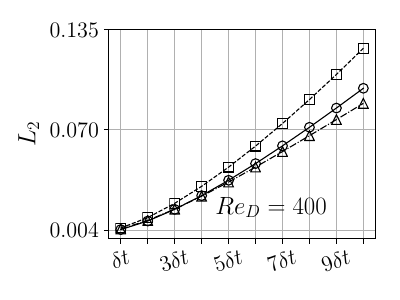}}
    \subfigure{\includegraphics[width = 0.315\linewidth,trim={1.10cm 0.3cm 0.3cm 0.3cm},clip]{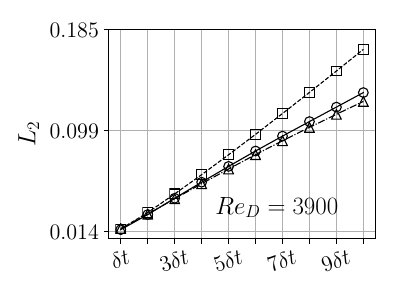}}
    \vspace{-0.3cm}

    \subfigure{\includegraphics[width = 0.345\linewidth,trim={0.3cm 0.3cm 0.3cm 0.3cm},clip]{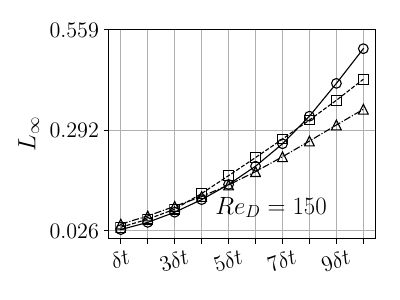}}
    \setcounter{subfigure}{1}%
    \subfigure[]{\includegraphics[width = 0.315\linewidth,trim={1.10cm 0.3cm 0.3cm 0.3cm},clip]{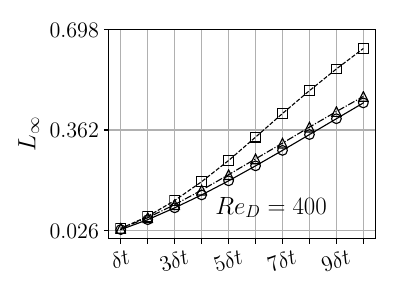}}
    \subfigure{\includegraphics[width = 0.315\linewidth,trim={1.10cm 0.3cm 0.3cm 0.3cm},clip]{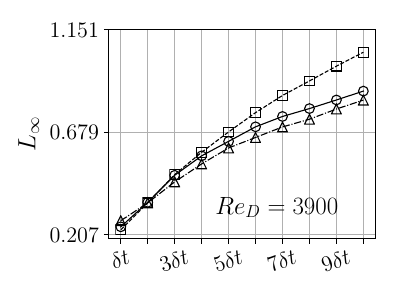}}
    \vspace{-0.3cm}

    \subfigure{\includegraphics[width = 0.345\linewidth,trim={0.3cm 0.3cm 0.3cm 0.3cm},clip]{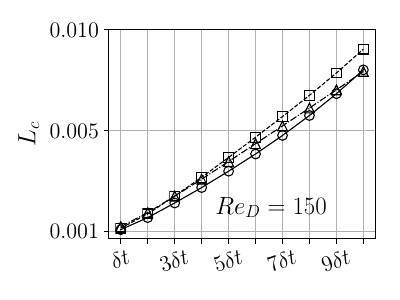}}
    \setcounter{subfigure}{2}%
    \subfigure[]{\includegraphics[width = 0.315\linewidth,trim={1.10cm 0.3cm 0.3cm 0.3cm},clip]{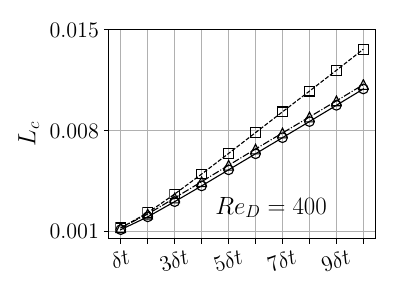}}
    \subfigure{\includegraphics[width = 0.315\linewidth,trim={1.10cm 0.3cm 0.3cm 0.3cm},clip]{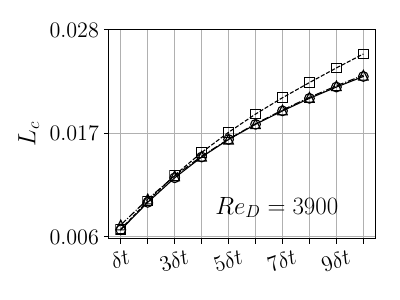}}
    \vspace{-0.3cm}

    \subfigure{\includegraphics[width = 0.345\linewidth,trim={0.3cm 0.3cm 0.3cm 0.3cm},clip]{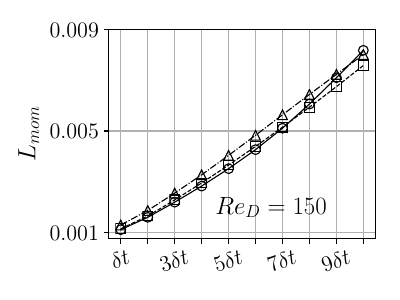}}
    \setcounter{subfigure}{3}%
    \subfigure[]{\includegraphics[width = 0.315\linewidth,trim={1.10cm 0.3cm 0.3cm 0.3cm},clip]{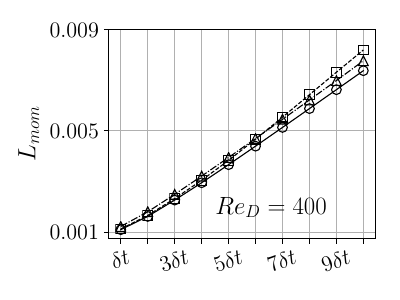}}
    \subfigure{\includegraphics[width = 0.315\linewidth,trim={1.10cm 0.3cm 0.3cm 0.3cm},clip]{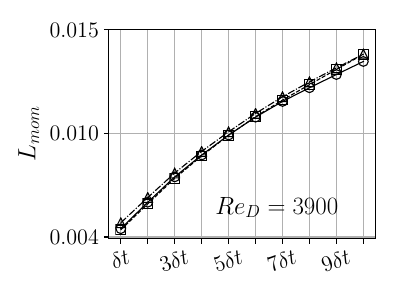}}
    \vspace{-0.3cm}
        \caption{Configuration dependency of the generator model.  (a) $L_{2}$, (b) $L_{\infty}$, (c) $L_{c}$, and (d) $L_{mom}$ errors from the multi-scale CNN without physical loss functions as a function of recursive prediction steps $\delta t$, where $\delta t = 20 \Delta t U_{\infty}/D = 0.1$.
    $\circ$ and solid line denote errors from $GM_{16}$;
    $\square$ and dashed line denote errors from $GM_{18}$;
    $\triangle$ and dash-dotted line denote errors from $GM_{20}$.}
    \label{fig:GM}
\end{figure}

Errors as a function of the number of convolution layers of the generator model are calculated by training three generator models with configurations of $GM_{16}$, $GM_{18}$, and $GM_{20}$ with the number set of $N_{128}$ (see table~\ref{tab:generator_size} for these configurations), while errors as a function of the number of feature maps of the generator model in multi-scale CNNs are calculated by training the generator model with number sets $N_{32}$, $N_{64}$, and $N_{128}$ with the configuration of $GM_{20}$.
All networks are trained with flow fields at $Re_{D}=300$ and $500$.
Magnitudes of errors in configurations considered in the present study are found not to be reduced monotonically with the increase of numbers of layers and feature maps. 
The configuration with the largest number of convolution layers ($GM_{20}$) tends to show smaller $L_{2}$ and $L_{\infty}$ errors, while shows $L_{c}$ and $L_{mom}$ errors of which magnitudes are similar to or smaller than those in configurations with smaller numbers of convolution layers ($GM_{16}$ and $GM{18}$) (figure~\ref{fig:GM}).

\begin{figure}
    \centering
    \subfigure{\includegraphics[width = 0.345\linewidth,trim={0.3cm 0.3cm 0.3cm 0.3cm},clip]{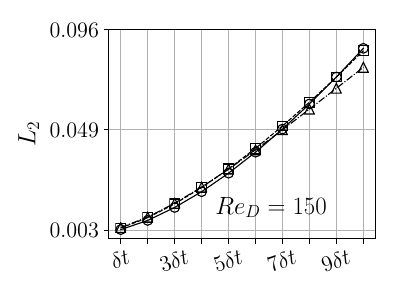}}
    \setcounter{subfigure}{0}%
    \subfigure[]{\includegraphics[width = 0.315\linewidth,trim={1.10cm 0.3cm 0.3cm 0.3cm},clip]{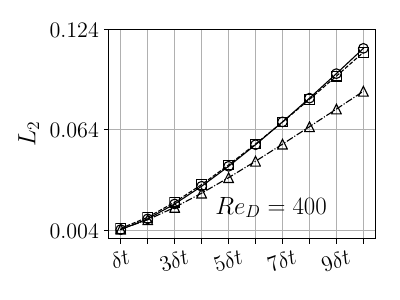}}
    \subfigure{\includegraphics[width = 0.315\linewidth,trim={1.10cm 0.3cm 0.3cm 0.3cm},clip]{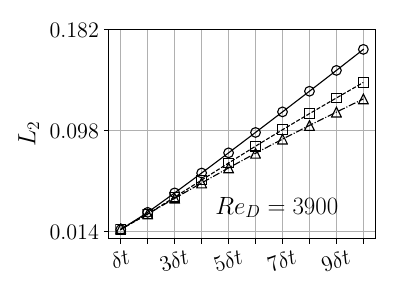}}
    \vspace{-0.3cm}

    \subfigure{\includegraphics[width = 0.345\linewidth,trim={0.3cm 0.3cm 0.3cm 0.3cm},clip]{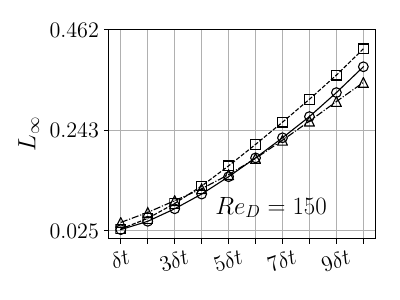}}
    \setcounter{subfigure}{1}%
    \subfigure[]{\includegraphics[width = 0.315\linewidth,trim={1.10cm 0.3cm 0.3cm 0.3cm},clip]{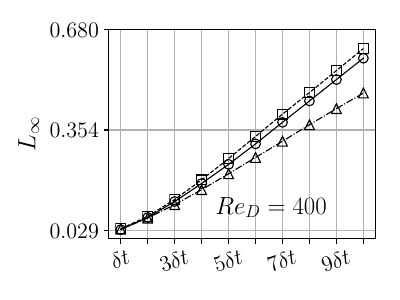}}
    \subfigure{\includegraphics[width = 0.315\linewidth,trim={1.10cm 0.3cm 0.3cm 0.3cm},clip]{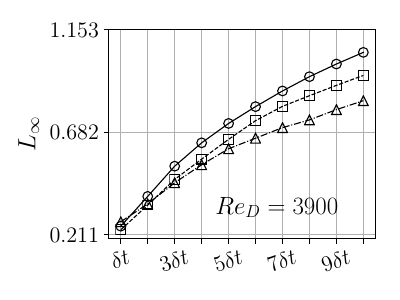}}
    \vspace{-0.3cm}

    \subfigure{\includegraphics[width = 0.345\linewidth,trim={0.3cm 0.3cm 0.3cm 0.3cm},clip]{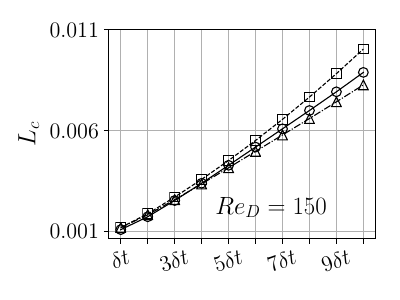}}
    \setcounter{subfigure}{2}%
    \subfigure[]{\includegraphics[width = 0.315\linewidth,trim={1.10cm 0.3cm 0.3cm 0.3cm},clip]{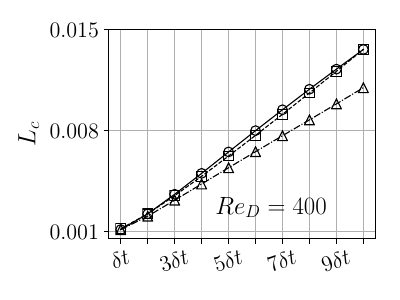}}
    \subfigure{\includegraphics[width = 0.315\linewidth,trim={1.10cm 0.3cm 0.3cm 0.3cm},clip]{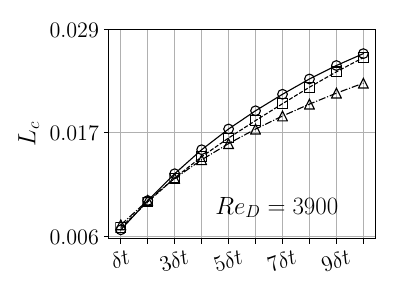}}
    \vspace{-0.3cm}

    \subfigure{\includegraphics[width = 0.345\linewidth,trim={0.3cm 0.3cm 0.3cm 0.3cm},clip]{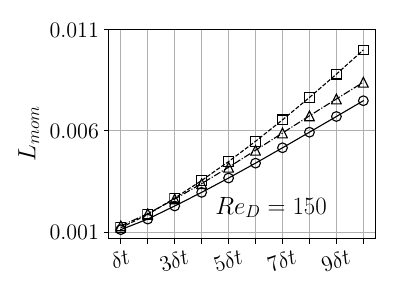}}
    \setcounter{subfigure}{3}%
    \subfigure[]{\includegraphics[width = 0.315\linewidth,trim={1.10cm 0.3cm 0.3cm 0.3cm},clip]{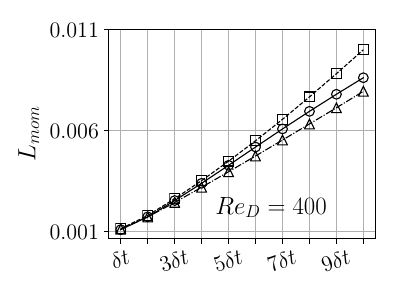}}
    \subfigure{\includegraphics[width = 0.315\linewidth,trim={1.10cm 0.3cm 0.3cm 0.3cm},clip]{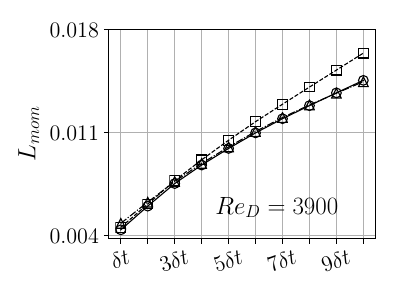}}
    \vspace{-0.3cm}
    \caption{Number set dependency of (a) $L_{2}$, (b) $L_{\infty}$, (c) $L_{c}$, and (d) $L_{mom}$ errors from the multi-scale CNN without physical loss functions as a function of recursive prediction steps $\delta t$, where $\delta t = 20 \Delta t U_{\infty}/D = 0.1$.
    $\circ$ and solid line denote errors from $N_{32}$;
    $\square$ and dashed line denote errors from $N_{64}$;
    $\triangle$ and dash-dotted line denote errors from $N_{128}$.}
    \label{fig:NS}
\end{figure}

The generator model with the largest number set $N_{128}$ tends to show smaller errors (except for the $L_{mom}$ error at $Re_D= 150$) on recursive prediction steps compared to smaller number sets models ($N_{32}$ and $N_{64}$) (figure~\ref{fig:NS}).
Therefore, the present study utilizes generator models with the configuration of $GM_{20}$ and with the number set of $N_{128}$.
 
\begin{figure}
    \centering
    \subfigure{\includegraphics[width = 0.35\linewidth,trim={0.2cm 0.3cm 0.3cm 0.3cm},clip]{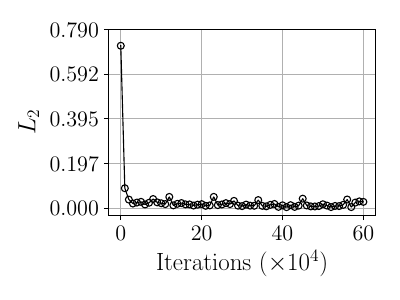}}
    \subfigure{\includegraphics[width = 0.35\linewidth,trim={0.2cm 0.3cm 0.3cm 0.3cm},clip]{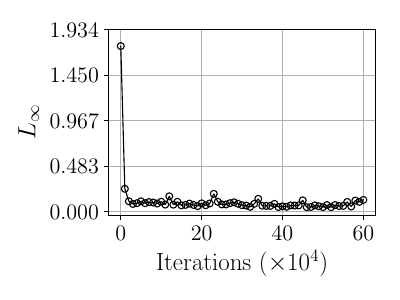}}
    \vspace{-0.3cm}
 
    \subfigure{\includegraphics[width = 0.35\linewidth,trim={0.2cm 0.3cm 0.3cm 0.3cm},clip]{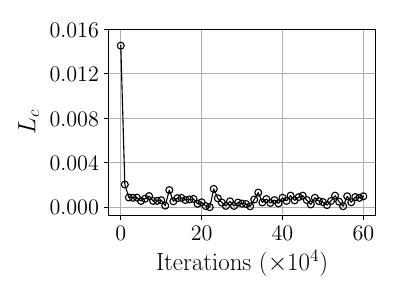}}
    \subfigure{\includegraphics[width = 0.35\linewidth,trim={0.2cm 0.3cm 0.3cm 0.3cm},clip]{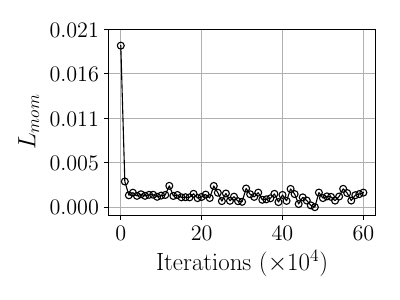}}
    \vspace{-0.3cm}
    \caption{Errors from the multi-scale CNN without physical loss functions as a function of the number of training iteration. The network is trained with flow fields at $Re_{D}=300$ and $500$. The errors are evaluated for flow predictions at $Re_{D}=400$.}
    \label{fig:iter}
\end{figure}

Figure~\ref{fig:iter} shows variations of $L_2$, $L_\infty$, $L_c$, and $L_{mom}$ errors as a function of training iteration number for the multi-scale CNN without physical loss functions. All errors are found to converge without overfitting. 

\subsection{Effects of $\lambda_{adv}$}\label{appendix:adv}
\begin{figure}
    \centering
    \subfigure{\includegraphics[width=0.235\linewidth,trim={0.3cm 0.3cm 0.3cm 0.3cm},clip]{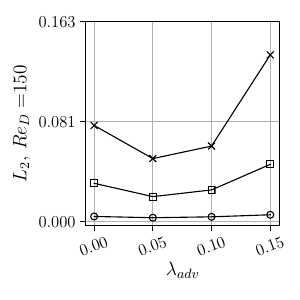}}
    \subfigure{\includegraphics[width=0.235\linewidth,trim={0.3cm 0.3cm 0.3cm 0.3cm},clip]{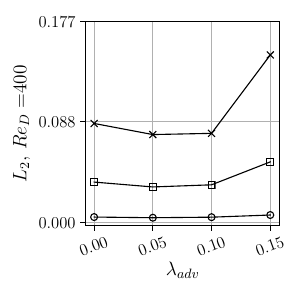}}
    \subfigure{\includegraphics[width=0.235\linewidth,trim={0.3cm 0.3cm 0.3cm 0.3cm},clip]{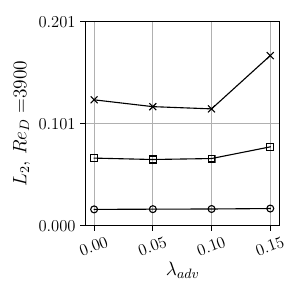}}
 
    \subfigure{\includegraphics[width=0.235\linewidth,trim={0.3cm 0.3cm 0.3cm 0.3cm},clip]{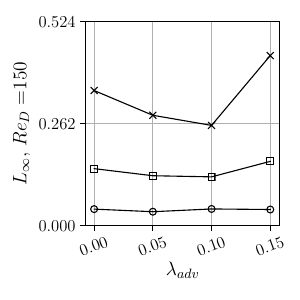}}
    \subfigure{\includegraphics[width=0.235\linewidth,trim={0.3cm 0.3cm 0.3cm 0.3cm},clip]{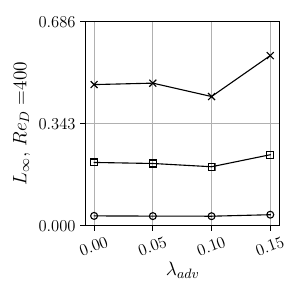}}
    \subfigure{\includegraphics[width=0.235\linewidth,trim={0.3cm 0.3cm 0.3cm 0.3cm},clip]{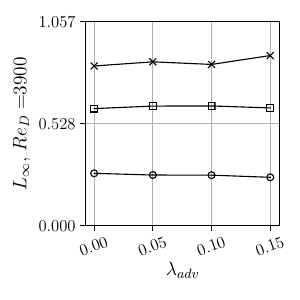}}
 
    \subfigure{\includegraphics[width=0.235\linewidth,trim={0.3cm 0.3cm 0.3cm 0.3cm},clip]{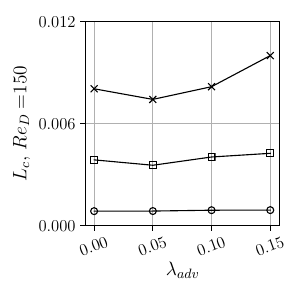}}
    \subfigure{\includegraphics[width=0.235\linewidth,trim={0.3cm 0.3cm 0.3cm 0.3cm},clip]{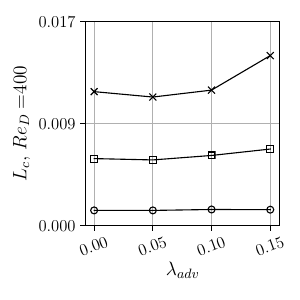}}
    \subfigure{\includegraphics[width=0.235\linewidth,trim={0.3cm 0.3cm 0.3cm 0.3cm},clip]{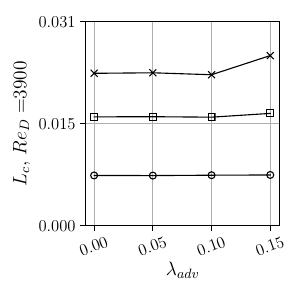}}
 
    \subfigure{\includegraphics[width=0.235\linewidth,trim={0.3cm 0.3cm 0.3cm 0.3cm},clip]{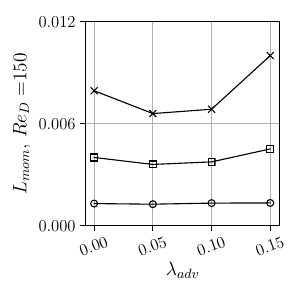}}
    \subfigure{\includegraphics[width=0.235\linewidth,trim={0.3cm 0.3cm 0.3cm 0.3cm},clip]{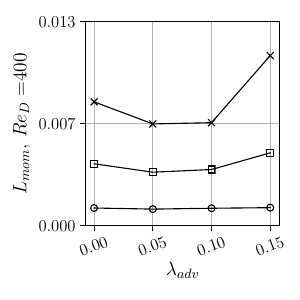}}
    \subfigure{\includegraphics[width=0.235\linewidth,trim={0.3cm 0.3cm 0.3cm 0.3cm},clip]{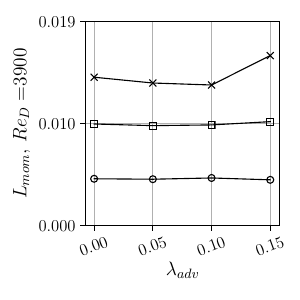}}
    \caption{Errors as a function of $\lambda_{adv}$. $\circ$, $\square$, and $\times$ indicate errors after $1\delta t$, $4 \delta t$, and $10\delta t$, respectively.}
    \label{fig:ADV}
\end{figure}
$L_{2}$, $L_{\infty}$, $L_{c}$, and $L_{mom}$ errors from the GAN without physical loss functions using different adversarial training coefficients ($\lambda_{adv}=$  0.00, 0.05, 0.10, 0.15) are compared in figure~\ref{fig:ADV}. For the present parameter study, $\lambda_{l2}$ and $\lambda_{gdl}$ are fixed to 1 and $\lambda_{phy}$ is fixed to 0.
The GAN is trained with flow fields at $Re_{D}=300$ and $500$ and tested on flow fields at $Re_{D}=150$, $400$, and $3900$.
$\lambda_{adv}$ of $0.10$ is selected for the present analysis in the result section because the case shows small $L_{\infty}$ errors in all Reynolds numbers and smallest $L_{2}$, $L_{c}$, and $L_{mom}$ errors at $Re_{D}=3900$.

\subsection{Effects of $\lambda_{phy}$}\label{appendix:phy}
\begin{figure}
    \centering
    \subfigure{\includegraphics[width=0.235\linewidth,trim={0.3cm 0.3cm 0.3cm 0.3cm},clip]{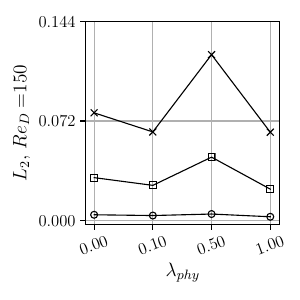}}
    \subfigure{\includegraphics[width=0.235\linewidth,trim={0.3cm 0.3cm 0.3cm 0.3cm},clip]{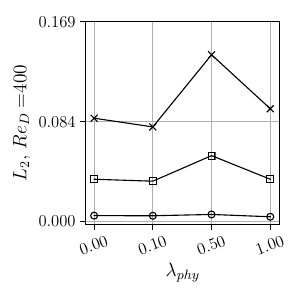}}
    \subfigure{\includegraphics[width=0.235\linewidth,trim={0.3cm 0.3cm 0.3cm 0.3cm},clip]{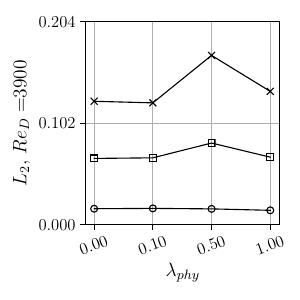}}

    \subfigure{\includegraphics[width=0.235\linewidth,trim={0.3cm 0.3cm 0.3cm 0.3cm},clip]{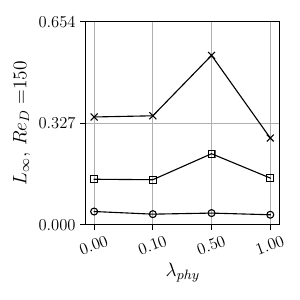}}
    \subfigure{\includegraphics[width=0.235\linewidth,trim={0.3cm 0.3cm 0.3cm 0.3cm},clip]{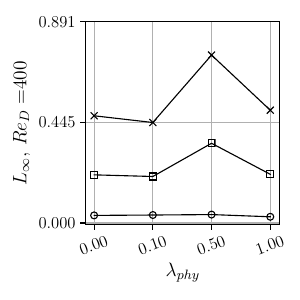}}
    \subfigure{\includegraphics[width=0.235\linewidth,trim={0.3cm 0.3cm 0.3cm 0.3cm},clip]{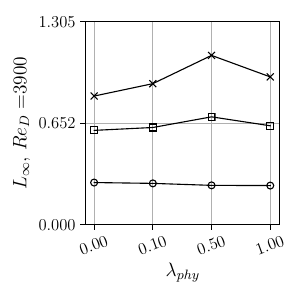}}

    \subfigure{\includegraphics[width=0.235\linewidth,trim={0.3cm 0.3cm 0.3cm 0.3cm},clip]{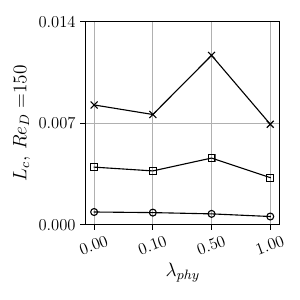}}
    \subfigure{\includegraphics[width=0.235\linewidth,trim={0.3cm 0.3cm 0.3cm 0.3cm},clip]{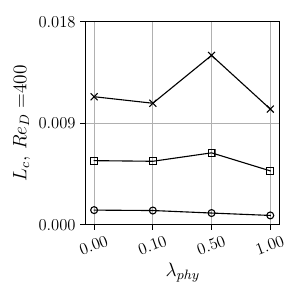}}
    \subfigure{\includegraphics[width=0.235\linewidth,trim={0.3cm 0.3cm 0.3cm 0.3cm},clip]{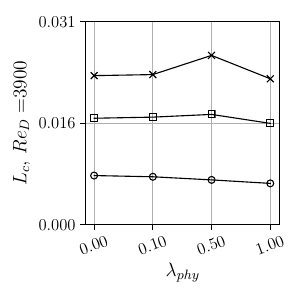}}

    \subfigure{\includegraphics[width=0.235\linewidth,trim={0.3cm 0.3cm 0.3cm 0.3cm},clip]{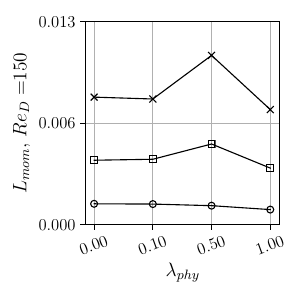}}
    \subfigure{\includegraphics[width=0.235\linewidth,trim={0.3cm 0.3cm 0.3cm 0.3cm},clip]{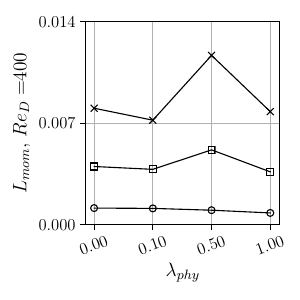}}
    \subfigure{\includegraphics[width=0.235\linewidth,trim={0.3cm 0.3cm 0.3cm 0.3cm},clip]{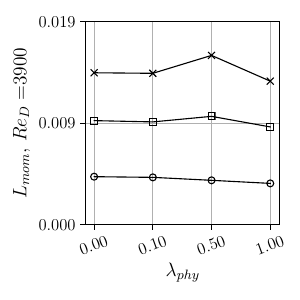}}
    \caption{Errors as a function of $\lambda_{phy}$. $\circ$, $\square$, and $\times$ indicate errors after $1\delta t$, $4 \delta t$, and $10\delta t$, respectively.}
    \label{fig:PHY}
\end{figure}

$L_{2}$, $L_{\infty}$, $L_{c}$, and $L_{mom}$ errors from the multi-scale CNN with physical loss functions using different coefficients ($\lambda_{phy}=$  0.00, 0.10, 0.50, 1.00) are compared in figure~\ref{fig:PHY}. $\lambda_{l2}$ and $\lambda_{gdl}$ are fixed to 1 and $\lambda_{adv}$ is fixed to 0.
The multi-scale CNN is trained with flow fields at $Re_{D}=300$ and $500$ and tested on flow fields at $Re_{D}=150$, $400$, and $3900$.
$\lambda_{phy}$ of $1.00$ has been selected for the analysis in the result section because it shows relatively small $L_{c}$ and $L_{mom}$ errors at all Reynolds numbers (see figure~\ref{fig:PHY}).

\section{Flow fields predicted by the GAN trained with a small time-step interval}\label{appendix:small_fields}
\begin{figure}
    \centering
    \subfigure{\includegraphics[width=0.17\linewidth,trim={0.5cm 0.5cm 0.5cm 0.5cm},clip]{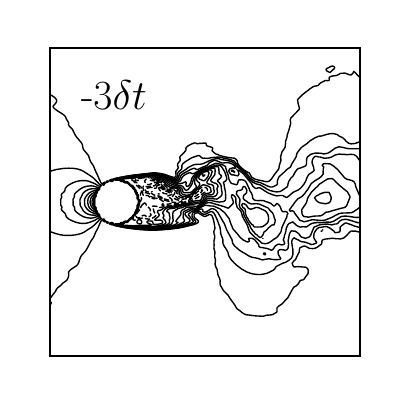}}
    \subfigure{\includegraphics[width=0.17\linewidth,trim={0.5cm 0.5cm 0.5cm 0.5cm},clip]{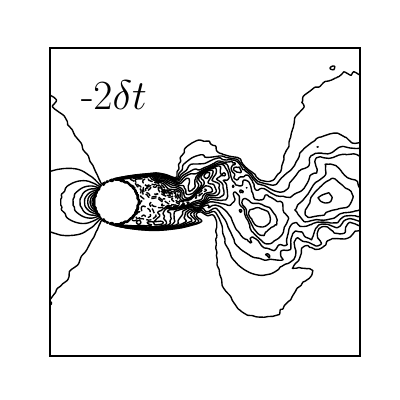}}
    \setcounter{subfigure}{0}%
    \subfigure[]{}\hspace{-1mm}
    \subfigure{\includegraphics[width=0.17\linewidth,trim={0.5cm 0.5cm 0.5cm 0.5cm},clip]{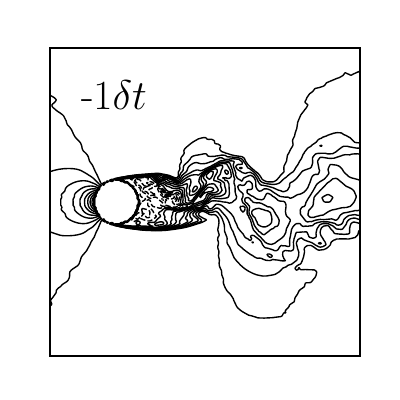}}
    \subfigure{\includegraphics[width=0.17\linewidth,trim={0.5cm 0.5cm 0.5cm 0.5cm},clip]{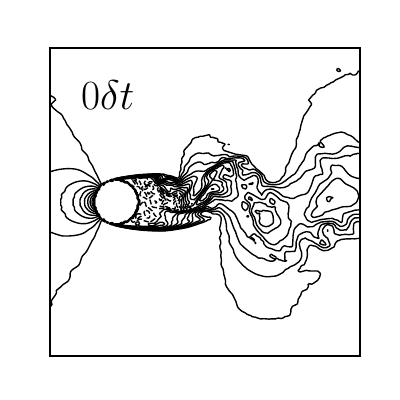}}

    \subfigure{\includegraphics[width=0.17\linewidth,trim={0.5cm 0.5cm 0.5cm 0.5cm},clip]{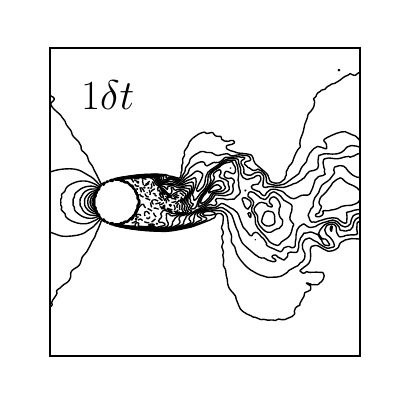}}
    \subfigure{\includegraphics[width=0.17\linewidth,trim={0.5cm 0.5cm 0.5cm 0.5cm},clip]{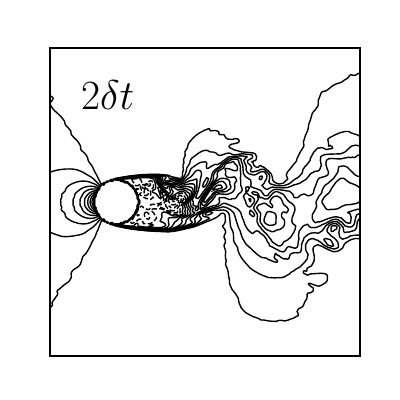}}
    \setcounter{subfigure}{1}%
    \subfigure[]{\includegraphics[width=0.17\linewidth,trim={0.5cm 0.5cm 0.5cm 0.5cm},clip]{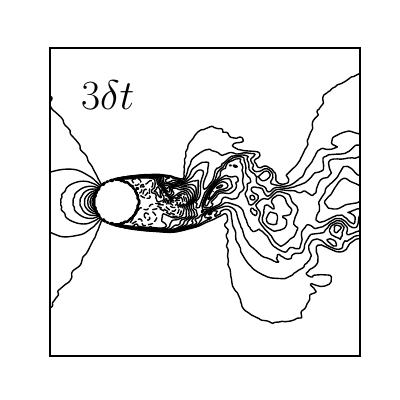}}
    \subfigure{\includegraphics[width=0.17\linewidth,trim={0.5cm 0.5cm 0.5cm 0.5cm},clip]{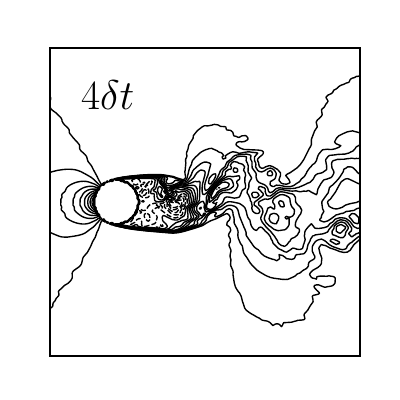}}
    \subfigure{\includegraphics[width=0.17\linewidth,trim={0.5cm 0.5cm 0.5cm 0.5cm},clip]{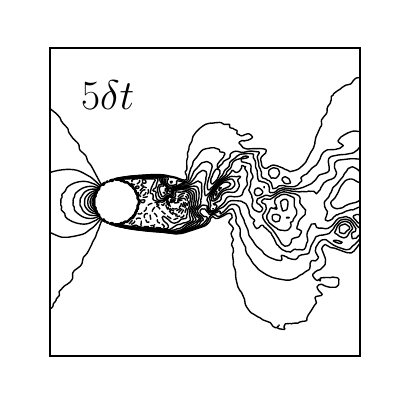}}

    \subfigure{\includegraphics[width=0.17\linewidth,trim={0.5cm 0.5cm 0.5cm 0.5cm},clip]{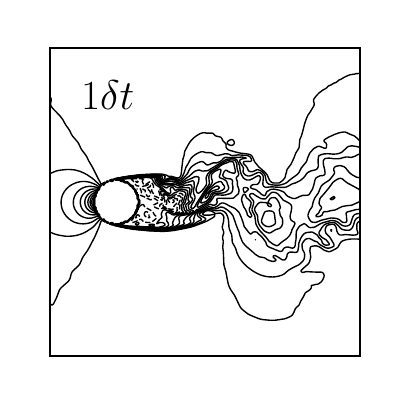}}
    \subfigure{\includegraphics[width=0.17\linewidth,trim={0.5cm 0.5cm 0.5cm 0.5cm},clip]{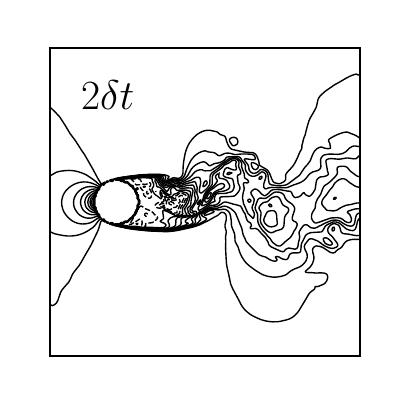}}
    \setcounter{subfigure}{2}%
    \subfigure[]{\includegraphics[width=0.17\linewidth,trim={0.5cm 0.5cm 0.5cm 0.5cm},clip]{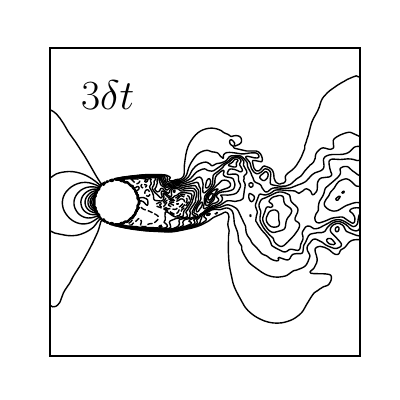}}
    \subfigure{\includegraphics[width=0.17\linewidth,trim={0.5cm 0.5cm 0.5cm 0.5cm},clip]{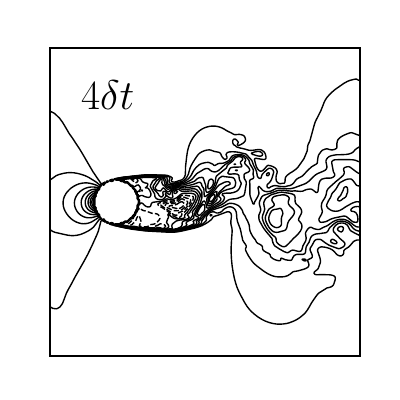}}
    \subfigure{\includegraphics[width=0.17\linewidth,trim={0.5cm 0.5cm 0.5cm 0.5cm},clip]{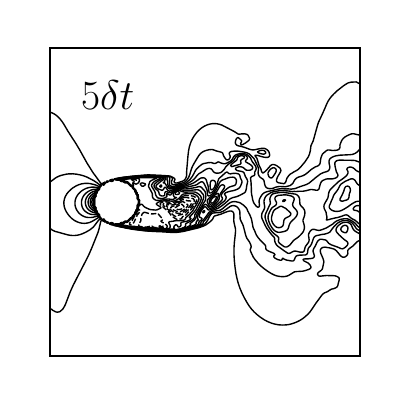}}
    \caption{
    Contour plots of the streamwise velocity ($u/U_{\infty}$) at $Re_{D}=3900$ after $1 \delta t$, $2 \delta t$, $3 \delta t$, $4 \delta t$, and $5 \delta t$, where $1 \delta t = 20 \Delta t U_{\infty}/D = 0.1$. Flow fields at $2 \delta t$, $3 \delta t$, $4 \delta t$, and $5 \delta t$ are recursively predicted (utilizing flow fields predicted prior time-steps as parts of the input).
    (a) Input set, (b) ground truth flow fields, and (c) flow fields predicted by the GAN.
    20 contour levels from -0.5 to 1.0 are shown.
     Solid lines and dashed lines indicate positive and negative contour levels, respectively.
    }
    \label{fig:u-R1R3R4}
\end{figure}

\begin{figure}
    \centering
    \subfigure{\includegraphics[width=0.17\linewidth,trim={0.5cm 0.5cm 0.5cm 0.5cm},clip]{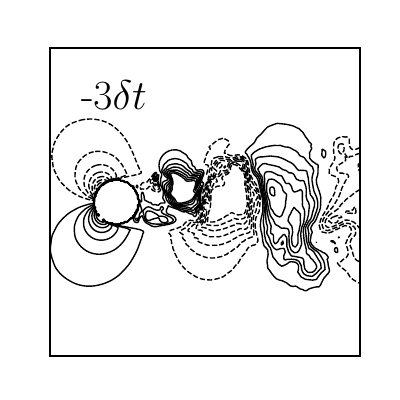}}
    \subfigure{\includegraphics[width=0.17\linewidth,trim={0.5cm 0.5cm 0.5cm 0.5cm},clip]{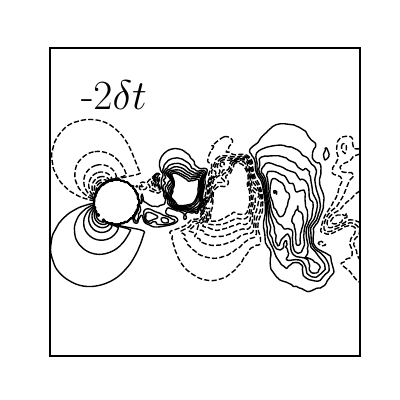}}
    \setcounter{subfigure}{0}%
    \subfigure[]{}\hspace{-1mm}
    \subfigure{\includegraphics[width=0.17\linewidth,trim={0.5cm 0.5cm 0.5cm 0.5cm},clip]{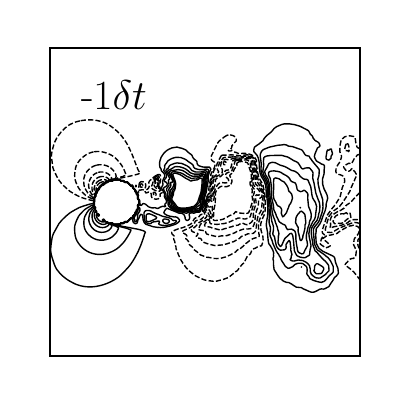}}
    \subfigure{\includegraphics[width=0.17\linewidth,trim={0.5cm 0.5cm 0.5cm 0.5cm},clip]{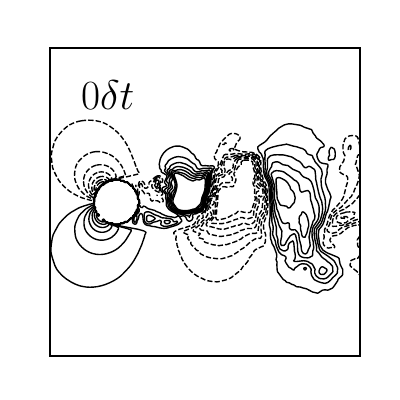}}

    \subfigure{\includegraphics[width=0.17\linewidth,trim={0.5cm 0.5cm 0.5cm 0.5cm},clip]{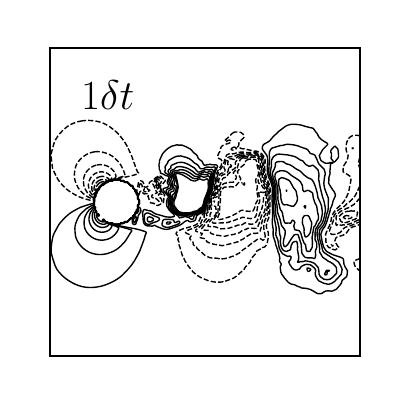}}
    \subfigure{\includegraphics[width=0.17\linewidth,trim={0.5cm 0.5cm 0.5cm 0.5cm},clip]{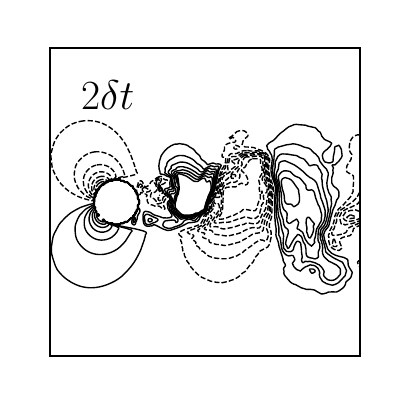}}
    \setcounter{subfigure}{1}%
    \subfigure[]{\includegraphics[width=0.17\linewidth,trim={0.5cm 0.5cm 0.5cm 0.5cm},clip]{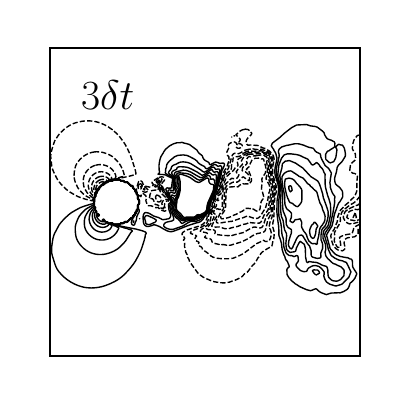}}
    \subfigure{\includegraphics[width=0.17\linewidth,trim={0.5cm 0.5cm 0.5cm 0.5cm},clip]{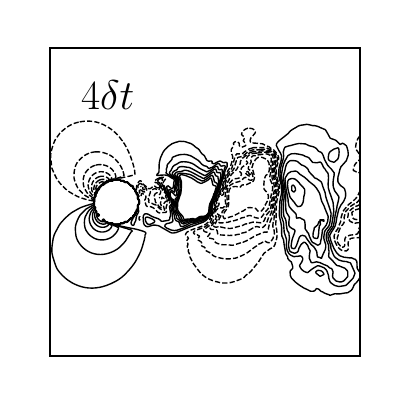}}
    \subfigure{\includegraphics[width=0.17\linewidth,trim={0.5cm 0.5cm 0.5cm 0.5cm},clip]{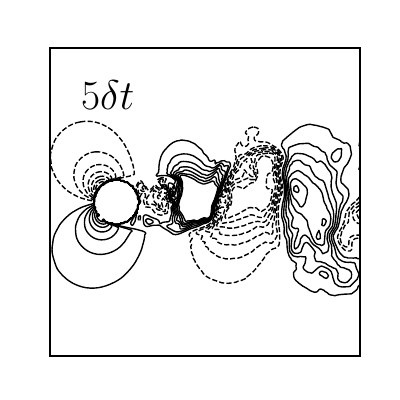}}

    \subfigure{\includegraphics[width=0.17\linewidth,trim={0.5cm 0.5cm 0.5cm 0.5cm},clip]{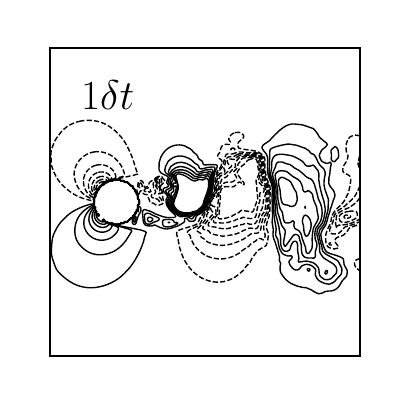}}
    \subfigure{\includegraphics[width=0.17\linewidth,trim={0.5cm 0.5cm 0.5cm 0.5cm},clip]{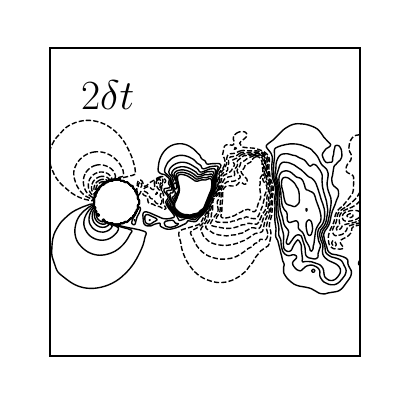}}
    \setcounter{subfigure}{2}%
    \subfigure[]{\includegraphics[width=0.17\linewidth,trim={0.5cm 0.5cm 0.5cm 0.5cm},clip]{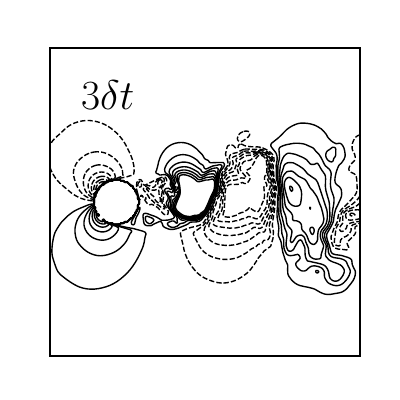}}
    \subfigure{\includegraphics[width=0.17\linewidth,trim={0.5cm 0.5cm 0.5cm 0.5cm},clip]{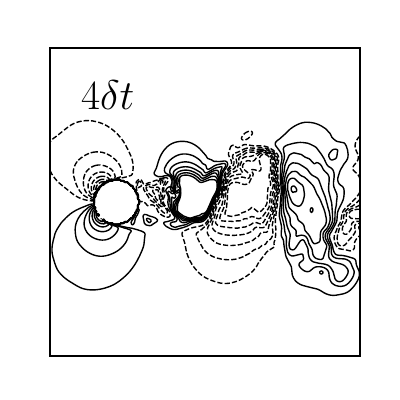}}
    \subfigure{\includegraphics[width=0.17\linewidth,trim={0.5cm 0.5cm 0.5cm 0.5cm},clip]{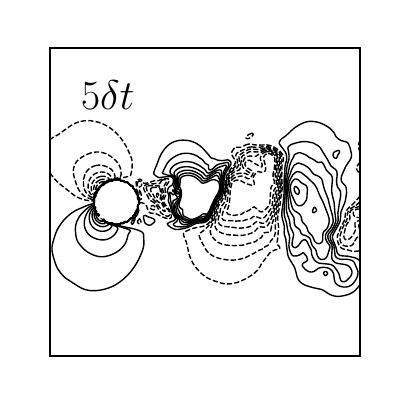}}
    \caption{
    Contour plots of the cross-stream velocity ($v/U_{\infty}$) at $Re_{D}=3900$ after $1 \delta t$, $2 \delta t$, $3 \delta t$, $4 \delta t$, and $5 \delta t$, where $1 \delta t = 20 \Delta t U_{\infty}/D = 0.1$. Flow fields at $2 \delta t$, $3 \delta t$, $4 \delta t$, and $5 \delta t$ are recursively predicted (utilizing flow fields predicted prior time-steps as parts of the input).
    (a) Input set, (b) ground truth flow fields, and (c) flow fields predicted by the GAN.
    20 contour levels from -0.5 to 1.0 are shown.
     Solid lines and dashed lines indicate positive and negative contour levels, respectively.
    }
    \label{fig:v-R1R3R4}
\end{figure}

\begin{figure}
    \centering
    \subfigure{\includegraphics[width=0.17\linewidth,trim={0.5cm 0.5cm 0.5cm 0.5cm},clip]{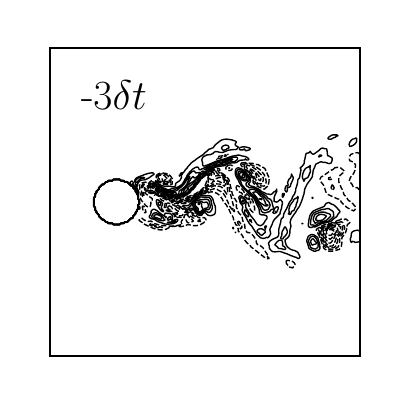}}
    \subfigure{\includegraphics[width=0.17\linewidth,trim={0.5cm 0.5cm 0.5cm 0.5cm},clip]{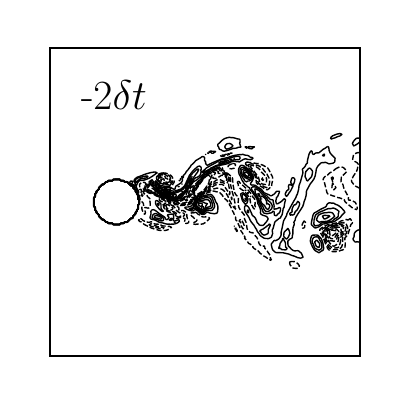}}
    \setcounter{subfigure}{0}%
    \subfigure[]{}\hspace{-1mm}
    \subfigure{\includegraphics[width=0.17\linewidth,trim={0.5cm 0.5cm 0.5cm 0.5cm},clip]{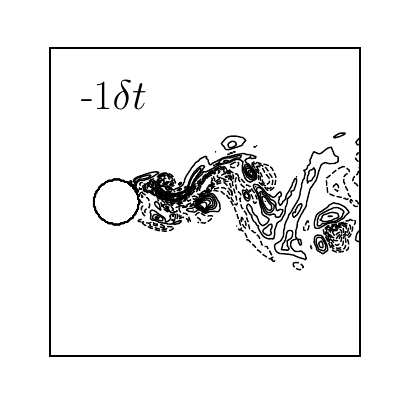}}
    \subfigure{\includegraphics[width=0.17\linewidth,trim={0.5cm 0.5cm 0.5cm 0.5cm},clip]{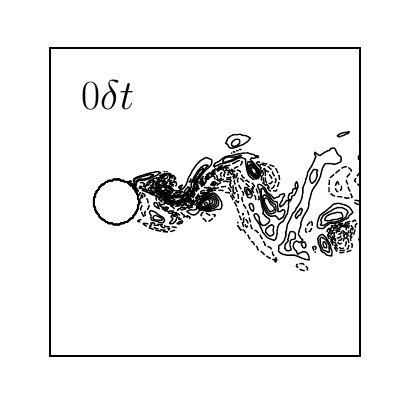}}

    \subfigure{\includegraphics[width=0.17\linewidth,trim={0.5cm 0.5cm 0.5cm 0.5cm},clip]{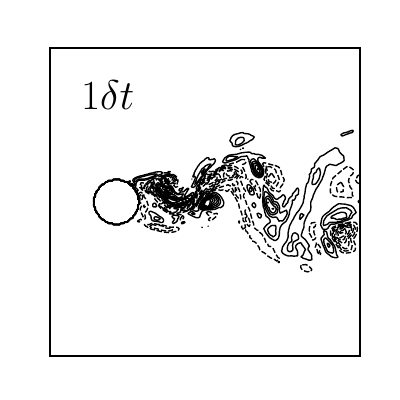}}
    \subfigure{\includegraphics[width=0.17\linewidth,trim={0.5cm 0.5cm 0.5cm 0.5cm},clip]{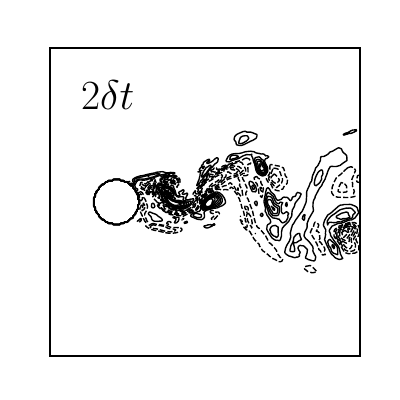}}
    \setcounter{subfigure}{1}%
    \subfigure[]{\includegraphics[width=0.17\linewidth,trim={0.5cm 0.5cm 0.5cm 0.5cm},clip]{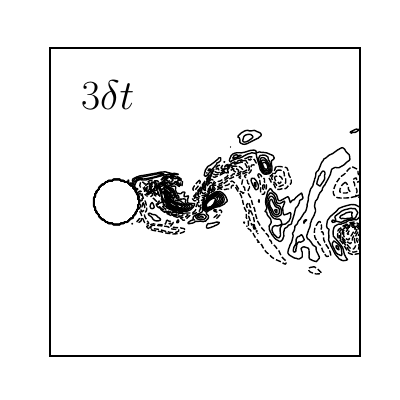}}
    \subfigure{\includegraphics[width=0.17\linewidth,trim={0.5cm 0.5cm 0.5cm 0.5cm},clip]{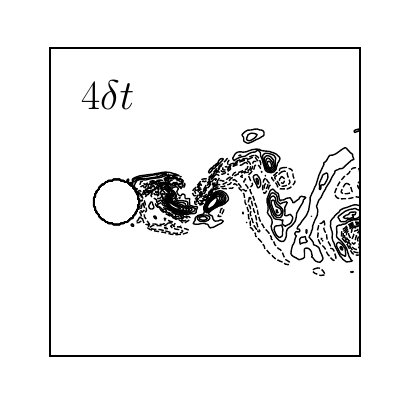}}
    \subfigure{\includegraphics[width=0.17\linewidth,trim={0.5cm 0.5cm 0.5cm 0.5cm},clip]{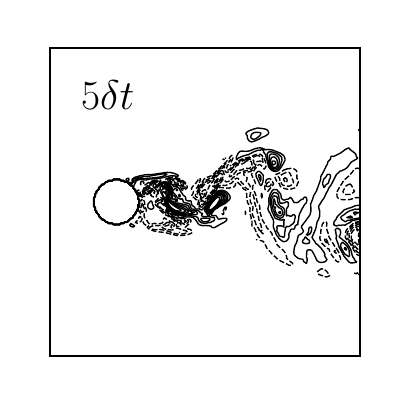}}

    \subfigure{\includegraphics[width=0.17\linewidth,trim={0.5cm 0.5cm 0.5cm 0.5cm},clip]{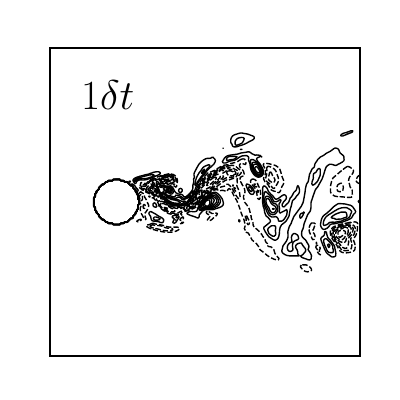}}
    \subfigure{\includegraphics[width=0.17\linewidth,trim={0.5cm 0.5cm 0.5cm 0.5cm},clip]{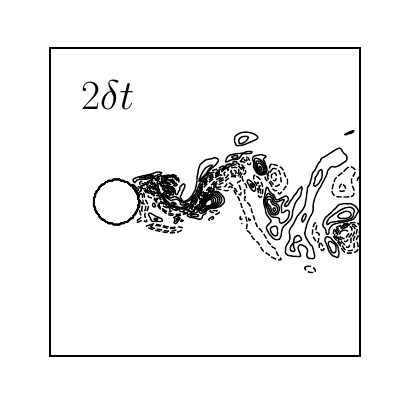}}
    \setcounter{subfigure}{2}%
    \subfigure[]{\includegraphics[width=0.17\linewidth,trim={0.5cm 0.5cm 0.5cm 0.5cm},clip]{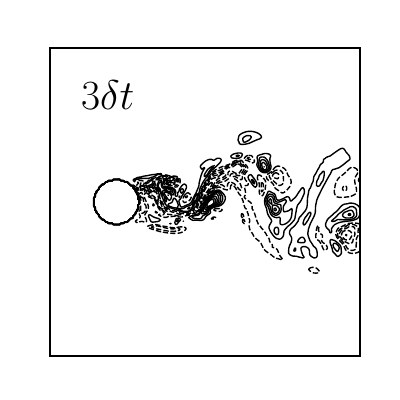}}
    \subfigure{\includegraphics[width=0.17\linewidth,trim={0.5cm 0.5cm 0.5cm 0.5cm},clip]{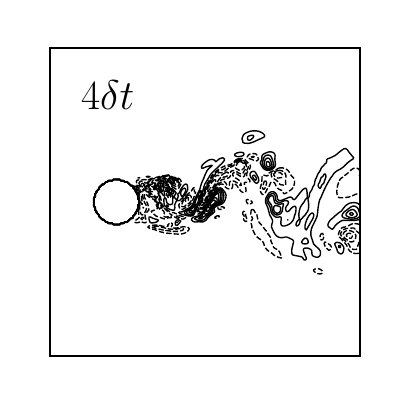}}
    \subfigure{\includegraphics[width=0.17\linewidth,trim={0.5cm 0.5cm 0.5cm 0.5cm},clip]{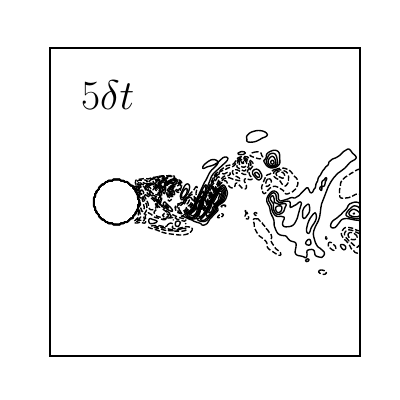}}
    \caption{
    Contour plots of the spanwise velocity ($w/U_{\infty}$) at $Re_{D}=3900$ after $1 \delta t$, $2 \delta t$, $3 \delta t$, $4 \delta t$, and $5 \delta t$, where $1 \delta t = 20 \Delta t U_{\infty}/D = 0.1$. Flow fields at $2 \delta t$, $3 \delta t$, $4 \delta t$, and $5 \delta t$ are recursively predicted (utilizing flow fields predicted prior time-steps as parts of the input).
    (a) Input set, (b) ground truth flow fields, and (c) flow fields predicted by the GAN.
    20 contour levels from -0.5 to 1.0 are shown.
     Solid lines and dashed lines indicate positive and negative contour levels, respectively.
    }
    \label{fig:w-R1R3R4}
\end{figure}

\begin{figure}
    \centering
    \subfigure{\includegraphics[width=0.17\linewidth,trim={0.5cm 0.5cm 0.5cm 0.5cm},clip]{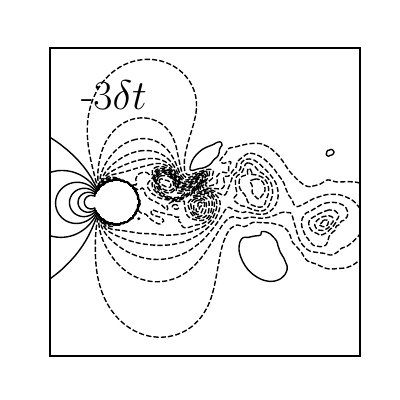}}
    \subfigure{\includegraphics[width=0.17\linewidth,trim={0.5cm 0.5cm 0.5cm 0.5cm},clip]{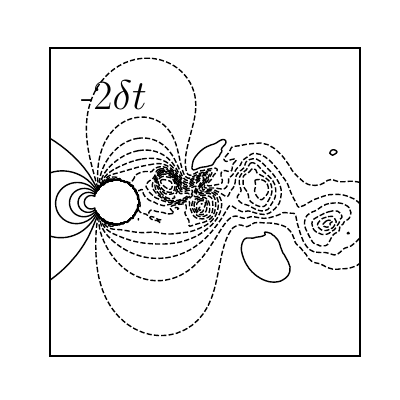}}
    \setcounter{subfigure}{0}%
    \subfigure[]{}\hspace{-1mm}
    \subfigure{\includegraphics[width=0.17\linewidth,trim={0.5cm 0.5cm 0.5cm 0.5cm},clip]{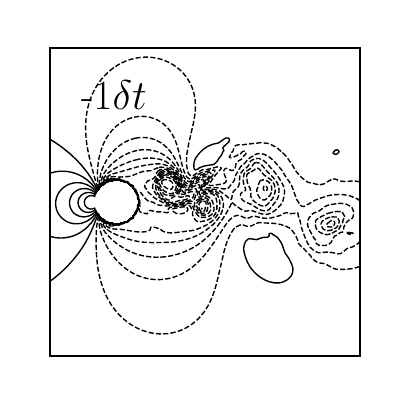}}
    \subfigure{\includegraphics[width=0.17\linewidth,trim={0.5cm 0.5cm 0.5cm 0.5cm},clip]{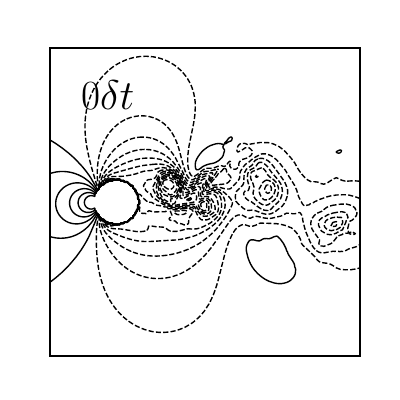}}

    \subfigure{\includegraphics[width=0.17\linewidth,trim={0.5cm 0.5cm 0.5cm 0.5cm},clip]{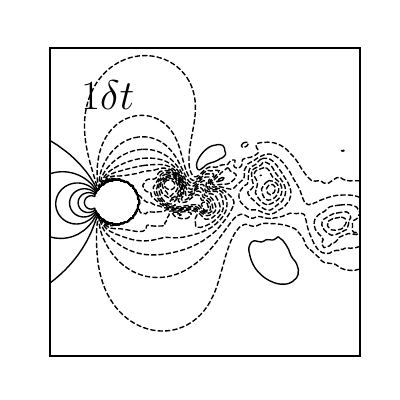}}
    \subfigure{\includegraphics[width=0.17\linewidth,trim={0.5cm 0.5cm 0.5cm 0.5cm},clip]{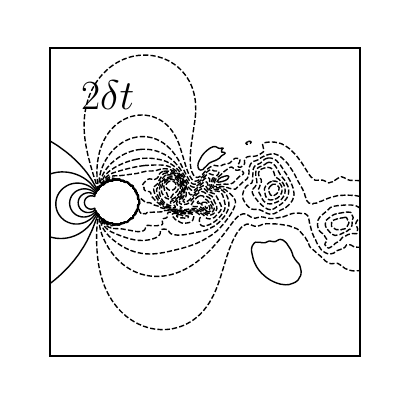}}
    \setcounter{subfigure}{1}%
    \subfigure[]{\includegraphics[width=0.17\linewidth,trim={0.5cm 0.5cm 0.5cm 0.5cm},clip]{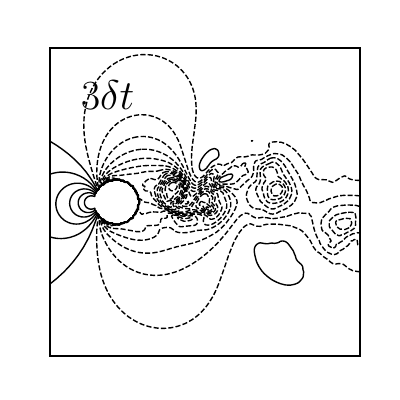}}
    \subfigure{\includegraphics[width=0.17\linewidth,trim={0.5cm 0.5cm 0.5cm 0.5cm},clip]{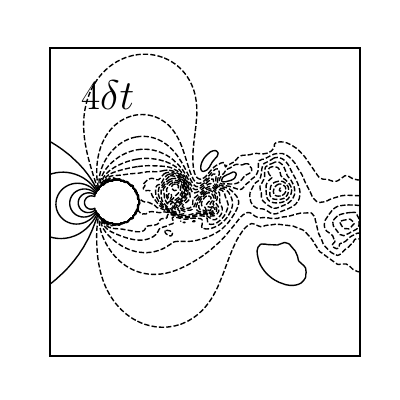}}
    \subfigure{\includegraphics[width=0.17\linewidth,trim={0.5cm 0.5cm 0.5cm 0.5cm},clip]{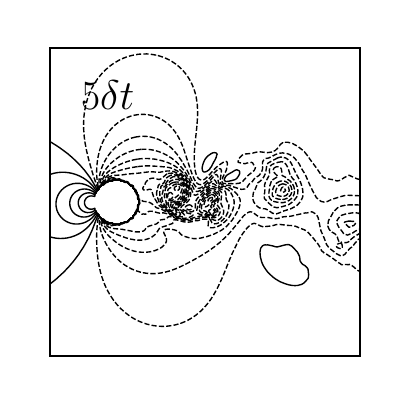}}

    \subfigure{\includegraphics[width=0.17\linewidth,trim={0.5cm 0.5cm 0.5cm 0.5cm},clip]{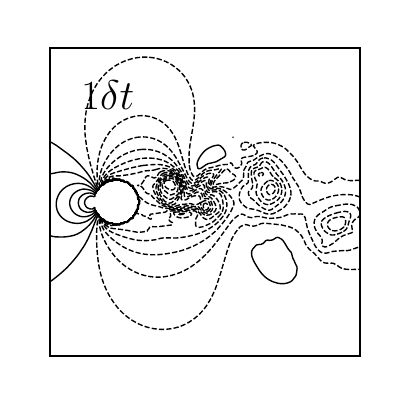}}
    \subfigure{\includegraphics[width=0.17\linewidth,trim={0.5cm 0.5cm 0.5cm 0.5cm},clip]{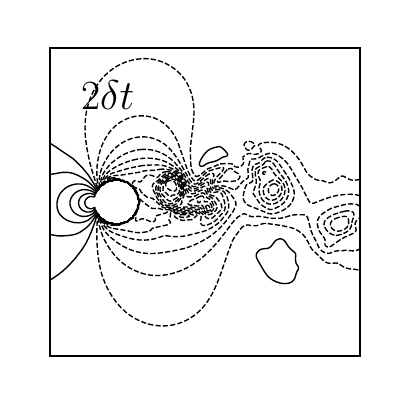}}
    \setcounter{subfigure}{2}%
    \subfigure[]{\includegraphics[width=0.17\linewidth,trim={0.5cm 0.5cm 0.5cm 0.5cm},clip]{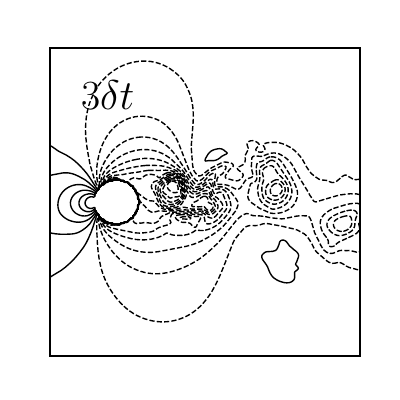}}
    \subfigure{\includegraphics[width=0.17\linewidth,trim={0.5cm 0.5cm 0.5cm 0.5cm},clip]{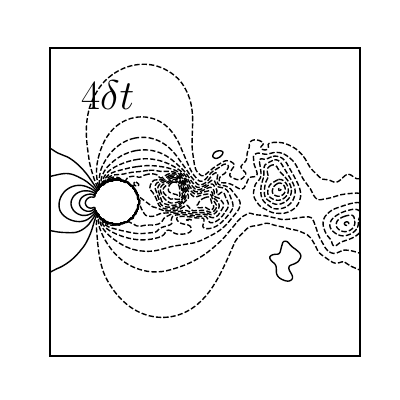}}
    \subfigure{\includegraphics[width=0.17\linewidth,trim={0.5cm 0.5cm 0.5cm 0.5cm},clip]{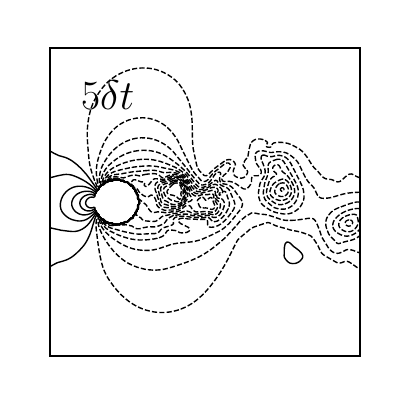}}
    \caption{
    Contour plots of the pressure ($p/\rho U^{2}_{\infty}$) at $Re_{D}=3900$ after $1 \delta t$, $2 \delta t$, $3 \delta t$, $4 \delta t$, and $5 \delta t$, where $1 \delta t = 20 \Delta t U_{\infty}/D = 0.1$. Flow fields at $2 \delta t$, $3 \delta t$, $4 \delta t$, and $5 \delta t$ are recursively predicted (utilizing flow fields predicted prior time-steps as parts of the input).
    (a) Input set, (b) ground truth flow fields, and (c) flow fields predicted by the GAN.
    20 contour levels from -1.0 to 0.4 are shown.
     Solid lines and dashed lines indicate positive and negative contour levels, respectively.
    }
    \label{fig:p-R1R3R4}
\end{figure}

Contour plots of the cross-stream velocity, the spanwise velocity, and the pressure predicted by the GAN at $Re_{D}=3900$ with prediction-step intervals of $1\delta t$ are shown in figures~\ref{fig:u-R1R3R4}-\ref{fig:p-R1R3R4}.

\section{Flow fields predicted by the GAN trained with a large time-step interval}\label{appendix:fields}
\begin{figure}
    \centering
    \subfigure{\includegraphics[width=0.17\linewidth,trim={0.5cm 0.5cm 0.5cm 0.5cm},clip]{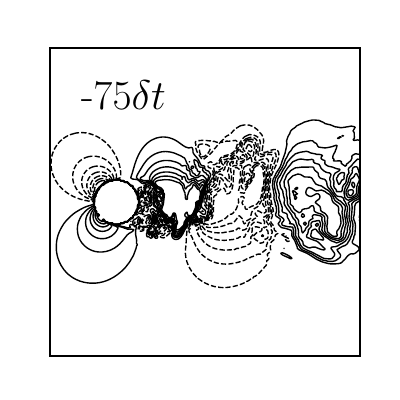}}
    \subfigure{\includegraphics[width=0.17\linewidth,trim={0.5cm 0.5cm 0.5cm 0.5cm},clip]{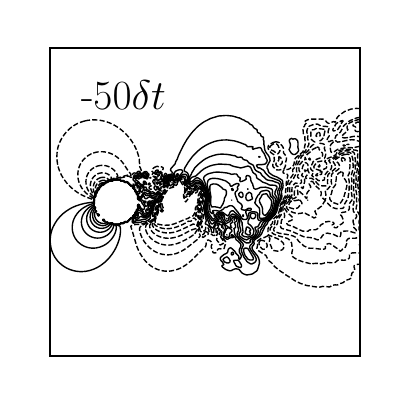}}
    \setcounter{subfigure}{0}%
    \subfigure[]{}\hspace{-1mm}
    \subfigure{\includegraphics[width=0.17\linewidth,trim={0.5cm 0.5cm 0.5cm 0.5cm},clip]{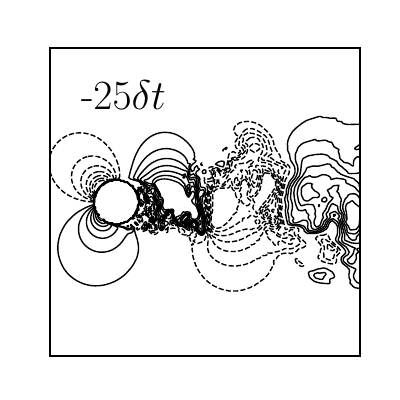}}
    \subfigure{\includegraphics[width=0.17\linewidth,trim={0.5cm 0.5cm 0.5cm 0.5cm},clip]{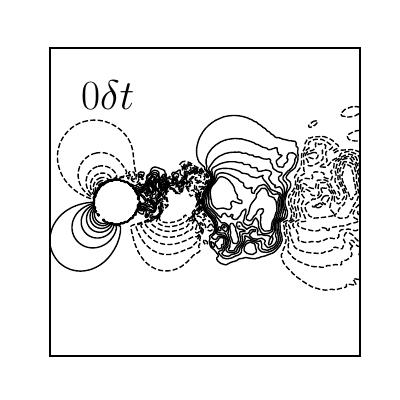}}

    \subfigure{\includegraphics[width=0.17\linewidth,trim={0.5cm 0.5cm 0.5cm 0.5cm},clip]{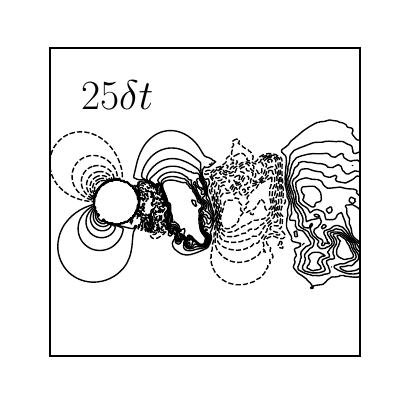}}
    \subfigure{\includegraphics[width=0.17\linewidth,trim={0.5cm 0.5cm 0.5cm 0.5cm},clip]{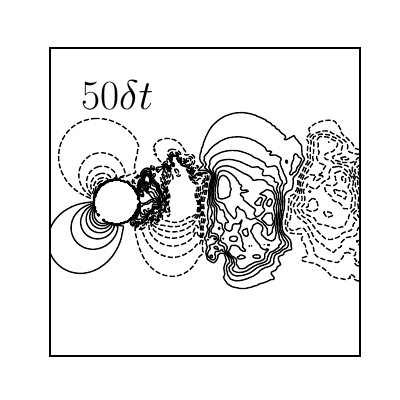}}
    \setcounter{subfigure}{1}%
    \subfigure[]{\includegraphics[width=0.17\linewidth,trim={0.5cm 0.5cm 0.5cm 0.5cm},clip]{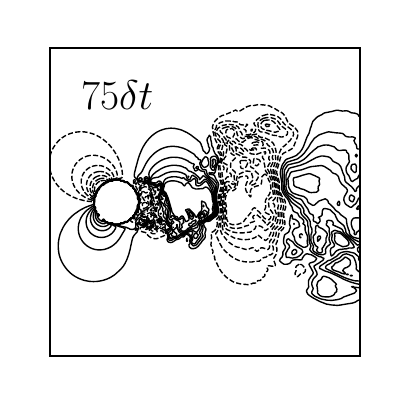}}
    \subfigure{\includegraphics[width=0.17\linewidth,trim={0.5cm 0.5cm 0.5cm 0.5cm},clip]{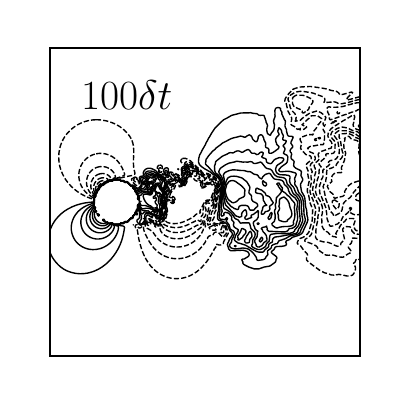}}
    \subfigure{\includegraphics[width=0.17\linewidth,trim={0.5cm 0.5cm 0.5cm 0.5cm},clip]{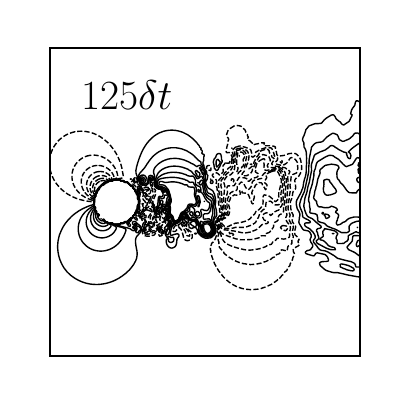}}

    \subfigure{\includegraphics[width=0.17\linewidth,trim={0.5cm 0.5cm 0.5cm 0.5cm},clip]{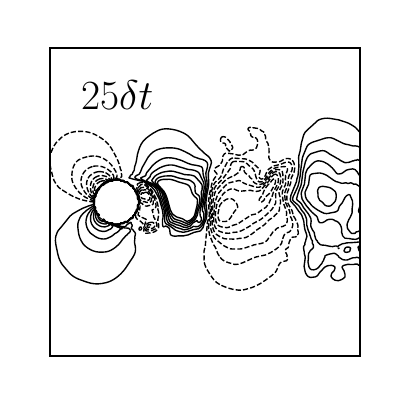}}
    \subfigure{\includegraphics[width=0.17\linewidth,trim={0.5cm 0.5cm 0.5cm 0.5cm},clip]{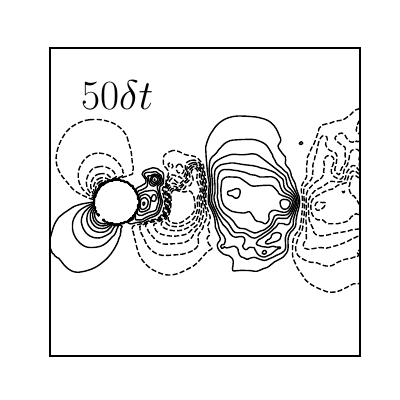}}
    \setcounter{subfigure}{2}%
    \subfigure[]{\includegraphics[width=0.17\linewidth,trim={0.5cm 0.5cm 0.5cm 0.5cm},clip]{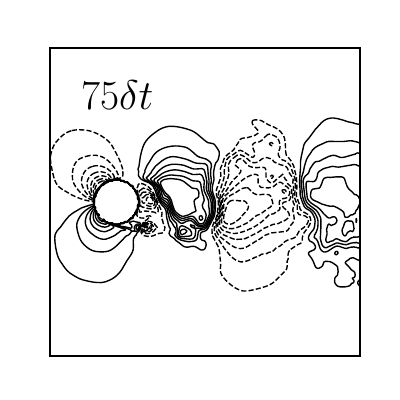}}
    \subfigure{\includegraphics[width=0.17\linewidth,trim={0.5cm 0.5cm 0.5cm 0.5cm},clip]{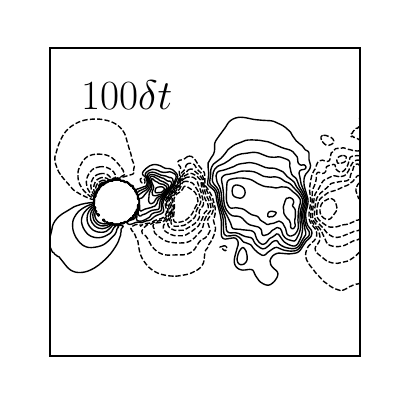}}
    \subfigure{\includegraphics[width=0.17\linewidth,trim={0.5cm 0.5cm 0.5cm 0.5cm},clip]{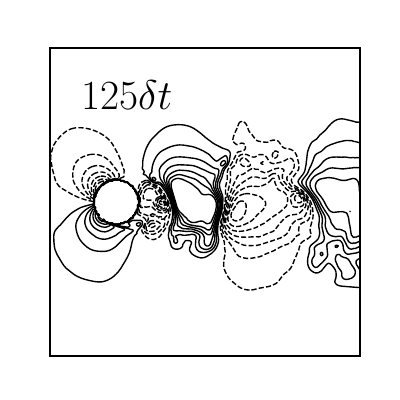}}
    \caption{
    Contour plots of the cross-stream velocity ($v/ U_{\infty}$) at $Re_{D}=3900$ after $25\delta t$, $50 \delta t$, $75 \delta t$, $100 \delta t$, and $125 \delta t$, where $1 \delta t = 20 \Delta t U_{\infty}/D = 0.1$. Flow fields at $50 \delta t$, $75 \delta t$, $100 \delta t$, and $125 \delta t$ are recursively predicted (utilizing flow fields predicted prior time-steps as parts of the input).
    (a) Input set, (b) ground truth flow fields, and (c) flow fields predicted by the GAN.
    14 contour levels from -0.7 to 0.7 are shown.
     Solid lines and dashed lines indicate positive and negative contour levels, respectively.
    }
    \label{fig:v-T25}
\end{figure}

\begin{figure}
    \centering
    \subfigure{\includegraphics[width=0.17\linewidth,trim={0.5cm 0.5cm 0.5cm 0.5cm},clip]{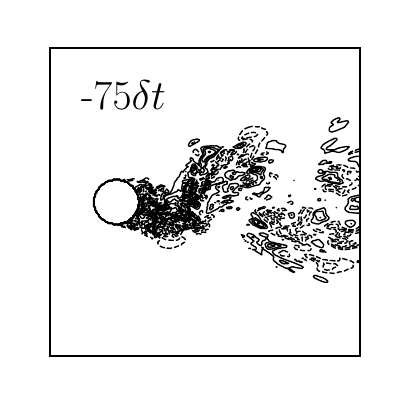}}
    \subfigure{\includegraphics[width=0.17\linewidth,trim={0.5cm 0.5cm 0.5cm 0.5cm},clip]{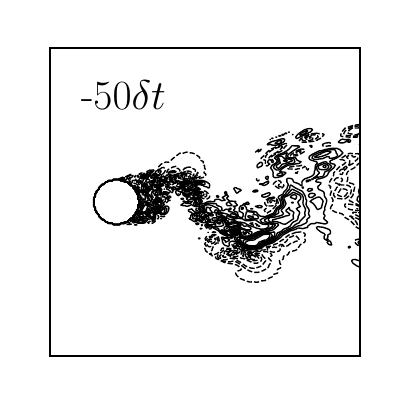}}
    \setcounter{subfigure}{0}%
    \subfigure[]{}\hspace{-1mm}
    \subfigure{\includegraphics[width=0.17\linewidth,trim={0.5cm 0.5cm 0.5cm 0.5cm},clip]{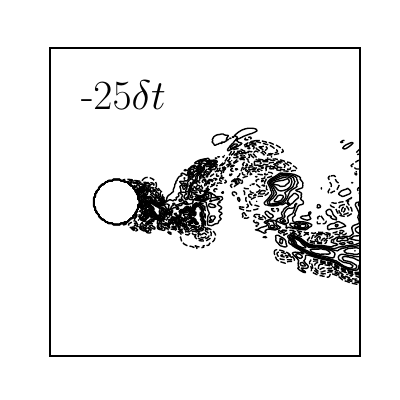}}
    \subfigure{\includegraphics[width=0.17\linewidth,trim={0.5cm 0.5cm 0.5cm 0.5cm},clip]{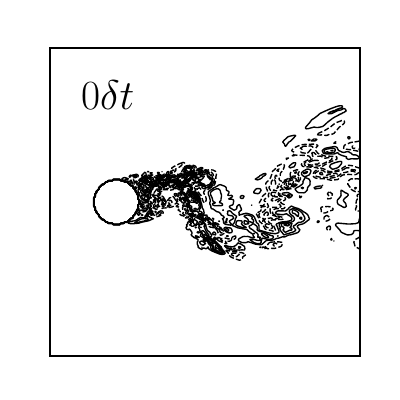}}

    \subfigure{\includegraphics[width=0.17\linewidth,trim={0.5cm 0.5cm 0.5cm 0.5cm},clip]{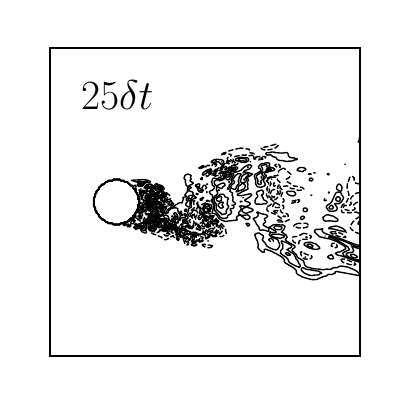}}
    \subfigure{\includegraphics[width=0.17\linewidth,trim={0.5cm 0.5cm 0.5cm 0.5cm},clip]{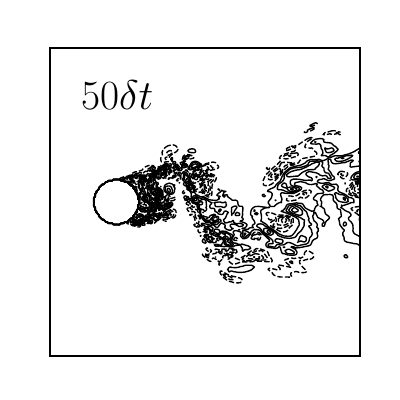}}
    \setcounter{subfigure}{1}%
    \subfigure[]{\includegraphics[width=0.17\linewidth,trim={0.5cm 0.5cm 0.5cm 0.5cm},clip]{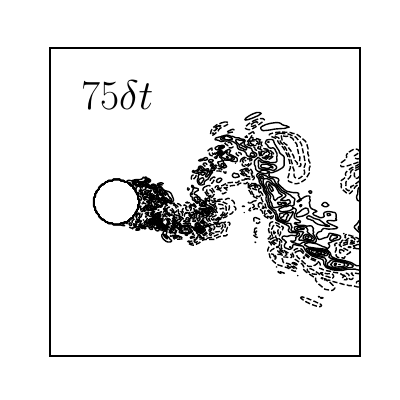}}
    \subfigure{\includegraphics[width=0.17\linewidth,trim={0.5cm 0.5cm 0.5cm 0.5cm},clip]{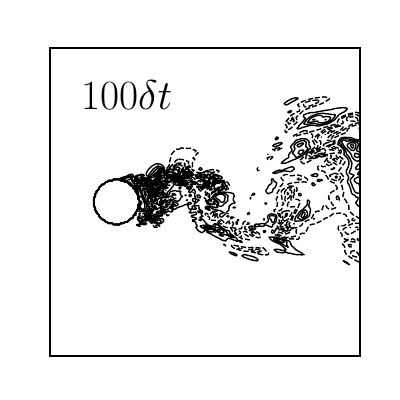}}
    \subfigure{\includegraphics[width=0.17\linewidth,trim={0.5cm 0.5cm 0.5cm 0.5cm},clip]{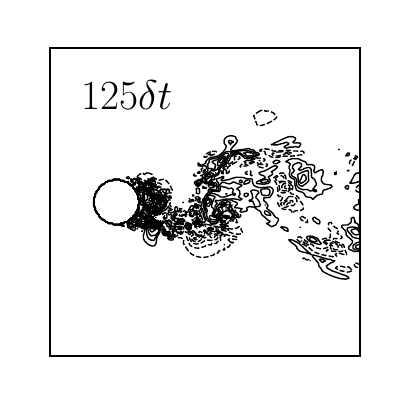}}

    \subfigure{\includegraphics[width=0.17\linewidth,trim={0.5cm 0.5cm 0.5cm 0.5cm},clip]{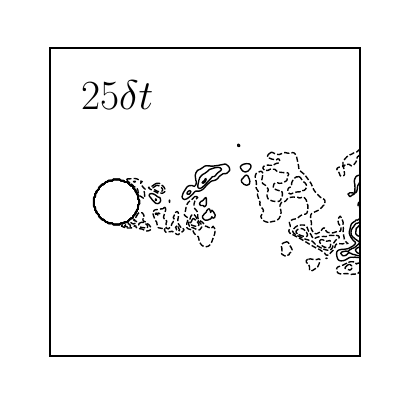}}
    \subfigure{\includegraphics[width=0.17\linewidth,trim={0.5cm 0.5cm 0.5cm 0.5cm},clip]{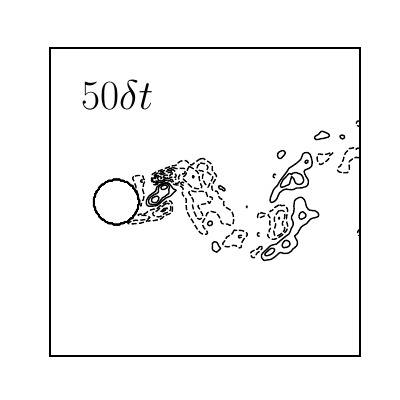}}
    \setcounter{subfigure}{2}%
    \subfigure[]{\includegraphics[width=0.17\linewidth,trim={0.5cm 0.5cm 0.5cm 0.5cm},clip]{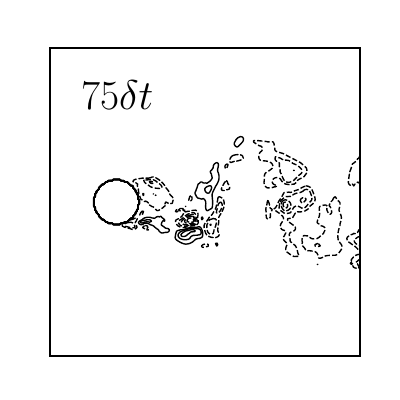}}
    \subfigure{\includegraphics[width=0.17\linewidth,trim={0.5cm 0.5cm 0.5cm 0.5cm},clip]{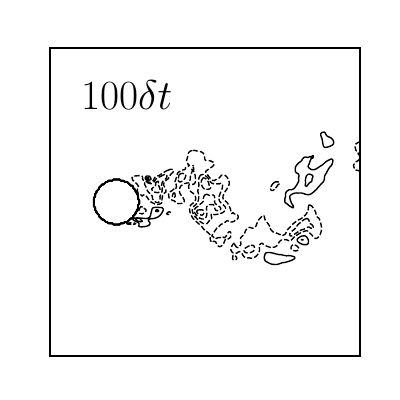}}
    \subfigure{\includegraphics[width=0.17\linewidth,trim={0.5cm 0.5cm 0.5cm 0.5cm},clip]{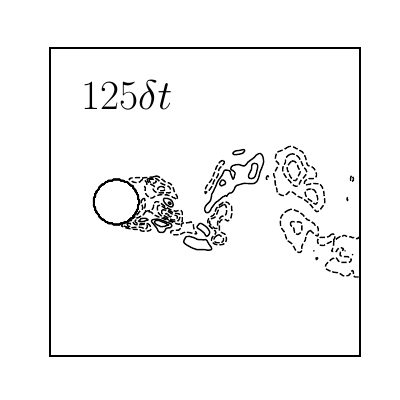}}
    \caption{
    Contour plots of the spanwise velocity ($w/ U_{\infty}$) at $Re_{D}=3900$ after $25\delta t$, $50 \delta t$, $75 \delta t$, $100 \delta t$, and $125 \delta t$, where $1 \delta t = 20 \Delta t U_{\infty}/D = 0.1$. Flow fields at $50 \delta t$, $75 \delta t$, $100 \delta t$, and $125 \delta t$ are recursively predicted (utilizing flow fields predicted prior time-steps as parts of the input).
    (a) Input set, (b) ground truth flow fields, and (c) flow fields predicted by the GAN.
    14 contour levels from -0.5 to 0.5 are shown.
     Solid lines and dashed lines indicate positive and negative contour levels, respectively.
    }
    \label{fig:w-T25}
\end{figure}

\begin{figure}
    \centering
    \subfigure{\includegraphics[width=0.17\linewidth,trim={0.5cm 0.5cm 0.5cm 0.5cm},clip]{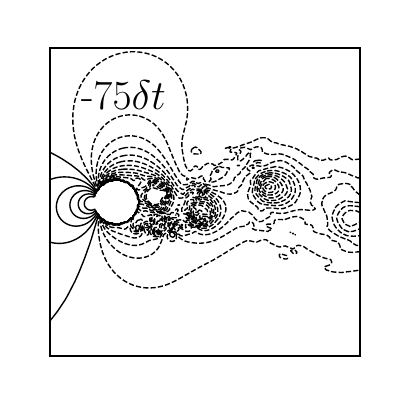}}
    \subfigure{\includegraphics[width=0.17\linewidth,trim={0.5cm 0.5cm 0.5cm 0.5cm},clip]{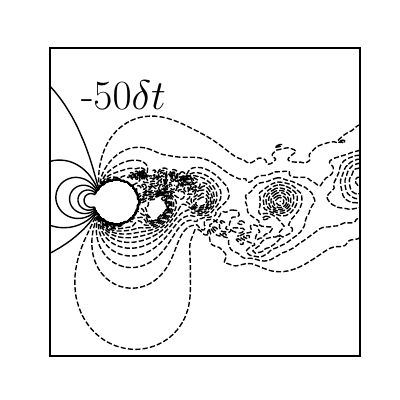}}
    \setcounter{subfigure}{0}%
    \subfigure[]{}\hspace{-1mm}
    \subfigure{\includegraphics[width=0.17\linewidth,trim={0.5cm 0.5cm 0.5cm 0.5cm},clip]{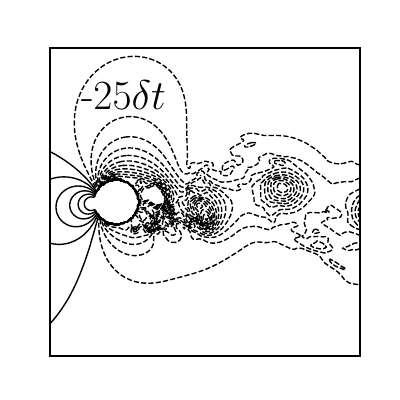}}
    \subfigure{\includegraphics[width=0.17\linewidth,trim={0.5cm 0.5cm 0.5cm 0.5cm},clip]{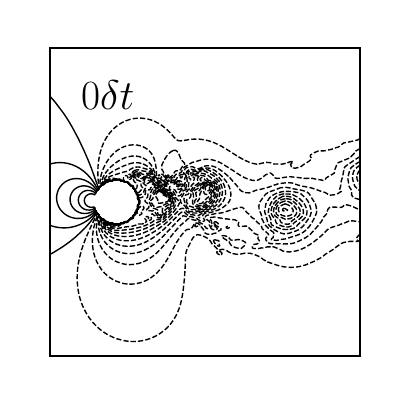}}

    \subfigure{\includegraphics[width=0.17\linewidth,trim={0.5cm 0.5cm 0.5cm 0.5cm},clip]{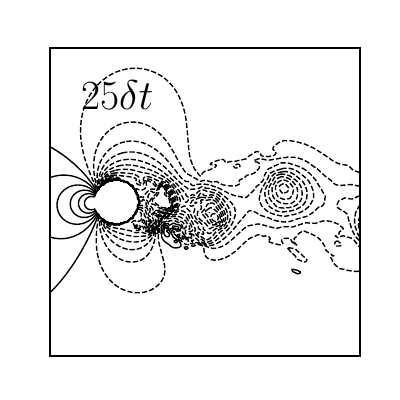}}
    \subfigure{\includegraphics[width=0.17\linewidth,trim={0.5cm 0.5cm 0.5cm 0.5cm},clip]{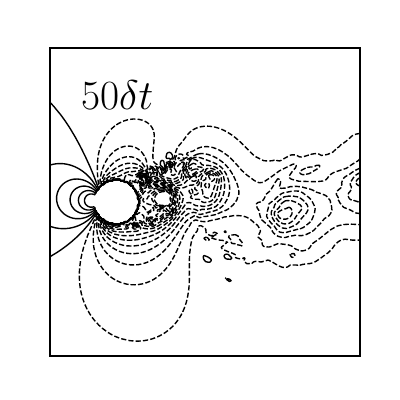}}
    \setcounter{subfigure}{1}%
    \subfigure[]{\includegraphics[width=0.17\linewidth,trim={0.5cm 0.5cm 0.5cm 0.5cm},clip]{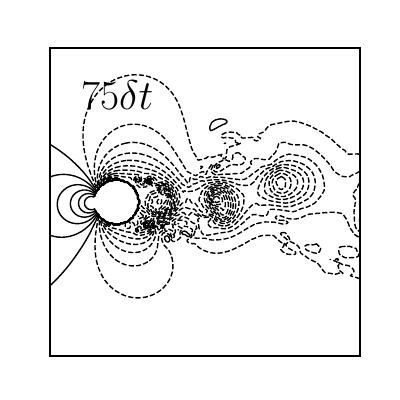}}
    \subfigure{\includegraphics[width=0.17\linewidth,trim={0.5cm 0.5cm 0.5cm 0.5cm},clip]{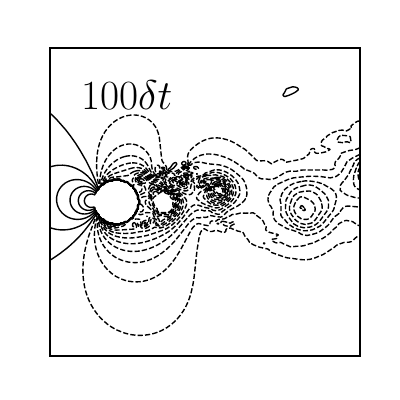}}
    \subfigure{\includegraphics[width=0.17\linewidth,trim={0.5cm 0.5cm 0.5cm 0.5cm},clip]{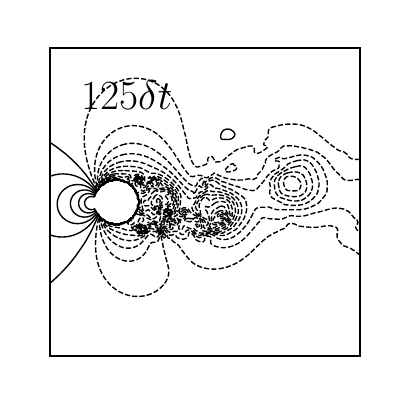}}

    \subfigure{\includegraphics[width=0.17\linewidth,trim={0.5cm 0.5cm 0.5cm 0.5cm},clip]{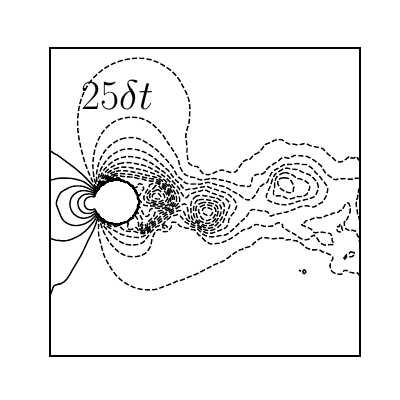}}
    \subfigure{\includegraphics[width=0.17\linewidth,trim={0.5cm 0.5cm 0.5cm 0.5cm},clip]{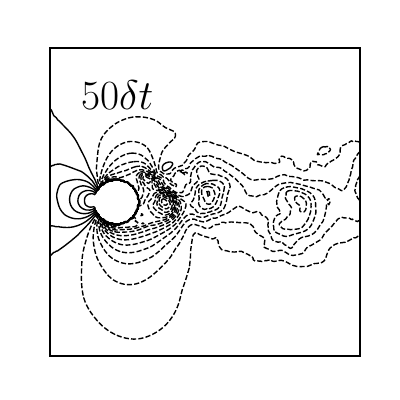}}
    \setcounter{subfigure}{2}%
    \subfigure[]{\includegraphics[width=0.17\linewidth,trim={0.5cm 0.5cm 0.5cm 0.5cm},clip]{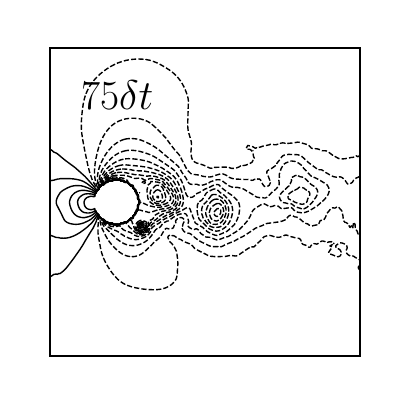}}
    \subfigure{\includegraphics[width=0.17\linewidth,trim={0.5cm 0.5cm 0.5cm 0.5cm},clip]{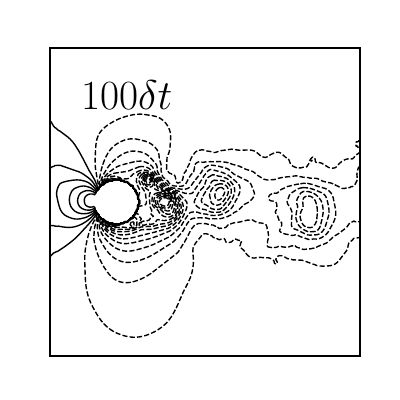}}
    \subfigure{\includegraphics[width=0.17\linewidth,trim={0.5cm 0.5cm 0.5cm 0.5cm},clip]{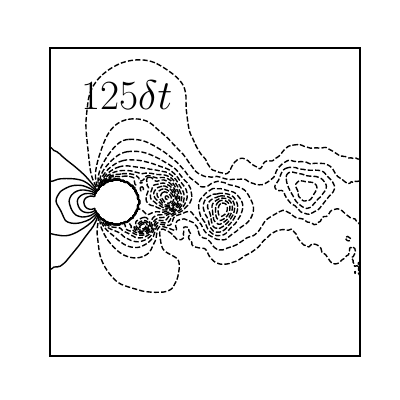}}
    \caption{
    Contour plots of the pressure ($p/\rho U^{2}_{\infty}$) at $Re_{D}=3900$ after $25\delta t$, $50 \delta t$, $75 \delta t$, $100 \delta t$, and $125 \delta t$, where $1 \delta t = 20 \Delta t U_{\infty}/D = 0.1$. Flow fields at $50 \delta t$, $75 \delta t$, $100 \delta t$, and $125 \delta t$ are recursively predicted (utilizing flow fields predicted prior time-steps as parts of the input).
    (a) Input set, (b) ground truth flow fields, and (c) flow fields predicted by the GAN.
    20 contour levels from -1.0 to 0.4 are shown.
     Solid lines and dashed lines indicate positive and negative contour levels, respectively.
    }
    \label{fig:p-T25}
\end{figure}

Contour plots of the cross-stream velocity, the spanwise velocity, and the pressure predicted by the GAN at $Re_{D}=3900$ with prediction-step intervals of $25\delta t$ are shown in figures~\ref{fig:v-T25}-\ref{fig:p-T25} (see figure~\ref{fig:u-T25} for contour plots of the streamwise velocity).

\end{document}